\documentstyle[12pt]{article}
\input epsf.sty
\begin{document}

\newcommand{\zcc}{$Z^0 \rightarrow c\bar{c}$}
\newcommand{\zbb}{$Z^0 \rightarrow b\bar{b}$}
\newcommand{\zuds}{$Z^0 \rightarrow u\bar{u},d\bar{d},s\bar{s}$}
\newcommand{\asz}{\alpha_s(M_Z^2)}
\newcommand{\llb}{$\Lambda^0/\bar{\Lambda}^0$}
\newcommand{\kkb}{$K^{*0}/\bar{K}^{*0}$}
\newcommand{\pnb}{differential cross section}

\thispagestyle{empty}

\renewcommand{\thefootnote}{\fnsymbol{footnote}}

\begin{flushright}
{\small
SLAC--PUB--7766\\
April 1998\\}
\end{flushright}

\vspace{0.5cm}

\begin{center}
{\large \bf  PRODUCTION OF $\pi^+$, $K^+$, $K^0$, $K^{*0}$, $\phi$,
p and $\Lambda^0$ IN HADRONIC $Z^0$ DECAYS\footnote{Work supported by
Department of Energy contracts:
DE-FG02-91ER40676, DE-FG03-91ER40618, DE-FG03-92ER40689, DE-FG03-93ER40788,
DE-FG02-91ER40672, DE-FG02-91ER40677, DE-AC03- 76SF00098, DE-FG02-92ER40715,
DE-FC02-94ER40818, DE-FG03-96ER40969, DE-AC03-76SF00515, DE-FG05-91ER40627,
DE-FG02-95ER40896, DE-FG02-92ER40704;
  National Science Foundation grants:
PHY-91-13428, PHY-89-21320, PHY-92-04239, PHY-95-10439, PHY-88-19316,
PHY-92-03212;
  The UK Particle Physics and Astronomy Research Council;
  The Istituto Nazionale di Fisica Nucleare of Italy;
  The Japan-US Cooperative Research Project on High Energy Physics;
  The Korea Science and Engineering Foundation.}}

\vspace{1.0cm}

 {\bf The SLD Collaboration$^{**}$}\\
Stanford Linear Accelerator Center,
Stanford University, Stanford, CA~94309

\end{center}

\vfill

\begin{center}
{\bf\large   
Abstract }
\end{center}

We have measured the differential production cross sections
as a function of scaled momentum $x_p=2p/E_{cm}$
of the identified hadron species
$\pi^+$, $K^+$, $K^0$, $K^{*0}$, $\phi$, p, $\Lambda^0$, and of the
corresponding antihadron species in inclusive hadronic $Z^0$ decays, as well as
separately for $Z^0$ decays into light ($u$, $d$, $s$), $c$ and $b$ flavors.
Clear flavor dependences are observed, consistent with expectations based upon
previously measured production and decay properties of heavy hadrons.
These results were used to test the QCD predictions of
Gribov and Lipatov, the predictions of QCD in the Modified Leading Logarithm
Approximation with the ansatz of Local Parton-Hadron Duality,
and the predictions of three fragmentation models.
Ratios of production of different hadron species were also measured as a
function of $x_p$
and were used to study the suppression of strange meson,
strange and non-strange baryon, and vector meson production in the jet
fragmentation process.
The light-flavor results provide improved tests of the above predictions,
as they remove the contribution of heavy hadron production and
decay from that of the rest of the fragmentation process.
In addition we have compared hadron and antihadron production as a function of
$x_p$ in light quark (as opposed to antiquark) jets.
Differences are observed at high $x_p$, providing direct evidence that
higher-momentum hadrons are more likely to contain a primary quark or
antiquark.
The differences for pseudoscalar and vector kaons provide new measurements of
strangeness suppression for high-$x_p$ fragmentation products.

\vfill

\begin{center}

Submitted to {\it Phys. Rev. D}\\

\end{center}

\vfill\eject

\pagestyle{plain}

\section{Introduction}

The production of jets of hadrons from hard partons produced in high
energy collisions is believed to proceed in three stages.
Considering the process $e^+e^- \rightarrow q\bar{q}$,
the first stage involves the radiation of gluons from the primary quark and
antiquark,
which in turn may radiate gluons or split into $q\bar{q}$ pairs until their
virtuality approaches the hadron mass scale.
This process is in principle calculable in perturbative QCD, and three
approaches have been taken so far:
i) differential cross sections have been calculated \cite{ert} for the
production of up to 4 partons to second order in the strong coupling $\alpha_s$,
and leading order calculations have been performed recently for as many as 6
partons (see e.g. \cite{moretti});
ii) certain parton distributions have been calculated 
to all orders in $\alpha_s$ in the Modified Leading Logarithm
Approximation (MLLA)~\cite{mlla};
iii) ``parton shower" calculations \cite{nlla} have been implemented
numerically; these consist of an arbitrary number of $q\!\rightarrow \! qg$,
$g\! \rightarrow \! gg$ and $g\! \rightarrow \! q\bar{q}$ branchings,
with each branching probability determined
from QCD in the Leading Logarithm Approximation.

In the second stage these partons transform into ``primary" hadrons.  This
``fragmentation" process is not understood quantitatively and there are few
theoretical predictions
that do not explicitly involve heavy ($c$ or $b$) quarks.
Using perturbative QCD, Gribov and Lipatov have studied \cite{glip} the
fragmentation of quarks produced in $e^+e^-$ collisions in the limit of high
hadron momentum fraction $x_p=p_{hadron}/E_{beam}$,
and have related it to the proton structure function at high
$x=E_{quark}/E_{proton}$.
They predict that as $x_p \rightarrow 1$ the distribution of $x_p$ for
baryons is proportional to $(1-x_p)^3$, and that for mesons is
proportional to $(1-x_p)^2$.
Another approach is to make 
the ansatz of local parton-hadron duality (LPHD) \cite{mlla}, that
inclusive distributions of primary hadrons are the same, up to a normalization
factor, as those for partons.
Calculations using MLLA QCD, cut off at a virtual parton mass comparable with
the mass of the hadron in question, have been used in
combination with LPHD to predict that the
shape of the distribution of $\xi=\ln (1/x_p)$ for a given primary
hadron species is approximately Gaussian within about one unit of the peak,
that the shape can be approximated over a wider $\xi$ range by a
Gaussian with the addition of small distortion terms, and
that the peak position depends inversely on the hadron mass and logarithmically
on the center-of-mass (c.m.) energy.
It is desirable to test the existing calculations experimentally
and to encourage deeper theoretical understanding of the fragmentation process.

In the third stage unstable primary hadrons decay into the stable particles that
traverse particle detectors.
This stage is understood inasmuch as proper lifetimes and decay branching
ratios have been measured for many hadron species.
However, these decays complicate fundamental fragmentation
measurements because a sizable fraction of the stable particles are decay
products rather than primary hadrons, and it is typically not possible to
determine the origin of each detected hadron.
Previous measurements at $e^+e^-$ colliders (see e.g. \cite{saxon,bohrer})
indicate that decays of vector mesons, strange baryons and decuplet baryons
produce roughly two-thirds of the stable particles;
scalar mesons, tensor mesons and radially excited baryons have also been
observed \cite{bohrer}, and there are large uncertainties on their
contributions.
Ideally one would measure every possible hadron species and distinguish
primary hadrons from decay products on a statistical basis.
A body of knowledge could be assembled by
reconstructing heavier and heavier states, and subtracting their known decay
products from the measured \pnb s of lighter hadrons.

Additional complications arise in jets initiated by heavy quarks,
since the leading heavy hadrons carry a large fraction of the beam energy,
restricting that available to other primary hadrons, and
their decays produce a sizable fraction of the stable particles in the jet.
Although decays of some $B$ and $D$ hadrons have been studied inclusively,
there are large uncertainties in heavy hadron production, $B^0_s$ and heavy
baryon decay, and the suppression of gluon radiation from heavy quarks.
The removal of heavy flavor events will therefore
simplify the study of the fragmentation of light quarks into hadrons.

A particularly interesting aspect of fragmentation is the question of what
happens to the quark or antiquark that initiated the jet.
A common prejudice is that the initial quark is ``contained" as a
valence constituent of a particular hadron, and that this ``leading"
hadron has on average a higher momentum than the other hadrons in the jet.
The highly polarized electron beam delivered by the SLAC Linear Collider (SLC)
gives a unique, high purity,
unbiased tag of quark vs. antiquark jets, via the large electroweak
forward-backward quark production asymmetry at the $Z^0$ resonance.
We have previously observed \cite{lpprl} evidence for the production of leading
baryons, $K^\pm$ and \kkb\ in light-flavor jets.
The quantification of leading particle effects
could lead to methods for identifying jets of specific light flavors,
which could
have a number of applications in $ep$ and hadron-hadron collisions as well as in
$e^+e^-$ annihilations.

There are several phenomenological models of jet fragmentation, which combine
modelling of all three stages of particle production; it is
important to test their predictions.
To simulate the parton production stage,
the HERWIG \cite{herwig}, JETSET \cite{jetset74} and UCLA \cite{ucla}
event generators use a combination of first order matrix elements and a
parton shower.
To simulate the fragmentation stage, the HERWIG model splits the gluons
produced in the first stage into $q\bar{q}$ pairs, and these quarks and
antiquarks are paired up locally to form colorless clusters that decay into
the primary hadrons.
The JETSET model takes a different approach, representing the color
field between the partons by a semi-classical string,
which is broken, according to an iterative algorithm,
into several pieces that correspond to primary hadrons.
In the UCLA model, whole events are generated according to weights derived
from the phase space available to their final states and
the relevant Clebsch-Gordan coefficients.
Each of these models contains arbitrary parameters that control various
aspects of fragmentation and have been tuned to reproduce data from
$e^+e^-$ annihilations.
The JETSET model includes a large number of parameters that control, on average,
the species of primary hadron produced at each string break,
giving it the potential to model the observed properties of identified hadron
species in great detail.
In the HERWIG model, clusters are decayed into pairs of primary hadrons
according to phase space, and the relative production of different hadrons is
effectively governed by two parameters controlling the distribution of cluster
masses.
In the UCLA model, there is only one such free parameter, which
controls the degree of locality of baryon-antibaryon pair formation.

In this paper we present an analysis of $\pi^{\pm}$, $K^{\pm}$, $K^0/\bar{K}^0$,
$K^{*0}/\bar{K}^{*0}$, $\phi$, p/$\bar{\rm p}$, and $\Lambda^0/\bar{\Lambda}^0$ 
production in hadronic $Z^0$ decays collected by the SLC Large Detector (SLD).
The analysis is based upon the approximately 150,000 hadronic events obtained 
in runs of the SLC between 1993 and 1995.
We measure differential production cross sections for these seven hadron
species in an inclusive sample of hadronic $Z^0$ decays and use
the results to test the QCD predictions of Gribov and Lipatov, the
predictions of MLLA QCD$+$LPHD, and the predictions of the three
fragmentation models just described, as well as to study the suppression of
strange hadrons, baryons, and vector mesons in the fragmentation process.
We also measure these \pnb s separately in $Z^0$ decays into light flavors
($u\bar{u}$, $d\bar{d}$ and $s\bar{s}$), $c\bar{c}$ and $b\bar{b}$,
which provide improved
tests of the QCD predictions, new tests of the fragmentation models that
separate the heavy hadron production and decay modelling from that of the rest
of the fragmentation process, and cleaner measurements of strangeness, baryon
and vector-meson suppression.
In addition we update our measurements of hadron and antihadron \pnb s in light
quark jets, and use the results to make additional new tests of the
fragmentation models and to make two new measurements of strangeness
suppression at high $x_p$.

In section 2 we describe the SLD, including a detailed description of the
Cherenkov Ring Imaging Detector, which is used to identify charged hadrons.
In section 3 we describe the selection of hadronic events of different primary
flavor, using impact parameters of charged tracks measured in the Vertex
Detector, and the selection of light quark and antiquark hemispheres, using the
large production asymmetry in polar angle induced by the polarization of the
SLC electron beam.
In section 4 we describe the hadron identification analyses and present
results for flavor-inclusive events.
In section 5 we present results separately for light- (\zuds),
$c$- (\zcc) and $b$-flavor (\zbb) events.
In section 6 we use the flavor-inclusive and light-flavor results to test the
QCD predictions of Gribov and Lipatov, and of MLLA QCD$+$LPHD.
In section 7 we extract total production cross sections of each hadron species 
per hadronic event.
In section 8 we update our measurements of leading particle production in
light-flavor jets.
In section 9 we present ratios of production of pairs of hadrons, and discuss
the suppression of
strange hadrons, baryons, and vector mesons in the fragmentation process.

\section{The SLD} 

This analysis of data from the SLD~\cite{sld}
used charged tracks measured in the Central Drift Chamber (CDC)~\cite{cdc} and
silicon Vertex Detector (VXD)~\cite{vxd}, and identified in the Cherenkov Ring
Imaging Detector (CRID) \cite{crid}.
The CDC consists of 80 layers of sense wires arranged in 10 axial or stereo
superlayers between 24 and 96 cm from the beam axis.
The outermost layer covers the solid angle range $|\cos\theta | < 0.68$.
The average spatial resolution for hits attached to charged tracks is 92 $\mu$m.
Momentum measurement is provided by a uniform axial magnetic field of 0.6 T.
The VXD and CRID are described in the following subsections.

Energy deposits reconstructed in the Liquid Argon Calorimeter (LAC)~\cite{lac}
were
used in the initial hadronic event selection and in the calculation of the
event thrust~\cite{thrust} axis.
The LAC is a lead-liquid argon sampling calorimeter covering the solid angle
range $|\cos\theta | < 0.98$,
which is segmented into 33$\times$36 mrad projective towers, each 
comprising two electromagnetic sections
and two hadronic sections, for a total thickness of 2.8 interaction lengths.
The energy resolution is measured to be $\sigma = 15\% \sqrt{E}$ for
electromagnetic showers and $60\% \sqrt{E}$ for hadronic showers, where $E$ is
the energy in GeV.

\subsection{The SLD Vertex Detector} 

Flavor tagging of events for this analysis was accomplished with the original
SLD Vertex Detector~\cite{vxd},
which was composed of 480 charge-coupled devices containing a total of
120 million 22$\times$22 $\mu$m$^2$ pixels,
arranged in four concentric layers of radius between 2.9 and 4.2 cm.
The outermost layer covered the solid angle range $|\cos\theta | < 0.75$, and
the azimuthal arrangement was such that a track would always encounter one of
the two innermost layers and one of the two outermost layers;
the average number of reconstructed hits per track was 2.3.
The 3-D spatial resolution for these hits was measured to be 5.5 $\mu$m.

\begin{figure}
 \hspace*{10.0cm}   
   \epsfxsize=4.0in
   \begin{center}\mbox{\epsffile{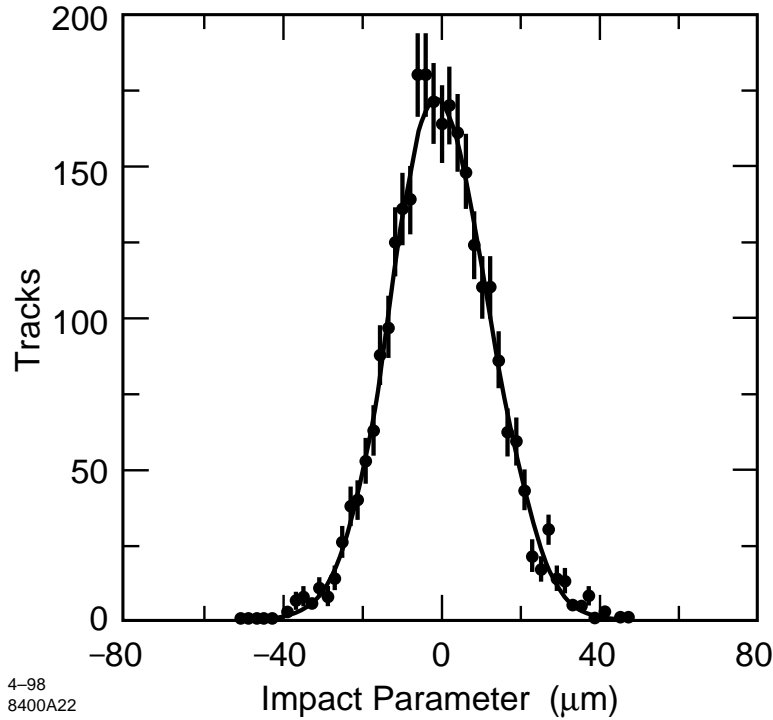}}\end{center}
  \caption{ 
 \label{vxdfig}
Distribution of transverse impact parameters of tracks in
$e^+e^- \rightarrow \mu^+ \mu^-$ events with respect to the primary interaction
point measured in hadronic events.
}
\end{figure}

Here we used only the information in the plane transverse to the beam axis.
The impact parameter resolution in this plane was measured \cite{homer} from the
distribution of miss
distances between the two tracks in $Z^0 \rightarrow \mu^+ \mu^-$ events to be
11 $\mu$m for 45.6 GeV/c muons reconstructed including at least one hit in the
VXD.
The transverse position of the primary interaction point (IP) was measured using
tracks in sets of $\sim$30 sequential hadronic $Z^0$ decays,
with a resolution measured from the distribution of impact parameters in the
statistically independent $\mu$-pair event sample (see fig. \ref{vxdfig})
of $7\pm 2 \, \mu$m.
The impact parameter resolution for lower momentum tracks was determined using
tracks in hadronic $Z^0$ decays, corrected for the contributions from decays of
heavy hadrons.
Including the uncertainty on the IP, a resolution of 
11$\oplus$70/$(p_{\perp} \sin^{3/2}\theta)$ $\mu$m was obtained,
where $p_{\perp}$ is the track momentum transverse to the beam axis in GeV/c and
$\theta$ is the polar angle of the track with respect to the beam axis.

\subsection{The SLD Cherenkov Ring Imaging Detector} 

Identification of charged tracks is accomplished with the barrel
CRID~\cite{crid}, which covers the solid angle range $|\cos\theta|<0.68$.
Through the combined use of liquid C$_6$F$_{14}$ and gaseous C$_5$F$_{12}+$N$_2$
radiators, the barrel CRID is designed to perform efficient separation of
charged pions, kaons and protons over most of the momentum range in
$e^+e^-$ annihilations at the $Z^0$, $0.3<p<46$ GeV/c.
A charged particle that passes through a radiator of refractive index $n$
with velocity $\beta$ above Cherenkov threshold, $\beta > \beta_0 = 1/n$,
emits photons at an angle $\theta_c = \cos^{-1} (1/\beta n)$ with respect to
its flight direction.
In the SLD, a charged particle exiting the CDC encounters a 1 cm thick 
liquid radiator, contained in one of 40 radiator trays.
If the momentum of the particle is above its liquid Cherenkov threshold, UV
photons are emitted in a cone about the particle flight direction.  This 1-cm
thick cone
expands over a standoff distance of $\sim$12 cm and each photon can enter 
one of 40 time projection chambers (TPCs) through an inner quartz window.

The TPCs contain a photosensitive gas, ethane with $\sim$0.1\% TMAE \cite{crid}.
The resulting single photoelectrons drift along the beam direction to a wire
chamber where the conversion point of each Cherenkov photon
is measured in three dimensions using drift time, wire address and
charge division.
These positions are used to reconstruct a Cherenkov angle with respect to the
extrapolated charged track.
Liquid rings span 2--3 TPCs in azimuth and can be split between TPCs
in the forward and backward hemispheres.

The particle may then continue through a TPC, where it ionizes the drift gas,
saturating the readout electronics, which were designed for single-electron
detection, on 2--7 anode wires and effectively deadening
$\sim$5 cm$^2$ of detection area.
Following the TPC, the particle passes through $\sim$40 cm of the gas
radiator volume.
Radiated Cherenkov photons are focussed by one of 400 spherical mirrors onto
the outer quartz window of a TPC.
Gas rings are typically 2.5 cm in radius at the TPC surface, and the
mirrors are positioned such that no ring is focussed near an edge of a TPC
or near the region saturated by its own track.
The mirror arrangement and the large size of the liquid rings make the
identification performance largely independent of the proximity of the track to
any jet axis.

The average liquid (gas) Cherenkov angle resolution was measured from the data
to be 16 (4.5) mrad, including the effects of residual misalignments of the
TPCs, radiator trays and mirrors, and track extrapolation resolution.
The local or intrinsic resolution was measured to be 13 (3.8) mrad, consistent
with the design value.
The average number of detected photons per full ring for tracks with $\beta=1$
was measured in $\mu$-pair events to be 16.1 (10.0).
For hadronic events, a set of cuts was applied to reduce backgrounds from
spurious hits and cross-talk from saturating hits, resulting in an average
of 12.8 (9.2) accepted hits per ring.
The average reconstructed Cherenkov angle for $\beta=1$ tracks
was 675 (58.6) mrad,
corresponding to an index of refraction of 1.281 (1.00172), and Cherenkov
thresholds of 0.17 (2.4) GeV/c for charged pions, 0.62 (8.4) GeV/c for kaons and
1.17 (16.0) GeV/c for protons.
This index was found to be independent of position within the CRID and the
liquid index was found to be constant in time.
Time variations in the gas index of up to $\pm0.00007$ were tracked with an
online monitor and verified in the data.

Tracks were identified using a likelihood technique \cite{davea}. 
For each of the five stable charged particle hypotheses $i=e,\mu,\pi,K$, p, 
a likelihood $L_i$ was calculated based 
upon the number of detected photoelectrons and their measured angles, 
the expected number of photons, the expected Cherenkov angle, and a
background term.
The background included the effects of overlapping Cherenkov radiation from
other
tracks in the event as well as a constant term normalized to the number of
hits in the TPC in question that were not associated with any track.
Particle separation was based upon differences between
logarithms of these likelihoods, ${\cal L}_i = \ln L_i$.

\begin{figure}
 \hspace*{10.0cm}   
   \epsfxsize=4.4in
   \begin{center}\mbox{\epsffile{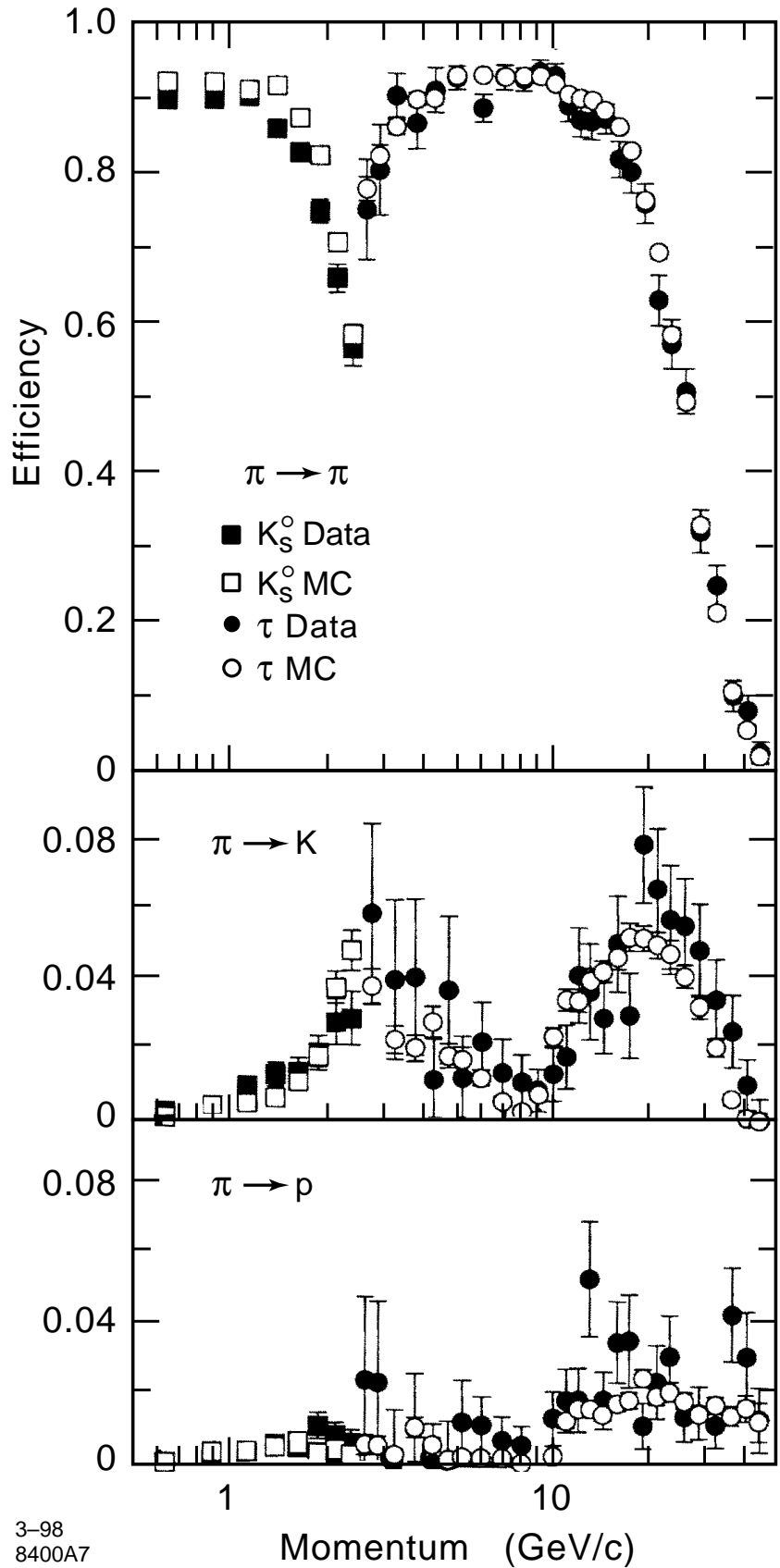}}\end{center}
  \caption{ 
 \label{effmlk}
Efficiencies for selected tracks from $K_s^0$ (squares) and $\tau$ (circles)
decays to be identified as each hadron species in the CRID.
The solid symbols represent the data and the open symbols the simulation.
}
\end{figure}

\begin{figure}
 \hspace*{0.5cm}   
   \epsfxsize=6.5in
   \begin{center}\mbox{\epsffile{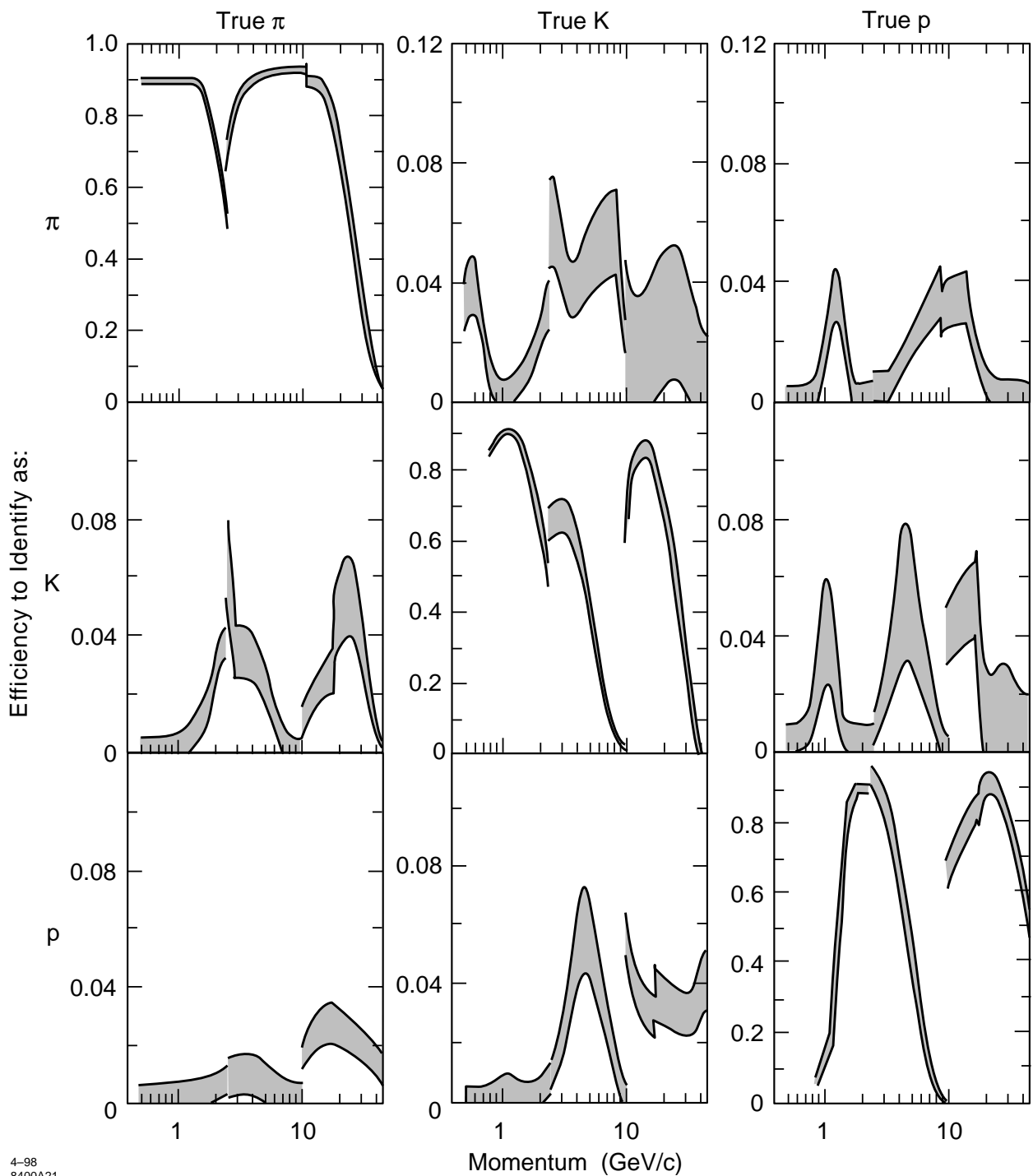}}\end{center}
  \caption{ 
 \label{effpar}
Calibrated identification efficiencies for tracks used in the charged hadron
fractions analysis.
The half-widths of the grey bands represent the systematic
uncertainties, which are strongly correlated between momenta.
Note the expanded vertical scale for the off-diagonal efficiencies.
    }
\end{figure}

The particle identification performance of the CRID depends on the track
selection and likelihood difference requirements for a given analysis.
Here we discuss the example of the hadron fractions analysis described in
section 4.1, where we consider only the three charged hadron hypotheses
$i=\pi$,$K$,p.
For tracks with $p<2.5$ ($p>2.5$) GeV/c, a particle was identified as species
$j$ if ${\cal L}_j$ exceeded both of the other log-likelihoods by at least 5 (3)
units.
We quantify the performance in terms of a momentum-dependent identification
efficiency matrix {\bf E}, each element $E_{ij}$ of which represents the
probability that a selected track from a true $i$-hadron is identified as a
$j$-hadron, with $i,j=\pi$,$K$,p.
The elements of this matrix were determined where possible from the data
\cite{tomp}.
For example, tracks from selected $K_s^0$ and $\tau$ decays were used as
``pion" test samples, having estimated kaon plus proton contents of 0.3\%
and 1.7\% respectively.
Figure \ref{effmlk} shows the probability for these tracks 
to be identified as pions, kaons and protons as a function of momentum.
Also shown are results of the same analysis of corresponding samples from a
detailed Monte Carlo (MC) simulation of the detector.
The MC describes the momentum dependence well and reproduces the
efficiencies to within $\pm$0.03.
Functional forms were fitted to the data,
chosen to describe the momentum dependence of both data and simulated test
samples, as well as that of simulated true pions in hadronic events.
The simulation
was used to correct the fitted parameters for non-pion content in the $K_s^0$
and $\tau$ samples and differences in tracking performance between tracks
in these samples and those from the IP in hadronic events.
The resulting identification efficiency functions, $E_{\pi \pi}$, $E_{\pi K}$
and $E_{\pi {\rm p}}$, are shown in the leftmost column of fig. \ref{effpar}.

A similar procedure using only $\pi$ and p likelihoods was used to measure the
$\pi$-p separation in the liquid (gas) system for $p>2$ (17) GeV/c, and the
simulation was used to convert that into $E_{\rm pp}$, shown in the bottom
right of fig. \ref{effpar}.
$E_{\rm pp}$ over the remaining momentum range, as well as the $\pi$-$K$
separation in the gas system below and near kaon threshold ($p<10$ GeV/c),
was measured using protons from decays of tagged lambda hyperons \cite{tomp}.
The remaining efficiencies in fig. \ref{effpar}
were derived from those measured, using the simulation.
For example, $E_{KK}$ is equal to $E_{\pi \pi}$ for momenta in the ranges
$1.5<p<2.5$ and $15<p<25$ GeV/c, since both species are well above the
relevant Cherenkov
threshold and their expected Cherenkov angles differ from that of the proton by
an amount large compared with the angular resolution.
Outside these ranges,
$E_{KK}$ was related to $E_{\pi \pi}$ by a function derived
from the simulation to account for the effects of the reduced photon yield near
the kaon Cherenkov threshold and the fact that the expected kaon ring radius
lies between those of the pion and proton.

The bands in fig. \ref{effpar} encompass the upper and lower 
systematic error bounds on the efficiencies.
The discontinuities correspond to the $\pi$ and $K$ Cherenkov thresholds in the
gas radiator.
For the diagonal elements, the systematic errors correspond to errors on the
fitted parameters and are strongly positively correlated across each of
the three momentum regions.
For the off-diagonal elements, representing misidentification rates, a more
conservative 25\% relative
error was assigned at all points to account for the limited experimental
constraints on the momentum dependence.  These errors are also strongly
positively correlated among momenta.
The identification efficiencies in fig. \ref{effpar} peak near or above
0.9 and
the pion coverage is continuous from 0.3 GeV/c up to approximately 35 GeV/c.
There is a gap in the kaon-proton separation between about 7 and 10 GeV/c
due to the
limited resolution of the liquid system and the fact that neither species is
far above Cherenkov threshold in the gas system.
The proton coverage extends to the beam momentum.
Misidentification rates are typically less than 0.03, with peak values of
up to 0.07.

\section{Event Selection} 

The trigger and initial selection of hadronic events are described
in~\cite{alr}.
The analysis presented here is based on charged tracks measured in the 
CDC and VXD.
A set of cuts was applied in order to select 
events well-contained within the detector acceptance.
Tracks were required to have
(i) a closest approach to the beam axis within 5 cm, and within
10 cm along the beam axis of the measured IP,
(ii) a polar angle $\theta$
with respect to the beam axis with $|\cos\theta|$ $<$ 0.80,
(iii) a momentum transverse to this axis $p_{\perp}$ $>$
150 MeV/$c$, and
(iv) a momentum $p$ $<$ 50 GeV/c.
Events were required:
to contain a minimum of seven such tracks;
to contain a minimum visible energy $E_{vis} > 18$ GeV,
calculated from the accepted tracks, assigned the charged pion mass;
to have a thrust axis polar angle $\theta_t$ with respect to the beam axis,
calculated from calorimeter clusters, with $|\cos\theta_t|$ $<$ 0.71; and
to have good VXD data \cite{homer} and a well-measured IP position.
A sample of 90,213 events passed these cuts.
For the analyses using the CRID, the additional requirements were made that the
CRID high voltage was on and that there was a good drift velocity measurement,
resulting in a sample of 79,711 events.
The non-hadronic background was estimated to be 0.1\%, dominated by
$Z^0\rightarrow \tau^+ \tau^-$ events.

Samples of events enriched in light and $b$ primary flavors were selected based
on signed impact parameters $\delta$ of charged tracks with respect to
the IP in the plane transverse to the beam.
For each event we define $n_{sig}$ to be the number of tracks passing a set of 
impact-parameter quality cuts~\cite{homer} that have
impact parameter greater than three times its estimated error,
$\delta > 3 \sigma_{\delta}$.
Events with $n_{sig}=0$
were assigned to the light-tagged sample and those with $n_{sig} \geq 3$
were assigned to the $b$-tagged sample.
The remaining events were classified as a $c$-tagged sample.
The light-, $c$- and $b$-tagged samples comprised 60.4\%, 24.5\% and 15.2\% of
the selected hadronic events, respectively.
The tagging efficiencies and sample purities were estimated from our
Monte Carlo simulation and are listed in table \ref{tlveff}.

\begin{table}
\begin{center}
 \begin{tabular}{|l||c|c|c||c|c|c|} \hline
       & \multicolumn{3}{c||}{ } & \multicolumn{3}{c|}{ } \\ [-.3cm]
       & \multicolumn{3}{c||}{Efficiency for $Z^0 \rightarrow$}
       & \multicolumn{3}{c|}{Composition} \\
       & $u\bar{u},d\bar{d},s\bar{s}$ & $c\bar{c}$ & $b\bar{b}$ 
       & $u\bar{u},d\bar{d},s\bar{s}$ & $c\bar{c}$ & $b\bar{b}$  \\ [.1cm] \hline 
 &&&&&&\\[-.3cm] 
light-tag  & 0.845 & 0.438 & 0.075 & 0.849 & 0.124 & 0.027 \\
$c$-tag    & 0.153 & 0.478 & 0.331 & 0.378 & 0.333 & 0.290 \\
$b$-tag    & 0.002 & 0.084 & 0.594 & 0.009 & 0.100 & 0.891 \\[.1cm] \hline 
 \end{tabular}
\caption{
\label{tlveff}
Efficiencies for simulated events in the three flavor categories to be
tagged as light, $c$ or $b$ events.
The three rightmost columns indicate the composition of each simulated tagged
sample assuming the Standard Model relative flavor production.}
\end{center}
\end{table}

Separate samples of hemispheres enriched in light-quark and light-antiquark jets
were selected from the light-tagged event sample by exploiting the large
electroweak forward-backward production asymmetry with respect to the beam
direction.
The event thrust axis was used to approximate the initial $q\bar{q}$ axis and
was signed such that its $z$-component was along the electron beam direction,
$\hat{t}_z>0$.
Events in the central region of the detector, where the production asymmetry is
small, were removed by the requirement $|\hat{t}_z|>0.2$, leaving 74\% of the
light-tagged events.
The quark-tagged hemisphere in events with left- (right-)handed electron beam
polarization was defined to comprise the set of tracks with
positive (negative) momentum
projection along the signed thrust axis.
The remaining tracks in each event
were defined to be in the antiquark-tagged hemisphere.
For the selected event sample, the average magnitude of the 
polarization was 0.73.
Using this value and assuming Standard Model couplings,
a tree-level calculation gives a quark (antiquark) purity of 0.73 in the
quark-(antiquark-)tagged sample.

\section{Hadron Identification Analysis} 

In the following subsections we discuss details of the analysis for three
categories of identified hadrons:  charged tracks identified as $\pi^\pm$,
$K^\pm$ or p/$\bar{\rm p}$ in the CRID;
$K^0_s$ and \llb\ reconstructed in their
charged decay modes and tagged by their long flight distance; and \kkb\ and
$\phi$ reconstructed in charged decay modes including one and two identified
$K^\pm$, respectively.
The resulting \pnb s for these seven hadron species in inclusive hadronic $Z^0$
decays are presented in the last subsection.

\subsection{Charged Hadron Fractions} 

Reconstructed charged tracks were identified as charged pions, kaons or protons
using information from only the CRID liquid (gas) radiator for tracks with
$p<2.5$ ($p>7.5$) GeV/c; in the overlap region, $2.5<p<7.5$ GeV/c,
liquid and gas information was combined.
Additional track selection cuts \cite{tomp}
were applied to remove tracks that interacted or
scattered through large angles before exiting the CRID and 
to ensure that the CRID performance was well-modelled by the simulation.
Tracks were required to have at least 40 CDC hits, at least one of which
was at a radius of at least 92 cm,
to extrapolate through an active region of the appropriate radiator(s),
and
to have at least 80 (100)\% of their expected liquid (gas) ring contained
within a sensitive region of the CRID TPCs.
The latter requirement included rejection of tracks with $p>2.5$ GeV/c for
which there was a saturated CRID hit 
within a 5 cm radius (twice the maximum ring radius) of the expected
gas ring center.
Tracks with $p<7.5$ GeV/c were required to have
a saturated hit within 1 cm of the extrapolated track, and
tracks with $p>2.5$ GeV/c were required to have either such a saturated hit or
the presence of at least four hits consistent with a liquid ring.
These cuts accepted 47\%, 28\% and 43\% of the tracks within the CRID acceptance
in the momentum ranges $p<2.5$, $2.5<p<7.5$ and $p>7.5$ GeV/c, respectively.
For momenta below 2 GeV/c, only negatively charged tracks were used in order
to reduce
the background from protons produced in particle interactions with the detector
material.

\begin{figure}
 \hspace*{0.5cm}   
   \epsfxsize=6.5in
   \begin{center}\mbox{\epsffile{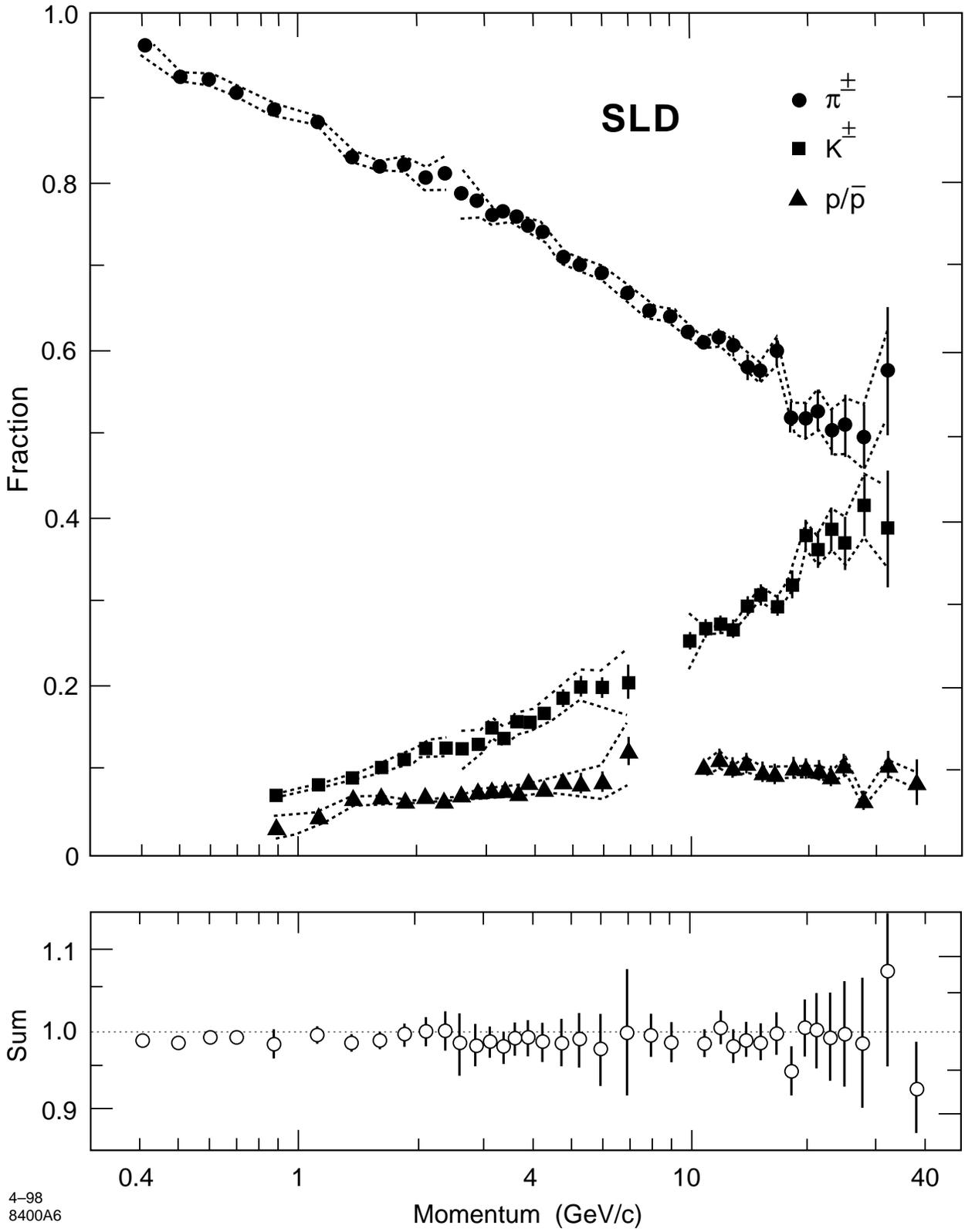}}\end{center}
  \caption{ 
 \label{fraxg}
Measured charged hadron production fractions in hadronic $Z^0$ decays. 
The circles represent the $\pi^\pm$ fraction, the squares the $K^\pm$ fraction,
the triangles the p/$\bar{\rm p}$ fraction, and the open circles the sum of the
three fractions.
The error bars in the upper plot are statistical only;
the dashed lines indicate the systematic errors,
which are strongly correlated between momenta.
The error bars on the sum are statistical and systematic added in quadrature.
    }
\end{figure}

In each momentum bin we measured the fractions of the selected tracks
that were identified as pions, kaons and protons.
The observed fractions were related to the true production fractions by an
efficiency matrix, composed of the values shown in fig.~\ref{effpar}.
This matrix was inverted and used to unfold our observed identified hadron
fractions.
This analysis procedure does not require that the sum 
of the charged hadron fractions be unity; instead the sum was used as a 
consistency check, which was found to be satisfied at all momenta (see fig.
\ref{fraxg}). 
In some momentum regions we cannot distinguish two of the three hadron species,
so the procedure was reduced to a 2$\times$2 matrix analysis and we present
only the fraction of the identified species, i.e. protons
above 35 GeV/c and pions below 0.75 GeV/c and between 7.5 and 9.5 GeV/c.

Electrons and muons were not distinguished from pions;
this background was estimated from the simulation to be about
5\% of the tracks in the inclusive flavor sample,
predominantly from $c$- and $b$-flavor events.
The fractions were corrected using the simulation for
the lepton backgrounds, as well as for the effects of beam-related
backgrounds, particles interacting in the detector material, and particles
decaying outside the tracking volume.
The conventional definition of a final-state charged hadron was used,
namely a charged pion, kaon or proton that is either
from the primary interaction or a direct decay product of a hadron that has 
proper lifetime less than 3$\times10^{-10}$s and is itself a primary or
a decay product of a primary hadron.

The measured charged hadron fractions in inclusive hadronic $Z^0$ decays
are shown in fig.~\ref{fraxg} and listed in tables \ref{pifraxa}--\ref{pfraxa}.
The systematic errors were determined by propagating the errors on the
calibrated efficiency matrix (see sec. 2.2) and correspond to uncertainties in
the average number of photons detected per track and the average
resolution on the measured Cherenkov angles.
They are therefore strongly positively correlated across each of the
three momentum regions, $p<2.5$, $2.5<p<7.5$ and $p>7.5$ GeV/c, and are
indicated by the pairs of dashed lines in fig. \ref{fraxg}.
The errors on the points below $\sim$6 GeV/c are dominated by the systematic
uncertainties;
for the points above $\sim$15 GeV/c the errors have roughly equal
statistical and systematic contributions.

\begin{table}
\begin{center}
\begin{tabular}{|r@{--}r|c|r@{$\pm$}r@{$\pm$}r|r@{$\pm$}l@{$\pm$}l|} \hline
\multicolumn{2}{|c|}{$x_p$ Range} & $<\! x_p \!>$ &
\multicolumn{3}{c|}{$f_{\pi}$} & \multicolumn{3}{c|}{1/N d$n_{\pi}$/d$x_p$}
\\ \hline
0.008 & 0.010 & 0.009 & 0.963 &0.004 &0.014 & 482.3 &   2.3 &   7.2 \\
0.010 & 0.012 & 0.011 & 0.924 &0.004 &0.006 & 439.0 &   2.3 &   3.7 \\
0.012 & 0.014 & 0.013 & 0.921 &0.003 &0.006 & 400.5 &   2.0 &   3.3 \\
0.014 & 0.016 & 0.015 & 0.906 &0.004 &0.006 & 356.1 &   1.9 &   3.0 \\
0.016 & 0.022 & 0.019 & 0.886 &0.002 &0.006 & 292.8 &   1.0 &   2.4 \\
0.022 & 0.027 & 0.025 & 0.872 &0.003 &0.006 & 228.5 &   1.0 &   1.9 \\
0.027 & 0.033 & 0.030 & 0.831 &0.003 &0.006 & 176.6 &   0.9 &   1.4 \\
0.033 & 0.038 & 0.036 & 0.820 &0.004 &0.006 & 144.4 &   0.8 &   1.2 \\
0.038 & 0.044 & 0.041 & 0.823 &0.004 &0.010 & 121.7 &   0.8 &   1.6 \\
0.044 & 0.049 & 0.047 & 0.806 &0.006 &0.015 & 102.5 &   0.9 &   1.9 \\
0.049 & 0.055 & 0.052 & 0.812 &0.008 &0.020 &  89.2 &   0.9 &   2.2 \\
0.055 & 0.060 & 0.058 & 0.788 &0.007 &0.029 &  75.3 &   0.8 &   2.8 \\
0.060 & 0.066 & 0.063 & 0.779 &0.007 &0.016 &  66.0 &   0.7 &   1.4 \\
0.066 & 0.071 & 0.069 & 0.763 &0.007 &0.010 & 57.81 &  0.60 &  0.81 \\
0.071 & 0.077 & 0.074 & 0.767 &0.007 &0.009 & 51.63 &  0.56 &  0.60 \\
0.077 & 0.082 & 0.079 & 0.761 &0.007 &0.009 & 45.95 &  0.52 &  0.54 \\
0.082 & 0.088 & 0.085 & 0.750 &0.007 &0.008 & 41.35 &  0.49 &  0.49 \\
0.088 & 0.099 & 0.093 & 0.743 &0.006 &0.008 & 35.24 &  0.32 &  0.42 \\
0.099 & 0.110 & 0.104 & 0.714 &0.006 &0.008 & 28.12 &  0.29 &  0.35 \\
0.110 & 0.121 & 0.115 & 0.705 &0.007 &0.009 & 23.57 &  0.27 &  0.30 \\
0.121 & 0.143 & 0.131 & 0.695 &0.005 &0.009 & 18.32 &  0.17 &  0.24 \\
0.143 & 0.164 & 0.153 & 0.670 &0.006 &0.009 & 13.22 &  0.14 &  0.19 \\
0.164 & 0.186 & 0.175 & 0.651 &0.006 &0.009 &  9.84 &  0.11 &  0.15 \\
0.186 & 0.208 & 0.197 & 0.644 &0.007 &0.008 &  7.47 &  0.09 &  0.11 \\
0.208 & 0.230 & 0.219 & 0.625 &0.008 &0.007 & 5.711 & 0.083 & 0.080 \\
0.230 & 0.252 & 0.241 & 0.611 &0.009 &0.006 & 4.414 & 0.074 & 0.063 \\
0.252 & 0.274 & 0.263 & 0.618 &0.010 &0.010 & 3.612 & 0.068 & 0.072 \\
0.274 & 0.296 & 0.285 & 0.608 &0.011 &0.010 & 2.886 & 0.061 & 0.060 \\
0.296 & 0.318 & 0.307 & 0.583 &0.012 &0.011 & 2.206 & 0.054 & 0.049 \\
0.318 & 0.351 & 0.334 & 0.578 &0.012 &0.012 & 1.739 & 0.040 & 0.044 \\
0.351 & 0.384 & 0.366 & 0.603 &0.014 &0.015 & 1.350 & 0.036 & 0.040 \\
0.384 & 0.417 & 0.400 & 0.523 &0.017 &0.016 & 0.874 & 0.031 & 0.032 \\
0.417 & 0.450 & 0.432 & 0.520 &0.021 &0.020 & 0.670 & 0.029 & 0.029 \\
0.450 & 0.482 & 0.465 & 0.534 &0.024 &0.024 & 0.520 & 0.026 & 0.025 \\
0.482 & 0.526 & 0.503 & 0.508 &0.028 &0.027 & 0.355 & 0.021 & 0.020 \\
0.526 & 0.570 & 0.547 & 0.514 &0.036 &0.031 & 0.248 & 0.018 & 0.016 \\
0.570 & 0.658 & 0.609 & 0.501 &0.040 &0.038 & 0.146 & 0.012 & 0.012 \\
0.658 & 0.768 & 0.704 & 0.580 &0.076 &0.053 & 0.071 & 0.009 & 0.007
 \\ \hline \hline 
  \multicolumn{3}{|c}{Total Observed/Evt.}
 &\multicolumn{3}{c|}{  }& 14.52 & 0.02 & 0.27
 \\ \hline 
 \end{tabular}
\vspace{-0.2cm}
\caption{
 \label{pifraxa}
Charged pion fraction $f_{\pi}$ and differential cross section
(1/N)d$n_{\pi}$/d$x_p$ per hadronic $Z^0$ decay.
$<\!\! x_p\!\!>$ is the average $x_p$-value of charged tracks in each bin.
The last row gives the integral over the $x_p$ range of the measurement.
The first error is statistical, the second systematic.
A 1.7\% normalization uncertainty is included in the systematic error on the
integral, but not in those on the cross section.}
\end{center}
\end{table}

\begin{table}
\begin{center}
 \begin{tabular}{|c|c|r@{$\pm$}r@{$\pm$}r|r@{$\pm$}l@{$\pm$}l|} \hline
 $x_p$ Range & $<\! x_p\! >$ & \multicolumn{3}{c|}{$f_K$}  &
                               \multicolumn{3}{c|}{1/N d$n_{K}$/d$x_p$}   \\ \hline 
  & & \multicolumn{3}{c|}{ } & \multicolumn{3}{c|}{  }   \\[-.3cm]
0.016--0.022 & 0.019 & 0.067 &0.001 &0.002 & 22.28 &  0.47 &  0.53 \\
0.022--0.027 & 0.025 & 0.081 &0.002 &0.002 & 21.22 &  0.45 &  0.62 \\
0.027--0.033 & 0.030 & 0.090 &0.002 &0.003 & 19.10 &  0.43 &  0.64 \\
0.033--0.038 & 0.036 & 0.102 &0.002 &0.005 & 18.02 &  0.43 &  0.80 \\
0.038--0.044 & 0.041 & 0.111 &0.003 &0.006 & 16.45 &  0.45 &  0.94 \\
0.044--0.049 & 0.047 & 0.127 &0.004 &0.008 & 16.13 &  0.49 &  1.03 \\
0.049--0.055 & 0.052 & 0.127 &0.005 &0.010 & 13.98 &  0.53 &  1.14 \\
0.055--0.060 & 0.058 & 0.125 &0.006 &0.022 & 11.96 &  0.54 &  2.11 \\
0.060--0.066 & 0.063 & 0.130 &0.006 &0.015 & 11.03 &  0.49 &  1.27 \\
0.066--0.071 & 0.069 & 0.150 &0.006 &0.012 & 11.37 &  0.46 &  0.87 \\
0.071--0.077 & 0.074 & 0.139 &0.007 &0.012 &  9.38 &  0.44 &  0.79 \\
0.077--0.082 & 0.079 & 0.157 &0.007 &0.013 &  9.51 &  0.44 &  0.76 \\
0.082--0.088 & 0.085 & 0.157 &0.008 &0.013 &  8.68 &  0.44 &  0.72 \\
0.088--0.099 & 0.093 & 0.168 &0.007 &0.014 &  7.96 &  0.31 &  0.68 \\
0.099--0.110 & 0.104 & 0.187 &0.009 &0.016 &  7.37 &  0.34 &  0.63 \\
0.110--0.121 & 0.115 & 0.202 &0.011 &0.018 &  6.74 &  0.37 &  0.60 \\
0.121--0.143 & 0.131 & 0.199 &0.011 &0.023 &  5.24 &  0.29 &  0.61 \\
0.143--0.164 & 0.153 & 0.207 &0.020 &0.041 &  4.08 &  0.40 &  0.80 \\ \hline 
0.208--0.230 & 0.219 & 0.256 &0.009 &0.033 &  2.34 &  0.08 &  0.30 \\
0.230--0.252 & 0.241 & 0.269 &0.009 &0.007 & 1.947 & 0.065 & 0.057 \\
0.252--0.274 & 0.263 & 0.274 &0.009 &0.007 & 1.603 & 0.057 & 0.042 \\
0.274--0.296 & 0.285 & 0.270 &0.010 &0.006 & 1.281 & 0.050 & 0.034 \\
0.296--0.318 & 0.307 & 0.298 &0.011 &0.007 & 1.127 & 0.045 & 0.030 \\
0.318--0.351 & 0.334 & 0.310 &0.011 &0.008 & 0.933 & 0.034 & 0.027 \\
0.351--0.384 & 0.366 & 0.299 &0.012 &0.009 & 0.669 & 0.029 & 0.023 \\
0.384--0.417 & 0.400 & 0.324 &0.015 &0.012 & 0.541 & 0.026 & 0.023 \\
0.417--0.450 & 0.432 & 0.383 &0.019 &0.016 & 0.493 & 0.026 & 0.023 \\
0.450--0.482 & 0.465 & 0.366 &0.022 &0.019 & 0.357 & 0.023 & 0.020 \\
0.482--0.526 & 0.503 & 0.391 &0.025 &0.023 & 0.273 & 0.019 & 0.018 \\
0.526--0.570 & 0.547 & 0.374 &0.032 &0.028 & 0.180 & 0.016 & 0.014 \\
0.570--0.658 & 0.609 & 0.420 &0.037 &0.036 & 0.122 & 0.011 & 0.011 \\
0.658--0.768 & 0.704 & 0.392 &0.070 &0.049 & 0.048 & 0.009 & 0.006
 \\[.1cm] \hline \hline 
  \multicolumn{2}{|c}{   }
& \multicolumn{3}{c|}{ } & \multicolumn{3}{c|}{  }   \\[-.3cm]
  \multicolumn{2}{|c}{Total Observed/Evt.}
& \multicolumn{3}{c|}{ }& 1.800 & 0.016 & 0.124
 \\[.1cm] \hline 
 \end{tabular}
\caption{ \label{kafraxa}
Charged kaon fraction and differential cross section per hadronic
$Z^0$ decay.}
\end{center}
\end{table}

\begin{table}
\begin{center}
 \begin{tabular}{|c|c|r@{$\pm$}r@{$\pm$}r|r@{$\pm$}l@{$\pm$}l|} \hline
 $x_p$ Range & $<\! x_p\! >$ & \multicolumn{3}{c|}{$f_{\rm p}$}  &
                               \multicolumn{3}{c|}{1/N d$n_{\rm p}$/d$x_p$} \\ \hline 
  & & \multicolumn{3}{c|}{ } & \multicolumn{3}{c|}{  }   \\[-.3cm]
0.016--0.022 & 0.019 & 0.029 &0.005 &0.013 &  9.55 &  1.55 &  4.33 \\
0.022--0.027 & 0.025 & 0.041 &0.003 &0.008 & 10.79 &  0.84 &  2.09 \\
0.027--0.033 & 0.030 & 0.064 &0.002 &0.005 & 13.56 &  0.47 &  0.98 \\
0.033--0.038 & 0.036 & 0.065 &0.002 &0.004 & 11.54 &  0.35 &  0.63 \\
0.038--0.044 & 0.041 & 0.061 &0.002 &0.002 &  9.03 &  0.30 &  0.25 \\
0.044--0.049 & 0.047 & 0.067 &0.002 &0.002 &  8.52 &  0.29 &  0.23 \\
0.049--0.055 & 0.052 & 0.062 &0.002 &0.002 &  6.83 &  0.26 &  0.22 \\
0.055--0.060 & 0.058 & 0.072 &0.003 &0.005 &  6.85 &  0.28 &  0.48 \\
0.060--0.066 & 0.063 & 0.074 &0.003 &0.005 &  6.70 &  0.28 &  0.42 \\
0.066--0.071 & 0.069 & 0.075 &0.004 &0.005 &  5.69 &  0.27 &  0.40 \\
0.071--0.077 & 0.074 & 0.075 &0.004 &0.006 &  5.03 &  0.27 &  0.38 \\
0.077--0.082 & 0.079 & 0.072 &0.004 &0.006 &  4.33 &  0.27 &  0.38 \\
0.082--0.088 & 0.085 & 0.085 &0.005 &0.007 &  4.65 &  0.29 &  0.39 \\
0.088--0.099 & 0.093 & 0.077 &0.004 &0.009 &  3.64 &  0.20 &  0.41 \\
0.099--0.110 & 0.104 & 0.087 &0.006 &0.012 &  3.42 &  0.23 &  0.45 \\
0.110--0.121 & 0.115 & 0.084 &0.007 &0.015 &  2.80 &  0.25 &  0.49 \\
0.121--0.143 & 0.131 & 0.085 &0.008 &0.021 &  2.22 &  0.21 &  0.54 \\
0.143--0.164 & 0.153 & 0.123 &0.016 &0.039 &  2.42 &  0.32 &  0.77 \\ \hline 
0.230--0.252 & 0.241 & 0.106 &0.007 &0.010 & 0.767 & 0.048 & 0.074 \\
0.252--0.274 & 0.263 & 0.114 &0.007 &0.010 & 0.668 & 0.043 & 0.059 \\
0.274--0.296 & 0.285 & 0.105 &0.008 &0.009 & 0.497 & 0.036 & 0.044 \\
0.296--0.318 & 0.307 & 0.109 &0.008 &0.009 & 0.413 & 0.032 & 0.035 \\
0.318--0.351 & 0.334 & 0.099 &0.007 &0.009 & 0.296 & 0.022 & 0.026 \\
0.351--0.384 & 0.366 & 0.098 &0.008 &0.008 & 0.219 & 0.018 & 0.019 \\
0.384--0.417 & 0.400 & 0.105 &0.009 &0.007 & 0.175 & 0.015 & 0.013 \\
0.417--0.450 & 0.432 & 0.104 &0.010 &0.007 & 0.134 & 0.013 & 0.009 \\
0.450--0.482 & 0.465 & 0.103 &0.011 &0.006 & 0.101 & 0.011 & 0.006 \\
0.482--0.526 & 0.503 & 0.095 &0.011 &0.006 & 0.066 & 0.008 & 0.004 \\
0.526--0.570 & 0.547 & 0.110 &0.013 &0.006 & 0.053 & 0.006 & 0.003 \\
0.570--0.658 & 0.609 & 0.066 &0.010 &0.006 & 0.019 & 0.003 & 0.002 \\
0.658--0.768 & 0.704 & 0.107 &0.016 &0.007 & 0.013 & 0.002 & 0.001 \\
0.768--0.987 & 0.836 & 0.087 &0.027 &0.012 & 0.002 & 0.001 & 0.000
 \\[.1cm] \hline \hline 
  \multicolumn{2}{|c}{   }
& \multicolumn{3}{c|}{ } & \multicolumn{3}{c|}{  }   \\[-.3cm]
  \multicolumn{2}{|c}{Total Observed/Evt.}
& \multicolumn{3}{c|}{  }  & 0.864 & 0.015 & 0.106
 \\[.1cm] \hline 
 \end{tabular}
\caption{ \label{pfraxa}
Proton fraction and differential cross section per hadronic
$Z^0$ decay.}
\end{center}
\end{table}

Pions are seen to dominate the charged hadron production at low momentum,
and to decline steadily in fraction as momentum increases.
The kaon fraction rises steadily to about one-third at high momentum.
The proton fraction rises to a plateau value of about one-tenth at about
10 GeV/c.
Where the momentum coverage overlaps, these measured fractions were found to be 
consistent with an average of previous measurements at the
$Z^0$~\cite{delphi,opal,aleph}.
Measurements based on ring imaging and those based on ionization energy loss
rates cover complementary
momentum ranges and can be combined to provide continuous
coverage over the range $0.22<p<45.6$ GeV/c.

Differential production cross sections were obtained by multiplying these
fractions by our measured inclusive charged particle \pnb , corrected, using our
simulation, for the contribution from leptons.
The integral of this cross section was constrained to be 20.95 tracks per event,
an average \cite{dcone} of charged multiplicity measurements in $Z^0$ decays,
and the momentum-dependence of our track
reconstruction efficiency was checked by comparing the momentum distributions
of charged tracks in data and simulated $\tau^\pm$ decays.
We include a 1.7\% error on the average multiplicity as a
systematic normalization uncertainty, as well as a momentum-dependent
uncertainty of 0.11$\times |p-3.8$ GeV/c$|$\%, derived from the study of
$\tau^\pm$ decays.
The inclusive charged particle \pnb\ is listed in table \ref{chgxs}, and
the resulting \pnb s per hadronic event per unit $x_p$ for the identified
hadrons are listed in tables \ref{pifraxa}--\ref{pfraxa}.
The 1.7\% normalization uncertainty is not included in the systematic error
listed for any of the identified hadrons,
nor is it included in the error bars in any of the figures.

\begin{table}
\begin{center}
\begin{tabular}{|r@{--}r|c|r@{$\pm$}l@{$\pm$}l|} \hline
\multicolumn{2}{|c|}{ } && \multicolumn{3}{c|}{ } \\[-.4cm]
\multicolumn{2}{|c|}{$x_p$ Range} & $<\! x_p \!>$ &
\multicolumn{3}{c|}{1/N d$n_{chg}$/d$x_p$}
\\[.1cm] \hline
0.008 & 0.010 & 0.009 & 509.6 &   1.6 &   8.9 \\
0.010 & 0.012 & 0.011 & 481.9 &   1.6 &   8.4 \\
0.012 & 0.014 & 0.013 & 440.9 &   1.5 &   7.7 \\
0.014 & 0.016 & 0.015 & 398.0 &   1.4 &   6.9 \\
0.016 & 0.022 & 0.019 & 334.6 &   0.9 &   5.8 \\
0.022 & 0.027 & 0.025 & 265.2 &   0.8 &   4.6 \\
0.027 & 0.033 & 0.030 & 215.2 &   0.7 &   3.7 \\
0.033 & 0.038 & 0.036 & 178.6 &   0.6 &   3.1 \\
0.038 & 0.044 & 0.041 & 150.0 &   0.6 &   2.6 \\
0.044 & 0.049 & 0.047 & 129.2 &   0.5 &   2.2 \\
0.049 & 0.055 & 0.052 & 111.7 &   0.5 &   1.9 \\
0.055 & 0.060 & 0.058 &  97.2 &   0.5 &   1.7 \\
0.060 & 0.066 & 0.063 &  86.3 &   0.4 &   1.5 \\
0.066 & 0.071 & 0.069 &  77.2 &   0.4 &   1.3 \\
0.071 & 0.077 & 0.074 &  68.7 &   0.4 &   1.2 \\
0.077 & 0.082 & 0.079 &  61.6 &   0.4 &   1.0 \\
0.082 & 0.088 & 0.085 & 56.35 &  0.35 &  0.96 \\
0.088 & 0.099 & 0.093 & 48.53 &  0.23 &  0.83 \\
0.099 & 0.110 & 0.104 & 40.40 &  0.21 &  0.69 \\
0.110 & 0.121 & 0.115 & 34.32 &  0.20 &  0.59 \\
0.121 & 0.143 & 0.131 & 27.12 &  0.12 &  0.47 \\
0.143 & 0.164 & 0.153 & 20.35 &  0.11 &  0.35 \\
0.164 & 0.186 & 0.175 & 15.65 &  0.09 &  0.28 \\
0.186 & 0.208 & 0.197 & 12.05 &  0.08 &  0.22 \\
0.208 & 0.230 & 0.219 &  9.50 &  0.07 &  0.17 \\
0.230 & 0.252 & 0.241 &  7.54 &  0.07 &  0.14 \\
0.252 & 0.274 & 0.263 &  6.11 &  0.06 &  0.12 \\
0.274 & 0.296 & 0.285 & 4.969 & 0.053 & 0.098 \\
0.296 & 0.318 & 0.307 & 3.978 & 0.048 & 0.081 \\
0.318 & 0.351 & 0.334 & 3.163 & 0.035 & 0.067 \\
0.351 & 0.384 & 0.366 & 2.367 & 0.030 & 0.052 \\
0.384 & 0.417 & 0.400 & 1.767 & 0.026 & 0.041 \\
0.417 & 0.450 & 0.432 & 1.359 & 0.023 & 0.033 \\
0.450 & 0.482 & 0.465 & 1.028 & 0.019 & 0.026 \\
0.482 & 0.526 & 0.503 & 0.735 & 0.014 & 0.020 \\
0.526 & 0.570 & 0.547 & 0.503 & 0.012 & 0.015 \\
0.570 & 0.658 & 0.609 & 0.300 & 0.006 & 0.009 \\
0.658 & 0.768 & 0.704 & 0.123 & 0.003 & 0.004 \\
0.768 & 0.987 & 0.836 & 0.027 & 0.001 & 0.001 \\ \hline 
 \end{tabular}
\caption{
 \label{chgxs}
Differential cross section (1/N)d$n_{chg}$/d$x_p$ 
for inclusive charged particle production per hadronic $Z^0$ decay.
The first error is statistical, the second systematic.}
\end{center}
\end{table}

\newpage

\subsection{Neutral $K^0/\bar{K}^0$ and \llb\ Production} 

We reconstructed the charged decay modes $K_s^0 \rightarrow
\pi^+\pi^-$ and
$\Lambda^0(\bar{\Lambda}^0) \rightarrow $p$\pi^- (\bar{\rm p}\pi^+)$
 \cite{kenb},
collectively referred to as $V^0$ decays.
In order to ensure good invariant mass resolution tracks were required to have
a minimum transverse momentum of 150 MeV/c with respect to the beam direction,
at least 40 hits measured in the CDC,
and a polar angle satisfying $\left|\cos\theta\right|<0.8$.

Pairs of oppositely charged tracks satisfying these requirements were
combined to form $V^0$s if their separation was less than 15 mm at their
point of closest approach in 3 dimensions.
A $\chi^2$ fit of the two tracks to a common vertex was performed, and
to reject combinatoric background we required:
the confidence level of the $\chi^2$ to be greater than 2\%;
the vertex to be separated from the IP by at least 1 mm,
and by at least $5\sigma_l$, where $\sigma_l$ is the calculated error on the
separation length of the $V^0$;
and vertices reconstructed outside the Vertex Detector to have
at most one VXD hit assigned to each track.

The two invariant masses $m_{\pi\pi}$ and $m_{{\rm p}\pi}$ were calculated for
each $V^0$ with, in the latter case,
the proton (charged pion) mass assigned to the higher-(lower)-momentum track.
In the plane perpendicular to the beam, the angle between
the vector sum of the momenta of the two charged tracks and the line
joining the IP to the vertex was required to be
less than both 60 mrad and $k\cdot(2 + 20/p_{\perp} + 5/p_{\perp}^2)$ mrad.
Here, $p_{\perp}$ is the component of the vector sum momentum transverse to
the beam in units of GeV/c and
$k$=1.75 for \llb\ candidates and 2.5 for $K_s^0$ candidates.
For \llb\ candidates, a minimum vector-sum momentum of 500 MeV/c
was required.

Note that it is possible for one $V^0$ to be considered a
candidate for both the $K_s^0$ and \llb\ hypotheses.
Kinematic regions exist where the two hypotheses cannot be
distinguished without particle identification.
In addition there is background from other processes that occur away from the
IP, most notably $\gamma$-conversions into $e^+e^-$ pairs.
Depending upon the type of analysis,
such ``kinematic-overlaps'' may introduce important biases.
In this analysis,
the kinematic-overlap region was removed only when it distorted the relevant
invariant mass distribution.
For the $K_s^0$ analysis, the \llb\ background causes an asymmetric
bump in the $m_{\pi\pi}$ distribution, which
complicated the subsequent fitting procedure.
A cut on the $\pi^+$ helicity angle $\theta^*_{\pi}$,
defined as the angle between the $\pi^+$ momentum vector
in the $K_s^0$ rest frame and the $K_s^0$ flight direction, 
of $\left|\cos\theta^*_{\pi}\right| \leq 0.8$ was used to remove
the $\Lambda^0$, $\bar{\Lambda}^0$ and $\gamma$-conversion contamination.

For the \llb\ analysis, the shape of the $K_s^0$ background depends strongly
on momentum.
Above a $V^0$ momentum of a few GeV/c,
the $K_s^0 \rightarrow \pi^+\pi^-$ background is essentially
uniform in the peak region of the $m_{{\rm p}\pi}$ distribution
and no cuts were made to remove the $K_s^0$ overlap.
At sufficiently low momentum, the $K_s^0$
background becomes asymmetric under the \llb\, peak due to detector acceptance;
the softer $\pi$ fails to be reconstructed and thus the
$K_s^0$ is not found.
Therefore, \llb\ candidates with total momentum below 1.8 GeV/c
were required to have $m_{\pi\pi}$ more than $3\sigma$ away from the
$K_s^0$ mass, where $\sigma$ is the measured resolution on $m_{\pi\pi}$,
parameterized as
$\sigma_{\pi\pi}(p) = 4.6 - 0.27p +0.21p^2 - 0.01p^3$ MeV/c$^2$,
and $p$ is the $V^0$ momentum in GeV/c. 
In order to remove $\gamma$ conversions, the proton helicity angle was
required to satisfy $\cos \theta^*_{\rm p} \geq -0.95$.

\begin{figure}[p]
\centering
\epsfxsize=13.1cm
\leavevmode
\epsfbox{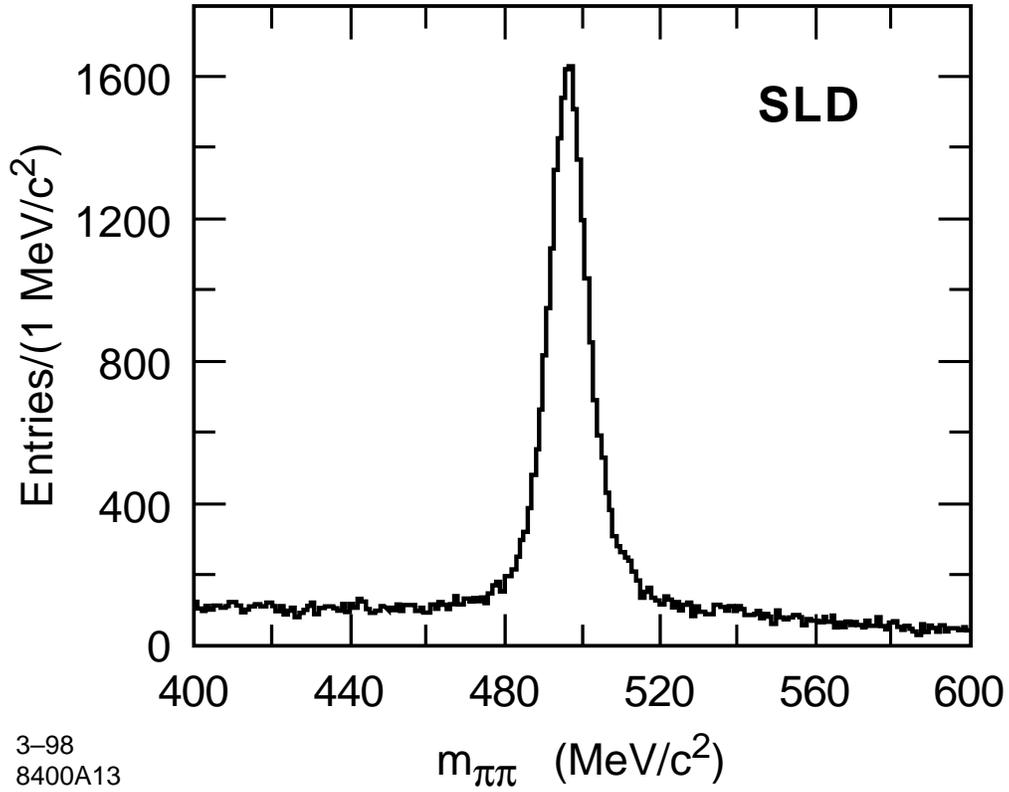}
\caption{\label{kspk}
Invariant mass distribution for all $K_s^0 \rightarrow \pi^+\pi^-$ candidates.
}
\end{figure}

\begin{figure}[p]
\centering
\epsfxsize=13.1cm
\leavevmode
\epsfbox{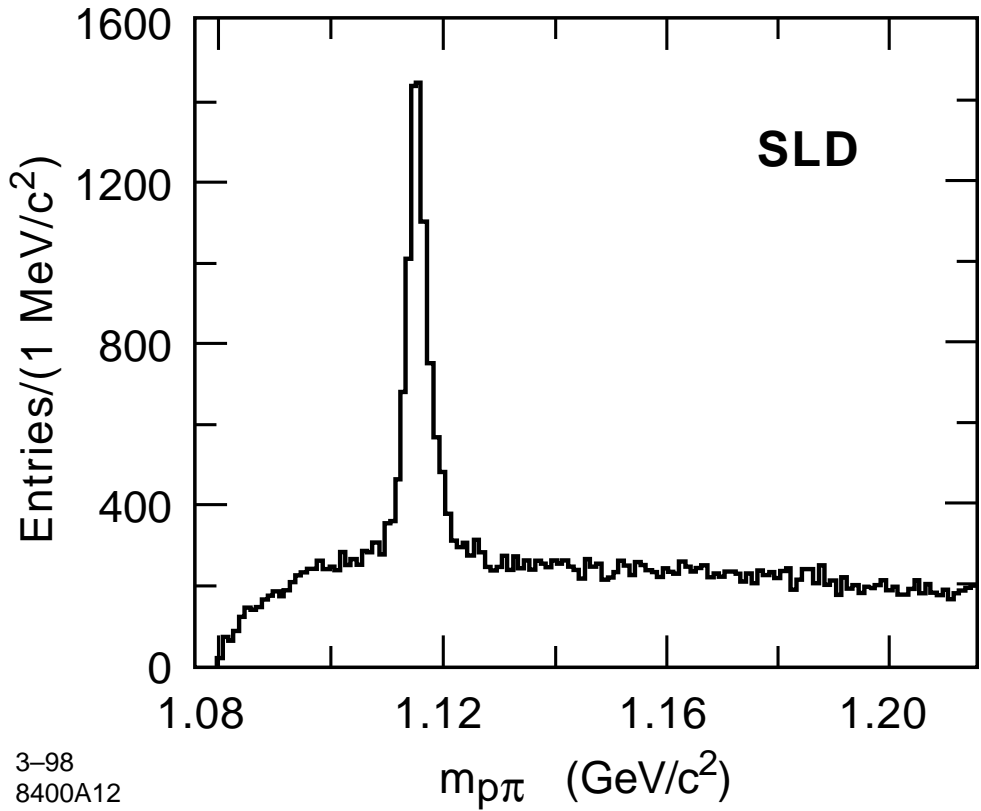}
\caption{\label{lampk}
Invariant mass distribution for all
$\Lambda^0 \rightarrow {\rm p}\pi^-$ and
$\bar{\Lambda}^0 \rightarrow \bar{\rm p}\pi^+$ candidates.
}
\end{figure}

The $m_{\pi\pi}$ and $m_{{\rm p}\pi}$ distributions for the remaining
candidates are shown in figs. \ref{kspk} and \ref{lampk}, respectively.
The $V^0$ candidates were binned in $x_p$, and the resulting
invariant mass distributions were fitted using a sum of signal and
background functions.
The function used for the signal peak was a Gaussian or a sum of two or three
Gaussians of common center, depending on $x_p$.
A single Gaussian was sufficient to describe the $K_s^0$ data in the
lowest-$x_p$ bin and the \llb\ data in the three lowest-$x_p$ bins.
However, the mass resolution is momentum-dependent and varies
substantially over the width of a typical $x_p$ bin; two Gaussians were
sufficient in most cases, with three being needed for both the $K_s^0$ and \llb\
data in the highest-$x_p$ bin.
The relative fractions and nominal widths of the Gaussians in the sum were
fixed from the MC simulation.
The normalization, common center, and a resolution scale-factor were
free parameters of the fit.
The fitted centers were consistent with world average mass values~\cite{pdg},
and the fitted scale factor was typically 1.1.
The background shape used for the $K_s^0$ fits was a quadratic polynomial;
for the
\llb\ fits a more complicated function was required due to the proximity
of the kinematic edge to the signal peak.
The function
$P_{bkg}(m) = a + b(m-m_\Lambda) + c(1-e^{d((m-m_\Lambda)-0.038)})$ was
found to be adequate in Monte Carlo studies, where $a$,$b$,$c$,$d$ were free
parameters.

\begin{figure}[t]
\centering
\epsfxsize=13.5cm
\leavevmode
\epsfbox{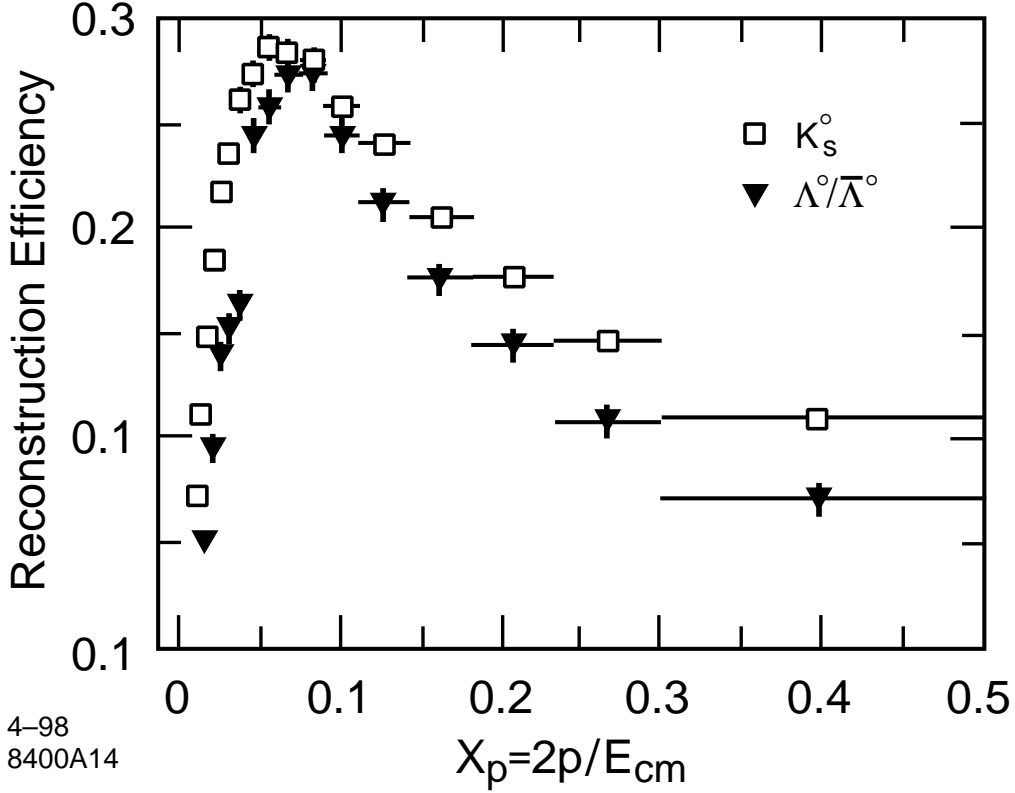}
\caption{\label{receff}
The simulated reconstruction efficiencies as a function of $x_p$
for $K_s^0$ (squares) and $\Lambda^0/\bar{\Lambda}^0$ (triangles).
The charged decay branching ratios are included in the efficiency.
The discontinuity in the \llb\ reconstruction efficiency at $x_p=0.04$ is
due to the invariant-mass cut to remove the low-momentum $K_s^0$ background.}
\end{figure}

The efficiencies for reconstructing true $K_s^0$ and \llb\ decays were
calculated, using
the simulation, by repeating the full selection and analysis on the simulated
sample and dividing by the number of generated $K_s^0$ or \llb .
Several checks were performed to verify the MC simulation, and thus
the $V^0$ reconstruction efficiency.
In particular, the proper lifetimes of the $K_s^0$ and $\Lambda^0$ were
measured,
yielding values consistent with the respective world averages.
The simulated reconstruction efficiencies are shown in fig. \ref{receff},
and were parametrized as functions of $x_p$.
The reconstruction efficiency is limited by the detector
acceptance of $\sim$0.67 and the charged decay branching fractions of 0.64 for
\llb\ and 0.68 for $K_s^0$.
The efficiency at high momentum
decreases due to finite detector size and two-track detector resolution, and
the efficiency at low-momentum is limited by the minimum $p_{\perp}$ and
flight distance requirements. The discontinuity in the
$\Lambda^0/\bar{\Lambda}^0$ reconstruction efficiency is due to the imposed
$K_s^0$ mass cut for low-$x_p$ candidates. 

The differential cross section 1/N d$n$/d$x_p$ per hadronic $Z^0$ decay was
then calculated in each bin by dividing the integrated area under the fitted
mass peak by the efficiency, the bin width and the number of observed hadronic
events corrected for trigger and selection efficiency.
As is conventional, the $K^0/\bar{K}^0$
cross section was obtained by multiplying the
measured $K_s^0$ cross section by a factor of 2 to account for the undetected
$K^0_L$ component.
The resulting \pnb s, including point-to-point systematic errors, discussed
below, are shown in fig. \ref{xsall} and listed in table \ref{xsgvee}. 

\begin{table}[t]
\begin{center}
 \begin{tabular}{|c||c|r@{$\pm$}l@{$\pm$}l||c|r@{$\pm$}l@{$\pm$}l|} \hline
 \multicolumn{9}{|c|}{   } \\[-.3cm]
 \multicolumn{9}{|c|}{Neutral $V^0$ Production} \\[.2cm]\hline\hline
  &&  \multicolumn{3}{c||}{  }  &&  \multicolumn{3}{c|}{  }  \\[-.3cm]
 $x_p$ Range & $<x_p>$  &  \multicolumn{3}{c||}{1/N d$n_{K^0}$/d$x_p$}   &
$<x_p>$  & \multicolumn{3}{c|}{1/N d$n_{\Lambda^0}$/d$x_p$}  \\[.1cm] \hline 
  &&  \multicolumn{3}{c||}{  }  &&  \multicolumn{3}{c|}{  }  \\[-.4cm]
0.009--0.011 & 0.010 & 18.1  & 1.7  & 2.4  &&  \multicolumn{3}{c|}{ }\\ \cline{6-9}
0.011--0.014 & 0.013 & 19.1  & 1.2  & 1.1  &&  \multicolumn{3}{c|}{ }\\
0.014--0.018 & 0.016 & 20.44 & 0.91 & 0.67 & 0.015 &  2.99 & 0.45 & 1.22 \\ \cline{6-9}
0.018--0.022 & 0.020 & 21.74 & 0.85 & 0.72 & 0.020 &  3.90 & 0.42 & 0.58 \\
0.022--0.027 & 0.025 & 20.51 & 0.70 & 0.53 & 0.025 &  4.10 & 0.30 & 0.23 \\
0.027--0.033 & 0.030 & 17.73 & 0.55 & 0.41 & 0.030 &  3.54 & 0.23 & 0.16 \\
0.033--0.041 & 0.037 & 16.20 & 0.46 & 0.34 & 0.037 &  3.34 & 0.20 & 0.14 \\
0.041--0.050 & 0.045 & 13.48 & 0.38 & 0.27 & 0.045 &  2.86 & 0.14 & 0.13 \\
0.050--0.061 & 0.055 & 11.40 & 0.31 & 0.21 & 0.055 &  2.39 & 0.11 & 0.13 \\
0.061--0.074 & 0.067 & 10.09 & 0.27 & 0.18 & 0.067 &  2.20 & 0.10 & 0.09 \\
0.074--0.091 & 0.082 &  8.12 & 0.23 & 0.15 & 0.082 &  1.63 & 0.08 & 0.06 \\
0.091--0.111 & 0.100 &  6.41 & 0.20 & 0.12 & 0.100 &  1.31 & 0.08 & 0.08 \\
0.111--0.142 & 0.126 &  4.95 & 0.16 & 0.09 & 0.125 &  0.98 & 0.06 & 0.05 \\
0.142--0.183 & 0.161 &  3.66 & 0.16 & 0.08 & 0.160 &  0.68 & 0.05 & 0.04 \\
0.183--0.235 & 0.206 &  2.53 & 0.17 & 0.07 & 0.205 &  0.51 & 0.05 & 0.04 \\
0.235--0.301 & 0.262 &  1.52 & 0.08 & 0.05 & 0.262 &  0.30 & 0.04 & 0.04 \\
0.301--0.497 & 0.371 &  0.60 & 0.05 & 0.02 & 0.368 &  0.15 & 0.02 & 0.03
 \\[.1cm] \hline \hline 
  \multicolumn{2}{|c|}{   }
&  \multicolumn{3}{c||}{  }  &&  \multicolumn{3}{c|}{  }  \\[-.3cm]
  \multicolumn{2}{|c|}{Total Observed/Evt.}
  &  1.90 & 0.02 & 0.07 &       &  0.37 & 0.01 & 0.02
 \\[.1cm] \hline 
 \end{tabular}
\caption{
\label{xsgvee}
Measured \pnb s of neutral $K^0/\bar{K}^0$-mesons and \llb -hyperons per
hadronic $Z^0$ decay.
A 3.4\% normalization uncertainty is included in the systematic errors on the
observed totals, but not in those on the cross sections.}
\end{center}
\end{table}

Several sources of systematic uncertainty were investigated for the $K_s^0$ and
\llb\ analysis.
An important contribution to the overall $V^0$ spectrum is the track
reconstruction efficiency of the detector, which was tuned using the world
average measured charged multiplicity in hadronic $Z^0$ decays.
We take the $\pm$1.7\% normalization uncertainty discussed above (sec. 4.1) 
as the uncertainty on our reconstruction efficiency,
which corresponds to a normalization error on the $K^0/\bar{K}^0$
and \llb\, \pnb s of 3.4\%.
This uncertainty is independent of momentum and is not shown in any of
the figures or included in the errors listed in table \ref{xsgvee}.
The momentum-dependent term discussed above and
a conservative 50\% variation of an ad hoc correction \cite{kenb} to the
simulated efficiency for $V^0$s that decayed near the outer layers of the VXD
were also included as systematic uncertainties due to detector modelling.

Each of the cuts used to select $V^0$ candidates was varied independently
\cite{kenb} and the analysis repeated.
For each bin the $rms$ of this set of measurements was calculated
and assigned as the systematic uncertainty due to modelling of the acceptance.
For both the $K^0/\bar{K}^0$ and the \llb\, candidates,
the signal and background shapes used in the fits were varied.
Single and multiple independent Gaussians, without common
centers or fixed widths, were used for the signal.
Alternative background shapes included constants
and polynomials of differing orders.
In each case the fits were repeated on both data and simulated invariant mass
distributions and the $rms$ of the resulting \pnb s was assigned as a systematic
uncertainty.
The MC statistical error on the calculated reconstruction efficiency
was also assigned as a systematic error.
These errors were added in quadrature to give the total systematic error.

\subsection{Neutral \kkb\ and $\phi$ Production} 

We reconstructed the strange vector mesons $\phi$ and \kkb\ in the
charged decay modes $\phi \rightarrow K^+K^-$ and
$K^{*0}/\bar{K}^{*0} \rightarrow K^{\pm}\pi^{\mp}$ \cite{mihaid}.
In order to ensure good invariant mass resolution, tracks were required to have
at least 40 hits measured in the CDC,
a track fit quality of $\chi^2$/dof$<7$,
and a polar angle satisfying $\left|\cos\theta\right|<0.8$.
Pairs of oppositely charged tracks satisfying these requirements were
combined to form neutral candidates if 
a $\chi^2$ fit of the two tracks to a common vertex converged.
The background from long-lived species was rejected by requiring the fitted
vertex to be within 10 cm or 9$\sigma_l$ of the IP in three dimensions,
and within 4 cm or 6$\sigma_l$ in the plane transverse to the beam direction.
The background from $\gamma$-conversions was rejected by assigning the
electron mass to both tracks and requiring $m_{ee}$ to be greater than
70 MeV/c$^2$.

To reject the high combinatoric background from $\pi^+\pi^-$ pairs we used the
CRID to identify charged kaon candidate tracks.
Only liquid (gas) information was used for tracks with $p<2.5 \, (>3.5)$
GeV/c, and
liquid and gas information was combined for the remaining tracks.
For this analysis a track was considered ``identifiable" if
it extrapolated through an active region of the appropriate CRID radiator(s);
it was considered identified as a kaon if
the log-likelihood difference between the kaon and pion
hypotheses, ${\cal L}_K - {\cal L}_\pi$, exceeded 3.
These cuts are considerably looser than those used in section 4.1, in order to
maximize the acceptance for the neutral vector mesons.
Efficiencies for identifying selected tracks as kaons by this definition were
calibrated using the data in a manner similar to that described in section 2.2.
The $K \rightarrow K$ efficiency was found to have a momentum dependence very
similar to the $\pi \rightarrow \pi$ efficiency shown in the upper left plot
of fig. \ref{effpar}, with about 12\% lower amplitude.
There is no dip in the 5--10 GeV/c region since no cut was made against protons.
The $\pi \rightarrow K$ misidentification rate averages 10\% and is
roughly independent of momentum;
the p $\rightarrow K$ misidentification rate is substantial, especially in the
3--10 GeV/c region,
but protons constitute only a small part of the combinatoric background.

A track pair was accepted as a $\phi \rightarrow K^+K^-$ candidate if
both tracks were identified as kaons.
A pair was accepted as a $K^{*0} \rightarrow K^+\pi^-$ candidate if
one track was identified as a kaon and the other was not.
Thus a track pair cannot be both a \kkb\ and $\phi$ candidate.

\begin{figure}[tbp]
\centering
\epsfxsize=7.2cm
\leavevmode
\epsfbox{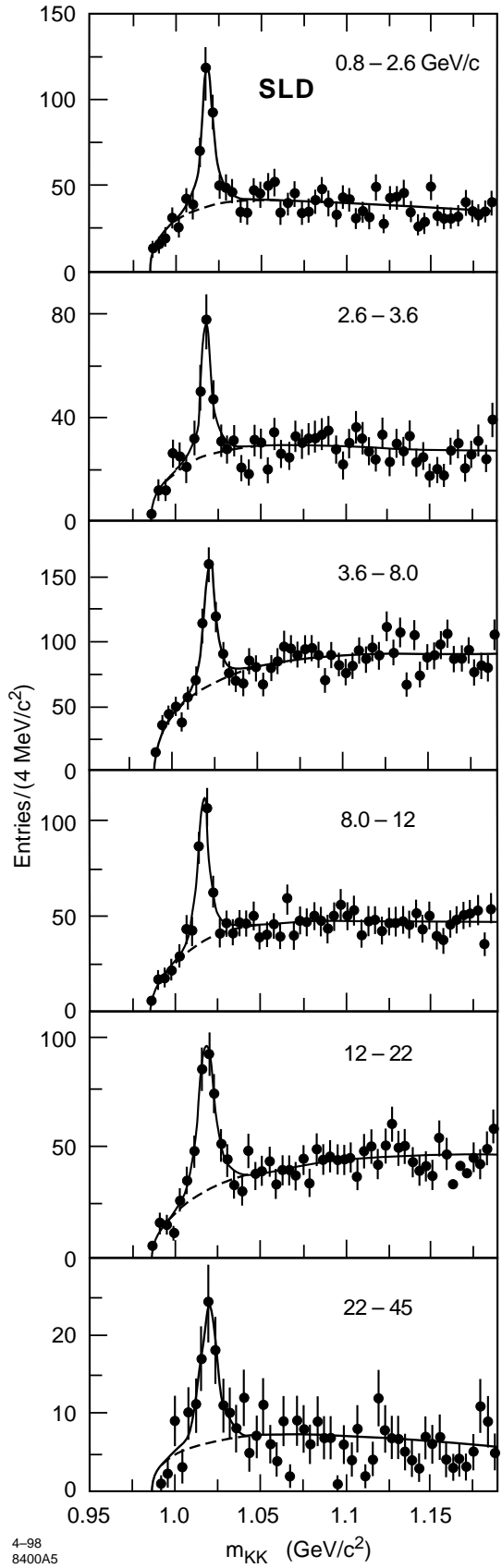}
\caption{\label{phipks}
Distributions of invariant mass $m_{KK}$ for $\phi$ candidates in six
momentum bins.
The points with error bars represent the data.
The solid curves represent the results of the fits described in the text;
the dashed curves represent the fitted background component.  }
\end{figure}

The $\phi$ candidates were binned in $x_p$, and the resulting
$m_{KK}$ distributions were fitted in a manner similar to that
described above for the $V^0$ candidates.
The signal shape was a sum of Gaussians of common center;
the center was fixed at the world-average mass value~\cite{pdg}, and
the amplitude and a resolution scale factor were free parameters.
A typical fitted scale factor was 1.08.
The background shape was parametrized as a threshold term
multiplied by a slowly decreasing exponential:
\begin{equation}
  P_{bkg}(x) = N   x^\gamma   e^{c_1 x + c_2 x^2
  +c_3 x^3 + c_4 x^4 +c_5 x^5}
  \label{eq:FBG}
\end{equation}
where $x = m_{KK} - 2m_K$, $N$ is an overall normalization factor,
and $\gamma$ and $c_{1...5}$ are free parameters.
Initial values of the background parameters were determined from fits to the
$m_{KK}$ distributions for simulated true combinatorial background
and for same-sign track pairs in the data.
The resulting parameters were consistent with each other and the functions
described the shape of the distribution for candidates in the data in the region
away from the signal peak.
The measured $m_{KK}$ distributions for the six $x_p$ bins are shown in fig.
\ref{phipks}, along with the results of the fits.

The case of the \kkb\ is considerably more complicated due to the natural
width of the $K^{*0}$ and the presence of
many reflections of resonances decaying into $\pi^+\pi^-(\pi^0)$.
The \kkb\ signal was parametrized using a relativistic Breit-Wigner with the
amplitude free and the center and width fixed to world-average
values \cite{pdg}.
The background was divided into combinatorial and resonant pieces.
The combinatorial piece was described by a polynomial parametrization similar to
that of the $\phi$ but with seven parameters.
Parameter values derived from fits to simulated combinatorial background and a
same-sign data test sample were found not to
agree with each other or with the opposite-sign data away from the peak,
and a search over a space of initial values was required in order to find the
best fit.

Knowledge of the resonant contributions to the background is essential,
since the $K^{*0}$ is a wide state and non-monotonic background variation
within its width can lead to systematic errors in the measured cross section.
We considered four classes of reflections:
\begin{itemize}
\item
$\rho^0 \rightarrow \pi^+\pi^-$, $K_s^0\rightarrow \pi^+\pi^-$, and
$\omega^0, \eta, \eta' \rightarrow N\pi$, where one of the charged pions is
misidentified as a $K^\pm$.
These backgrounds are large, even after reduction by a factor of about 5 by the
particle identification.
They are particularly important since the combination of $\rho$ and $\omega$
decays gives rise to a dip in the total background near the center of the signal
peak, and there is some uncertainty as to the shape of the $\rho$ resonance in
$Z^0$ decays (see ref. \cite{alephkst}).
\item
$\gamma$ conversions where one electron is misidentified as a kaon.
These are removed effectively by the $m_{ee}$ cut against $\gamma$ conversions
noted above.
\item
$\phi \rightarrow K^+K^-$, where one track is identified as a kaon but the other
is not.
This background is reduced substantially by the requirement that only one of
the tracks in the pair is identified as a kaon.
\item
$\Lambda^0 \rightarrow p\pi$, where the proton is misidentified as a kaon.
These are removed effectively by the cut against long-lived species noted above.
This and the last two categories give rise to a more pronounced shoulder
in the background just below the signal peak, so their removal is quite useful
in obtaining a robust fit.
 \end{itemize}

\begin{figure}[tbp]
\centering
\epsfxsize=9.0cm
\leavevmode
\epsfbox{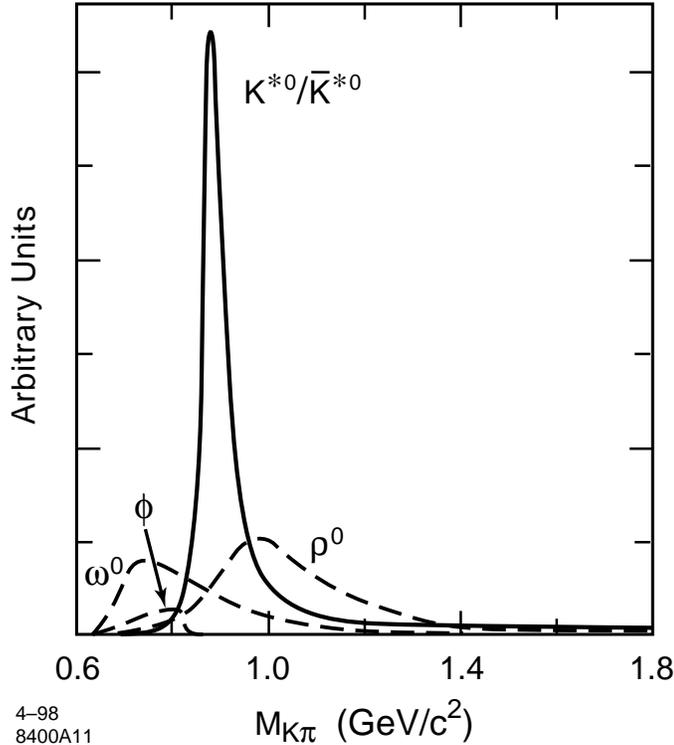}
\caption{\label{reson}
Simulated relative contributions of the $K^{*0}/\bar{K}^{*0}$ signal (line)
and of various resonant backgrounds (dashed lines) to the $m_{K\pi}$
distribution after all analysis cuts.
    }
\end{figure}

\begin{figure}[tbp]
\centering
\epsfxsize=14.9cm
\leavevmode
\epsfbox{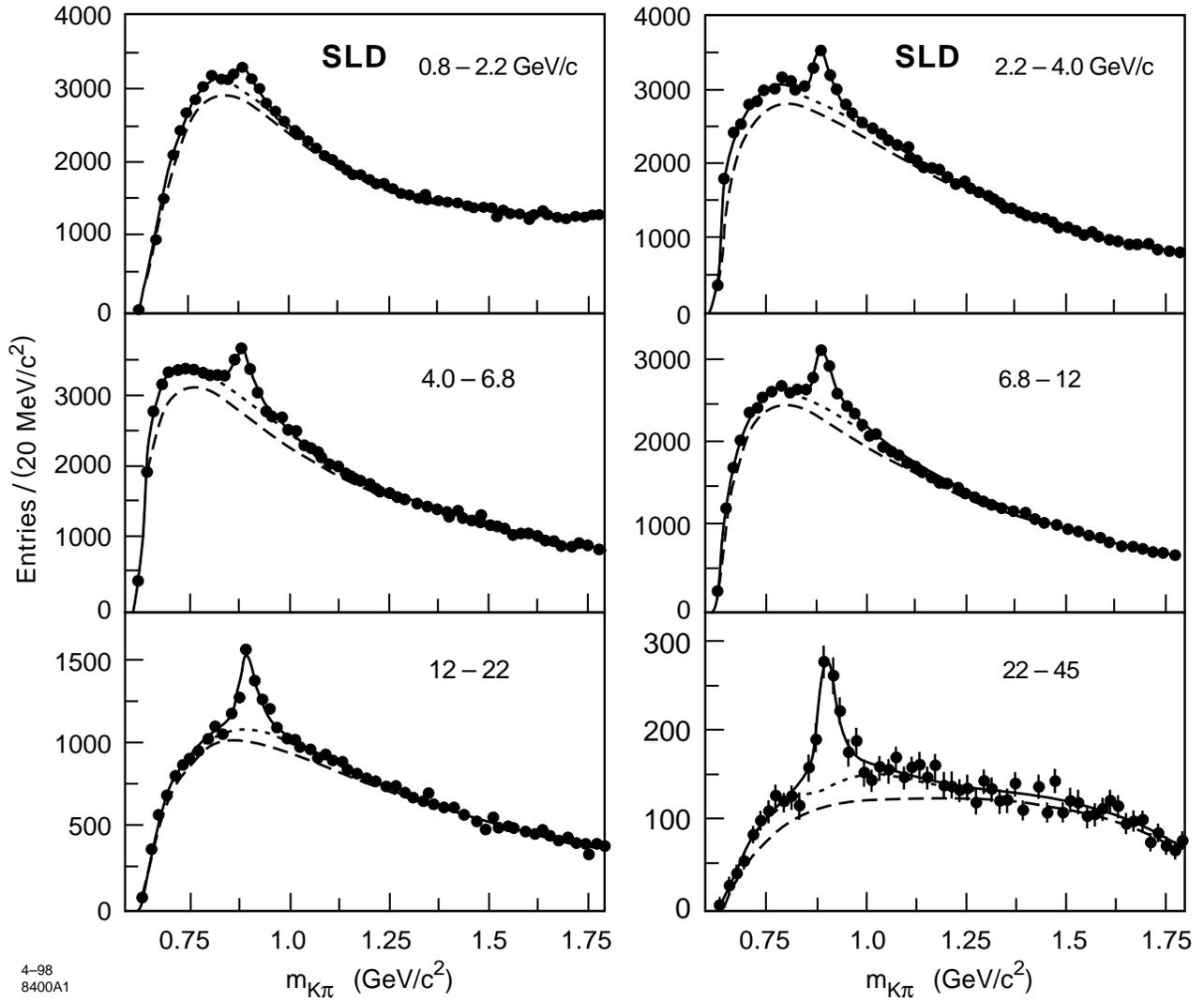}
\caption{\label{kstpks}
Distributions of invariant mass $m_{K\pi}$ for $K^{*0}/\bar{K}^{*0}$ candidates
in six momentum bins.
The points represent the data.
The solid curves represent the results of the fits described in the text;
the dotted and dashed curves represent the fitted total background and
combinatoric background components, respectively. }
\end{figure}

The shape of the $m_{K\pi}$ distribution for each reflection was parametrized by
a smooth function fitted to its simulated $m_{K\pi}$ distribution,
and its total production cross section was set to the world average
value~\cite{pdg} for $Z^0$ decays.
Figure \ref{reson} shows the simulated relative contributions from the main
resonant backgrounds along with the simulated signal, which was scaled
to match our measured total cross section (see below).
The set of reflection functions was added to the combinatorial function to
give the total background function.
A scale factor for each of the four categories of reflections was included as a
free parameter in the fit to
account for possible mismodelling of the misidentification rates;
their fitted values were consistent with unity.
Figure \ref{kstpks} shows the $m_{K\pi}$ distribution for each 
momentum bin, along with the results of the fits. 

\begin{figure}[tbp]
\centering
\epsfxsize=13.5cm
\leavevmode
\epsfbox{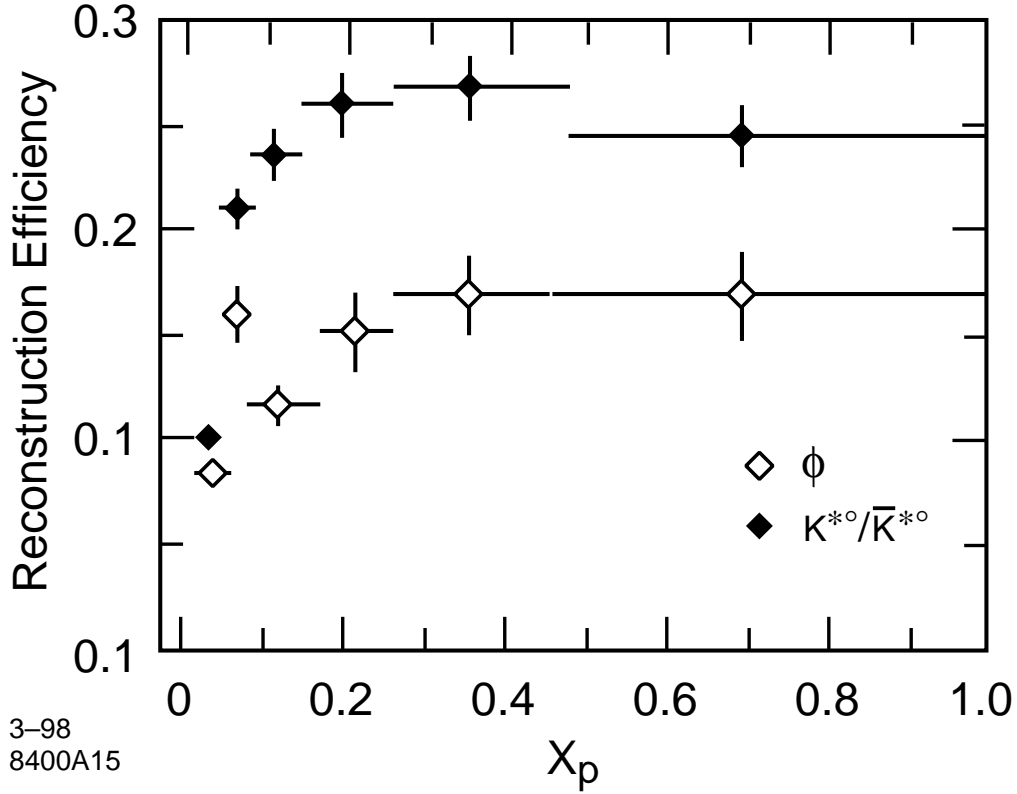}
\caption{
\label{effsksphi}
The simulated reconstruction efficiencies as a function of $x_p$
for $\phi$ (open diamonds) and \kkb\ (diamonds).
The charged decay branching ratios are included in the efficiency.
The dip in the $\phi$ efficiency at $x_p\approx 0.13$ reflects the dip in the
CRID $K$-$\pi$ separation at $p\approx 2.5$ GeV/c (see fig. 3, upper left).}
\end{figure}

As for the $K_s^0$ and \llb\, analysis, the $\phi$ and $K^{*0}/\bar{K}^{*0}$
reconstruction efficiencies were determined using the simulation, and are
shown in fig. \ref{effsksphi}.
Differential cross sections were calculated in the same way as for the
$K^0_s$ and \llb, and the results are shown in fig. \ref{xsall} and
listed in table \ref{xsgneu}.

\begin{table}
\begin{center}
 \begin{tabular}{|c|c|l@{$\pm$}l@{$\pm$}l||c|c|l@{$\pm$}l@{$\pm$}l|} \hline
 \multicolumn{10}{|c|}{   } \\[-.3cm]
 \multicolumn{10}{|c|}{Neutral Strange Meson Production} \\[.2cm]\hline\hline
  &&  \multicolumn{3}{c||}{  } & &&  \multicolumn{3}{c|}{  }  \\[-.3cm]
 $x_p$ Range & $<\! x_p\! >$ & \multicolumn{3}{c||}{1/N d$n_{K^{*0}}$/d$x_p$} &
 $x_p$ Range & $<\! x_p\! >$ & \multicolumn{3}{c|}{1/N d$n_{\phi}$/d$x_p$}  \\[.1cm] \hline 
  &&  \multicolumn{3}{c||}{  } &  &&  \multicolumn{3}{c|}{  }  \\[-.3cm]
0.018--0.048 & 0.033&4.69 &0.56 &0.33  & 0.018--0.057 & 0.037&0.744 &0.074 &0.048 \\
0.048--0.088 & 0.068&3.79 &0.21 &0.17  & 0.057--0.079 & 0.068&0.411 &0.055 &0.033\\
0.088--0.149 & 0.118&2.23 &0.13 &0.14  & 0.079--0.175 & 0.127&0.255 &0.026 &0.021\\
0.149--0.263 & 0.206&1.012&0.056&0.062 & 0.175--0.263 & 0.215&0.167 &0.018 &0.020 \\
0.263--0.483 & 0.342&0.343&0.019&0.019 & 0.263--0.483 & 0.357&0.0739&0.0068&0.0085\\
0.483--1.000 & 0.607&0.051&0.004&0.004 & 0.483--1.000 & 0.689&0.0089&0.0015&0.0011
 \\[.1cm] \hline \hline 
   \multicolumn{2}{|c|}{   }
&  \multicolumn{3}{c||}{  } &  &&  \multicolumn{3}{c|}{  }  \\[-.3cm]
  \multicolumn{2}{|c|}{Total Observed/Evt.}
    & 0.647 & 0.022 & 0.029 &              &      &0.0985&0.0046&0.0055
 \\[.1cm] \hline 
 \end{tabular}
\caption{
\label{xsgneu}
Measured \pnb s of \kkb\ and $\phi$ mesons per hadronic $Z^0$ decay.
A 3.4\% normalization uncertainty is included in the systematic errors on the
observed totals, but not in those on the cross sections.}
\end{center}
\end{table}

Systematic uncertainties for this analysis were grouped into efficiency and
fit-related categories.
The dominant contributions to the efficiency category were the uncertainty in
the track-finding efficiency (see above) and the uncertainty in
kaon identification efficiency,
for which the statistical error on the calibration from the data was used.
The total uncertainties on the reconstruction efficiencies were 
4--6\% for \kkb\ and 6--11\% for $\phi$, depending on momentum.

In the case of the $\phi$, fitting systematics were evaluated by varying the
signal shape as in the $V^0$ analysis.
In addition, fits were performed with the signal center shifted by plus and
minus the error on the world-average mass value.
The effect of background fluctuations was evaluated by taking the largest
variation in the result over a set of fits done with the background shape
parameters $c_i$ fixed to all combinations of their fitted values
$\pm$1$\sigma$.
The total fitting uncertainties were 2--8\%.

In the case of the \kkb , we considered the same variations, as well as
variation of the signal width by $\pm$1$\sigma$ from the world-average
value and several variations of the resonant background.
Fits were performed with the misidentification scale factors fixed to their
fitted values $\pm$50\% for the $\pi\pi$
category and $\pm$15\% for the others, corresponding to roughly twice the error
on our measured misidentification rates.
All 16 combinations were considered, and the largest
variation taken as a systematic error.
The cross section for production of each resonance was varied by the error on
the world-average value.
The sizes of the $\rho$ and $\omega$ contributions were varied in all four
combinations of $\pm$30\% and $\pm$10\%, respectively, and the largest variation
was taken as a systematic error.
Following \cite{alephkst} an error due to the uncertainty in the $\rho^0$
lineshape
was evaluated by shifting the $\rho$ reflection function down by 40 MeV/c$^2$.
The total fitting uncertainties were 2--6\%.

\subsection{Hadron Production in Inclusive Hadronic $Z^0$ Decays}

\begin{figure}
\vspace{-1.cm}
 \hspace*{0.5cm}   
   \epsfxsize=4.0in
   \begin{center}\mbox{\epsffile{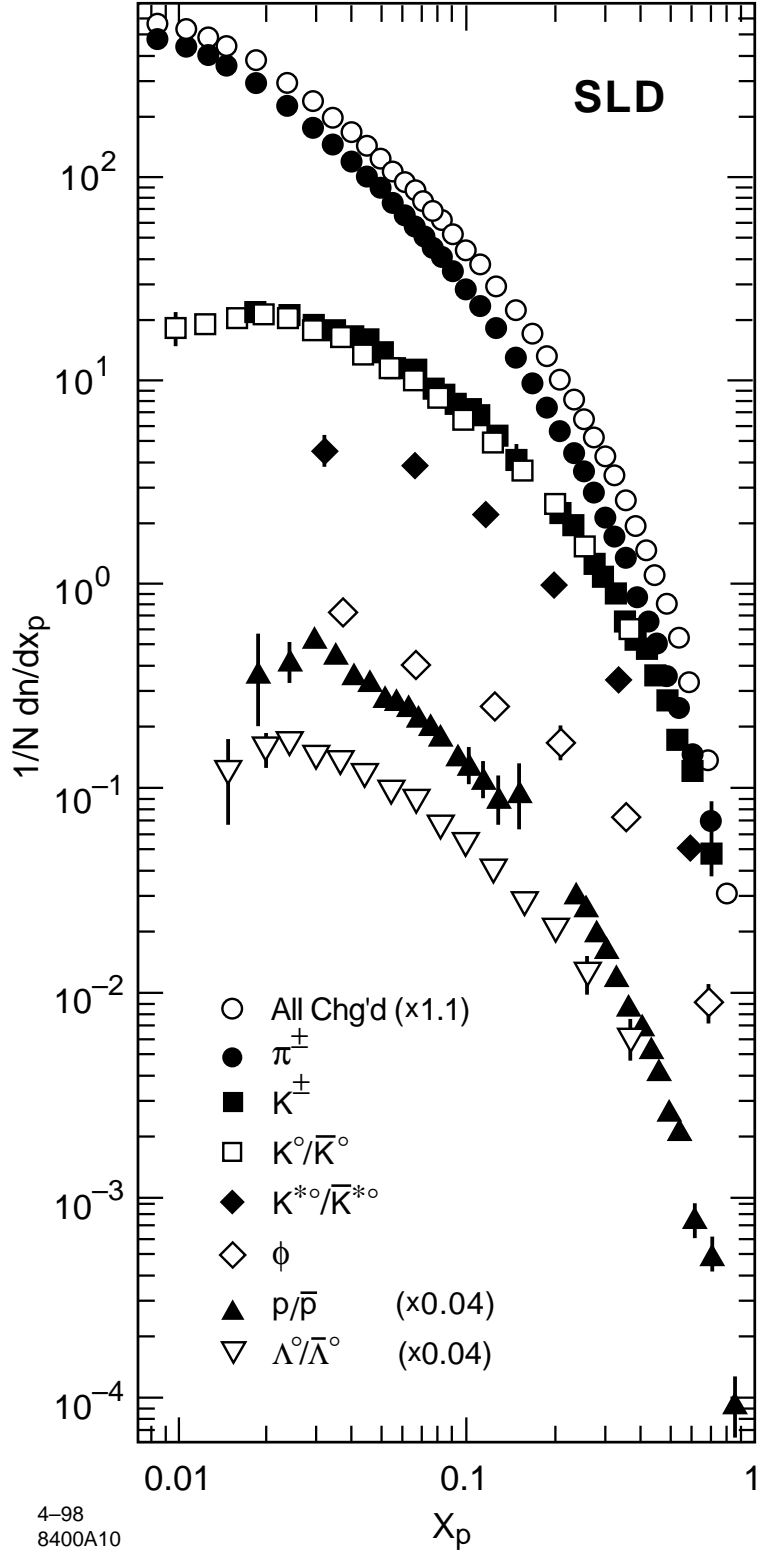}}\end{center}
  \caption{ 
 \label{xsall}
Differential cross sections per hadronic $Z^0$ decay per unit
$x_p$ for inclusive charged particles
(open circles),
$\pi^\pm$ (circles), $K^\pm$ (squares), $K^0/\bar{K}^0$ (open squares),
$K^{*0}/\bar{K}^{*0}$ (diamonds),
$\phi$ (open diamonds), p/$\bar{\rm p}$ (triangles), and
$\Lambda^0 / \bar{\Lambda}^0$ (open triangles).
The baryon and all-charged \pnb s have been scaled by 0.04 and 1.1,
respectively, for clarity.
The error bars represent statistical and systematic errors added in quadrature.
Each point is plotted at the average $x_p$ value of reconstructed particles
in that bin (see tables 2--7).
    }
\end{figure}

Our measured \pnb s per hadronic event of the seven hadron
species are shown as a function of $x_p$ in
fig. \ref{xsall}, along with that of inclusive charged particles.
At low $x_p$ pions are seen to dominate the hadrons produced in hadronic $Z^0$
decays.
For example, at $x_p \approx 0.03$, pseudoscalar $K^{\pm}$ and $K^0/\bar{K}^0$
are produced
at a rate about ten times lower than pions, vector $K^{*0}$ are suppressed by
an additional factor of $\sim$4, and the doubly strange vector $\phi$ by another
factor of $\sim$12.
The most commonly produced baryons, protons, are suppressed by a factor of
$\sim$25 relative to pions, and the strange baryon \llb\ by an additional
factor of $\sim$3.

These results are in general consistent with previous measurements from
experiments at LEP \cite{bohrer}, provided that the point-to-point
correlations in the systematic errors are taken into account.
However, although our proton \pnb\ for $x_p>0.35$ is consistent with that
measured by ALEPH \cite{aleph},
it is not consistent with that measured by OPAL \cite{opal}.

\begin{figure}
\vspace{-1.cm}
 \hspace*{0.5cm}   
   \epsfxsize=6.45in
   \begin{center}\mbox{\epsffile{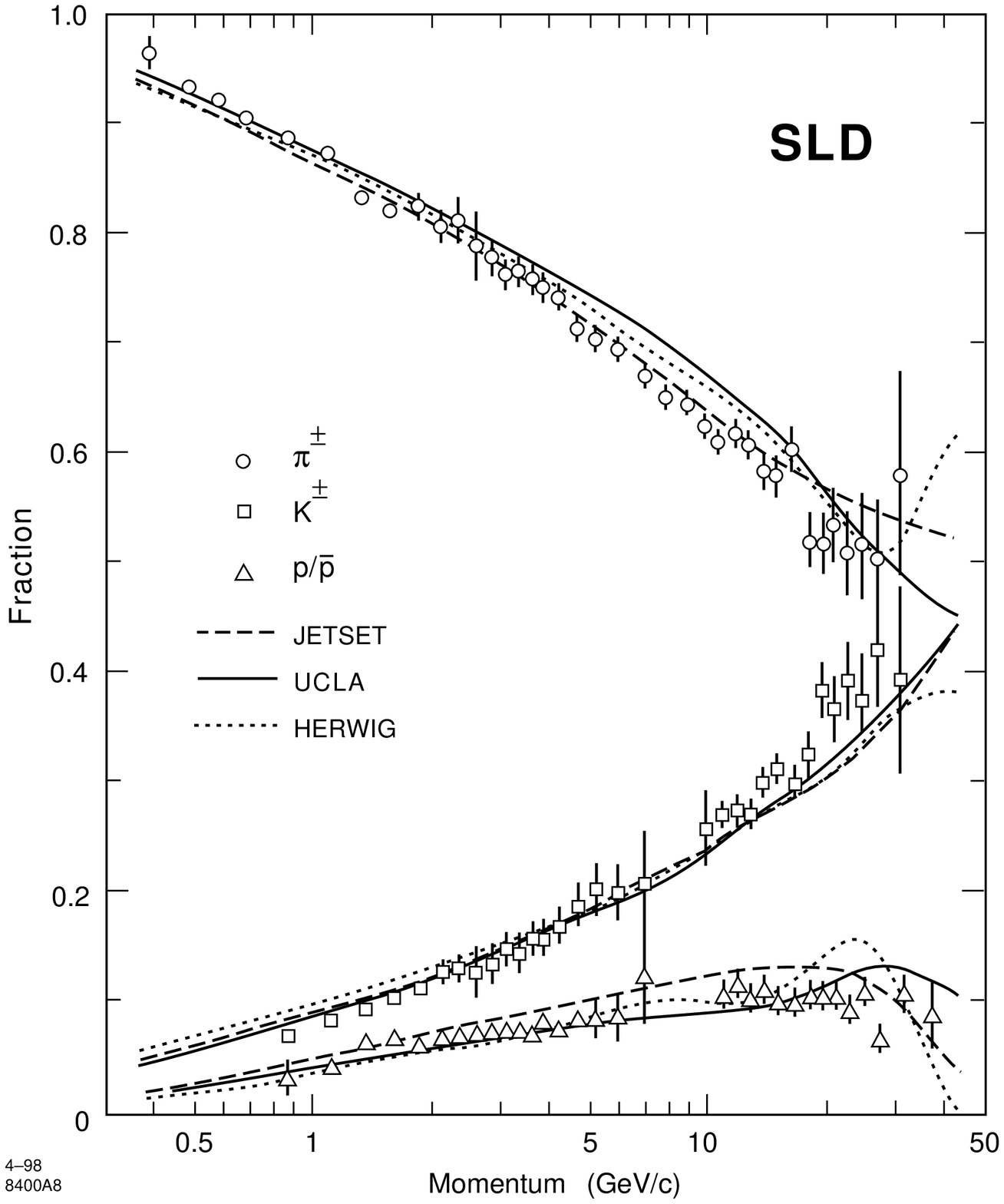}}\end{center}
  \caption{ 
 \label{fallmc}
Comparison of our measured charged hadron fractions (symbols)
with the predictions of the JETSET (dashed lines), UCLA (solid lines) and HERWIG
(dotted lines) fragmentation models.
    }
\end{figure} 

\begin{figure}
\vspace{-1.cm}
 \hspace*{0.5cm}   
   \epsfxsize=4.3in
   \begin{center}\mbox{\epsffile{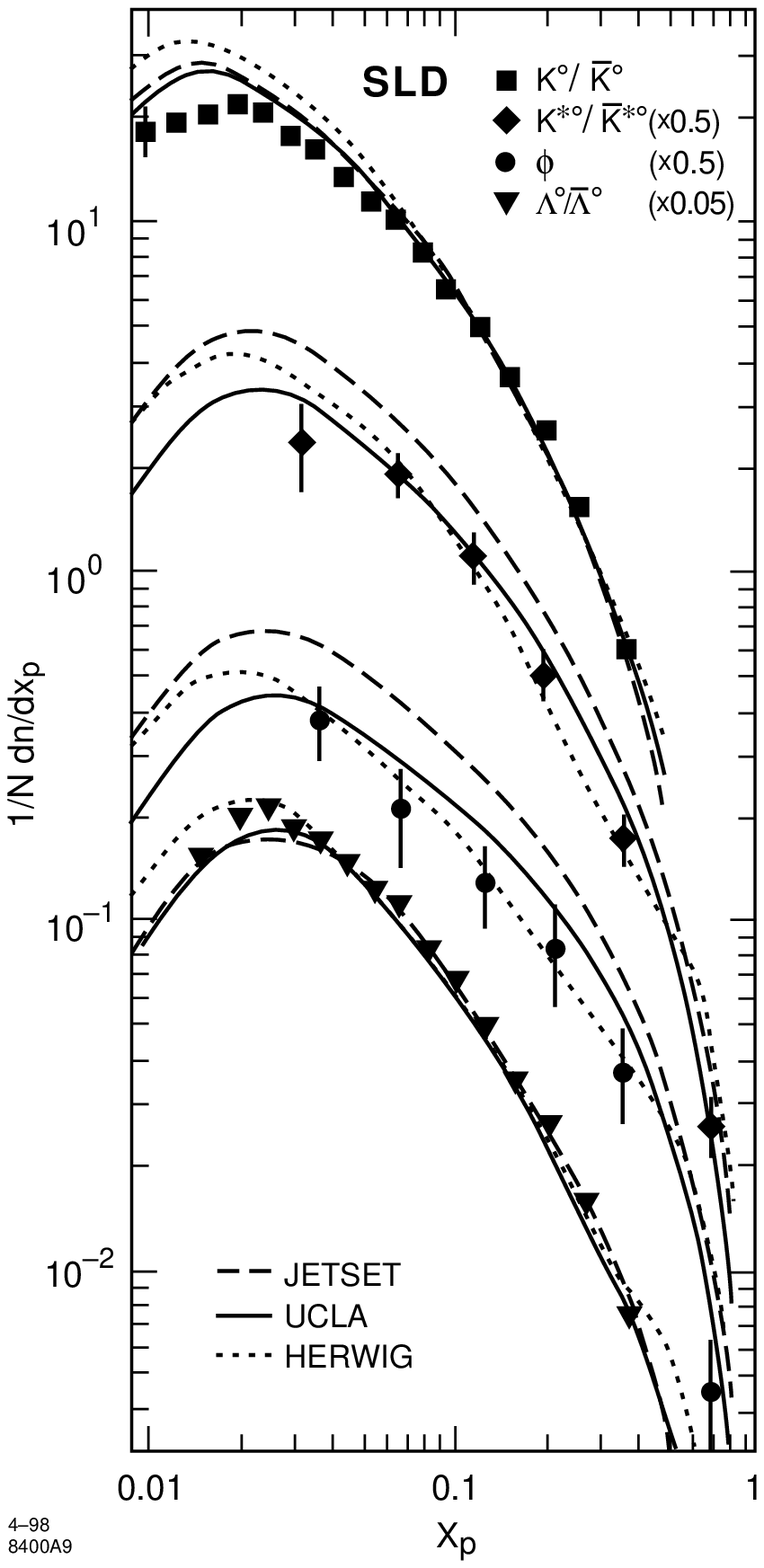}}\end{center}
  \caption{ 
 \label{xsallmc}
Comparison of our measured neutral hadron \pnb s
with the predictions of three fragmentation models.
    }
\end{figure} 

We compared our results with the predictions of the JETSET 7.4, UCLA 4.1 and
HERWIG 5.8
event generators described in section 1, using in all cases
the default parameters.
Figures \ref{fallmc} and \ref{xsallmc} show the
charged fractions and the neutral \pnb s, respectively,
along with the predictions of these three models.
The momentum dependence for each of the seven hadron species 
is reproduced qualitatively by all models.
For momenta below about 1.5 GeV/c, all models overestimate the kaon fraction
significantly and all except UCLA underestimate the pion fraction by about
2$\sigma$ (taking into account the correlation in the experimental errors).
In the 5--10 GeV/c range UCLA and HERWIG overestimate the pion fraction by
2--3$\sigma$.
For $p>10$ GeV/c, JETSET overestimates the proton fraction, but describes the
momentum dependence. 
In this momentum region, HERWIG and UCLA predict a momentum dependence in the
proton fraction that is inconsistent with the data.

In the case of $K^0/\bar{K}^0$,
all models describe the data well at high $x_p$, but
overestimate the cross section at low $x_p$ by as much as 50\%.
A similar excess was seen in the charged kaon fraction (see fig \ref{fallmc}).
In the case of \llb , JETSET and UCLA describe the data well except
for a 10\% shortfall near $x_p=0.02$.
HERWIG describes the data well except for the lowest and highest $x_p$ points,
where it overestimates the production.
The structure in the HERWIG prediction at very high $x_p$ is similar to that
seen in the proton fraction, and is also visible to varying degrees in the
predictions for the neutral strange mesons.
In the case of \kkb, JETSET is high by a roughly constant factor of 1.5 across
the $x_p$ range;
HERWIG and UCLA reproduce the data except at the lowest $x_p$ point.
In the case of $\phi$, JETSET is high by a factor of two over all $x_p$, UCLA is
high for $x_p>0.06$, and HERWIG describes the data except at the highest $x_p$
point.

\section{Flavor-Dependent Analysis}

The analyses described above were repeated on the light-, $c$- and $b$-tagged 
event samples described in section 3, to yield \pnb s $R^{ktag}_h$ for each
hadron species $h$ in each tagged sample.
True \pnb s $R^m_h$ in events of the three flavor types,
$k,m=l$, $c$, $b$, representing events of the types \zuds, \zcc, and \zbb,
respectively, were extracted by solving for each species
$h$ the relations:
\begin{equation}
R^{ktag}_h = \frac{\Sigma_m B^h_{mk} \epsilon_{mk} F_m R^m_h}
                   {\Sigma_m          \epsilon_{mk} F_m} .
\end{equation}
Here, $F_m$ is the fraction of hadronic $Z^0$ decays of flavor type $m$, taken
from the Standard Model, $\epsilon_{mk}$ is the event tagging efficiency matrix
(see table \ref{tlveff}), 
and $B_{mk}^h$ represents the momentum-dependent bias of tag $k$ toward
selecting events of flavor $m$ that contain hadrons of species $h$.
Ideally all biases would be unity in this formulation.
The biases were calculated from the MC simulation as
$B^h_{mk}=(n^h_{m,ktag}/N_{m,ktag})/(n^h_m/N_m)$, where $N_m$ $(n^h_m)$
is the number of simulated events (hadrons of species $h$ in events) of true
flavor $m$ and $N_{m,ktag}$ $(n^h_{m,ktag})$ is the number of ($h$-hadrons in)
those events that are tagged as flavor $k$.
The diagonal bias values \cite{tomp,kenb,mihaid}
are within a few percent of unity
for the charged hadrons, $\phi$ and $K^{*0}$,
reflecting a small multiplicity dependence of the flavor tags.
They deviate by as much as 10\% from unity for the $K^0/\bar{K}^0$ and \llb ,
since some tracks from
$V^0$ decays are included in the tagging track sample and have large impact
parameter.
The off-diagonal bias values deviate from unity by a larger amount, but
these have little effect on the unfolded results.

\begin{table}
\begin{center}
 \begin{tabular}{|r@{--}r|c|r@{$\pm$}l|r@{$\pm$}l|
                          r@{$\pm$}l|r@{$\pm$}l|r@{$\pm$}l|} \hline
\multicolumn{2}{|c|}{  } && \multicolumn{6}{|c|}{  } &
\multicolumn{4}{|c|}{  } \\[-.3cm]
\multicolumn{2}{|c|}{$x_p$} &&
\multicolumn{6}{|c|}{$\pi^{\pm}$ Production Cross Sections} &
\multicolumn{4}{|c|}{Ratios} \\
  \multicolumn{2}{|c|}{Range} & $<\! x_p\! >$
& \multicolumn{2}{|c|}{$u\bar{u}$, $d\bar{d}$, $s\bar{s}$}
& \multicolumn{2}{|c|}{$c\bar{c}$}
& \multicolumn{2}{|c|}{$b\bar{b}$}
& \multicolumn{2}{|c|}{$c$:$uds$}
& \multicolumn{2}{|c|}{$b$:$uds$} \\[.1cm]\hline
 0.008 & 0.010 & 0.009 &  467.2  &  9.0  &   493.  & 37.
                       &  508.1  & 10.6  &    1.05 &  0.09 &  1.09 &  0.03 \\
 0.010 & 0.012 & 0.011 &  428.1  &  8.2  &   413.  & 34.
                       &  481.2  &  9.7  &    0.96 &  0.09 &  1.12 &  0.03 \\
 0.012 & 0.014 & 0.013 &  383.2  &  7.3  &   403.  & 30.
                       &  441.3  &  8.6  &    1.05 &  0.09 &  1.15 &  0.03 \\
 0.014 & 0.016 & 0.015 &  337.1  &  6.6  &   375.  & 27.
                       &  388.4  &  7.9  &    1.11 &  0.09 &  1.15 &  0.03 \\
 0.016 & 0.022 & 0.019 &  274.7  &  4.6  &   301.  & 19.
                       &  333.6  &  4.8  &    1.10 &  0.08 &  1.21 &  0.02 \\
 0.022 & 0.027 & 0.025 &  214.5  &  3.7  &   230.  & 15.
                       &  264.4  &  4.1  &    1.07 &  0.08 &  1.23 &  0.03 \\
 0.027 & 0.033 & 0.030 &  165.5  &  3.1  &   178.  & 13.
                       &  205.4  &  3.6  &    1.08 &  0.09 &  1.24 &  0.03 \\
 0.033 & 0.038 & 0.036 &  137.2  &  2.7  &   141.  & 11.
                       &  166.9  &  3.3  &    1.03 &  0.09 &  1.22 &  0.03 \\
 0.038 & 0.044 & 0.041 &  117.2  &  2.5  &   111.  & 10.
                       &  141.4  &  3.2  &    0.95 &  0.10 &  1.21 &  0.04 \\
 0.044 & 0.049 & 0.047 &   98.4  &  2.4  &    96.  & 10.
                       &  118.6  &  3.3  &    0.97 &  0.11 &  1.20 &  0.04 \\
 0.049 & 0.055 & 0.052 &   83.6  &  2.4  &    86.  & 10.
                       &  106.3  &  3.5  &    1.03 &  0.13 &  1.27 &  0.06 \\
 0.055 & 0.066 & 0.060 &   66.9  &  1.4  &    65.8 &  5.9
                       &   84.2  &  2.0  &    0.98 &  0.10 &  1.26 &  0.04 \\
 0.066 & 0.077 & 0.071 &   52.8  &  1.1  &    48.8 &  4.8
                       &   64.0  &  1.6  &    0.93 &  0.10 &  1.21 &  0.04 \\
 0.077 & 0.088 & 0.082 &   41.61 &  0.95 &    43.4 &  4.0
                       &   49.2  &  1.4  &    1.04 &  0.11 &  1.18 &  0.04 \\
 0.088 & 0.099 & 0.093 &   34.11 &  0.81 &    32.3 &  3.5
                       &   40.6  &  1.2  &    0.95 &  0.11 &  1.19 &  0.04 \\
 0.099 & 0.110 & 0.104 &   28.74 &  0.72 &    23.6 &  3.1
                       &   30.1  &  1.1  &    0.82 &  0.11 &  1.05 &  0.04 \\
 0.110 & 0.132 & 0.120 &   21.64 &  0.46 &    21.3 &  2.1
                       &   22.72 &  0.76 &    0.99 &  0.10 &  1.05 &  0.04 \\
 0.132 & 0.164 & 0.147 &   15.26 &  0.31 &    12.4 &  1.4
                       &   13.54 &  0.51 &    0.81 &  0.10 &  0.89 &  0.04 \\
 0.164 & 0.186 & 0.175 &   10.76 &  0.26 &     8.8 &  1.1
                       &    8.26 &  0.42 &    0.82 &  0.11 &  0.77 &  0.04 \\
 0.186 & 0.208 & 0.197 &    8.44 &  0.22 &    6.66 &  0.90
                       &    5.57 &  0.34 &    0.79 &  0.11 &  0.66 &  0.04 \\
 0.208 & 0.230 & 0.219 &    6.29 &  0.19 &    6.03 &  0.77
                       &    3.93 &  0.29 &    0.96 &  0.13 &  0.62 &  0.05 \\
 0.230 & 0.274 & 0.251 &    4.81 &  0.12 &    3.77 &  0.48
                       &    2.52 &  0.18 &    0.78 &  0.11 &  0.52 &  0.04 \\
 0.274 & 0.318 & 0.294 &   2.932 & 0.090 &    2.62 &  0.36
                       &    1.39 &  0.13 &    0.89 &  0.13 &  0.47 &  0.05 \\
 0.318 & 0.384 & 0.348 &   1.815 & 0.059 &    1.69 &  0.23
                       &   0.695 & 0.084 &    0.93 &  0.14 &  0.38 &  0.05 \\
 0.384 & 0.471 & 0.421 &   0.915 & 0.037 &    0.42 &  0.14
                       &   0.380 & 0.053 &    0.46 &  0.16 &  0.42 &  0.06 \\
 0.471 & 0.603 & 0.529 &   0.376 & 0.023 &   0.146 & 0.084
                       &   0.108 & 0.031 &    0.39 &  0.22 &  0.29 &  0.08 \\
 0.603 & 0.768 & 0.654 &   0.145 & 0.017 &   0.027 & 0.054
                       &   0.006 & 0.015 &    0.18 &  0.37 &  0.04 &  0.10 \\
  [.1cm] \hline
  \end{tabular}
\caption{
\label{xsfpi}
Measured \pnb s (1/N)d$n_{\pi^{\pm}}$/d$x_p$ for the production of charged
pions per $Z^0$ decay into light ($u$, $d$, $s$), $c$ and $b$ primary flavors.
The errors are the sum in quadrature of statistical errors and those systematic
uncertainties arising from the unfolding procedure.  Systematic errors common
to the three flavors are not included.  The $<\! x_p \!>$ values for the three
flavor samples are consistent in each bin, and have been averaged.}
\end{center}
\end{table}

\begin{table}
\begin{center}
 \begin{tabular}{|r@{--}r|c|r@{$\pm$}l|r@{$\pm$}l|
                          r@{$\pm$}l|r@{$\pm$}l|r@{$\pm$}l|} \hline
\multicolumn{2}{|c|}{  } && \multicolumn{6}{|c|}{  } &
\multicolumn{4}{|c|}{  } \\[-.3cm]
\multicolumn{2}{|c|}{$x_p$} &&
\multicolumn{6}{|c|}{$K^{\pm}$ Production Cross Sections} &
\multicolumn{4}{|c|}{Ratios} \\
  \multicolumn{2}{|c|}{Range} & $<\!\! x_p\!\! >$
& \multicolumn{2}{|c|}{$u\bar{u}$, $d\bar{d}$, $s\bar{s}$}
& \multicolumn{2}{|c|}{$c\bar{c}$}
& \multicolumn{2}{|c|}{$b\bar{b}$}
& \multicolumn{2}{|c|}{$c$:$uds$}
& \multicolumn{2}{|c|}{$b$:$uds$} \\[.1cm]\hline
 0.016 & 0.022 & 0.019 &   22.6  &  1.2  &   19.5  &  5.0
                       &   24.3  &  1.7  &\multicolumn{2}{|c|}{ }&1.08 &0.09 \\
 0.022 & 0.027 & 0.025 &   19.2  &  1.1  &   26.8  &  4.7
                       &   22.3  &  1.6  &    1.11 &  0.18 &  1.16 &  0.11 \\
\cline{10-11}
 0.027 & 0.033 & 0.030 &   18.6  &  1.1  &   16.4  &  4.4
                       &   22.3  &  1.6  &\multicolumn{2}{|c|}{ }&1.20 &0.11 \\
 0.033 & 0.038 & 0.036 &   17.0  &  1.0  &   14.9  &  4.4
                       &   22.8  &  1.6  &    0.88 &  0.19 &  1.34 &  0.12 \\
\cline{10-11}
 0.038 & 0.044 & 0.041 &   14.6  &  1.1  &   18.5  &  4.5
                       &   19.7  &  1.6  &\multicolumn{2}{|c|}{ }&1.35 &0.15 \\
 0.044 & 0.049 & 0.047 &   15.3  &  1.2  &   13.6  &  4.9
                       &   19.9  &  1.8  &    1.08 &  0.24 &  1.30 &  0.15 \\
\cline{10-11}
 0.049 & 0.055 & 0.052 &   14.5  &  1.3  &    6.1  &   5.2
                       &   18.3  &  1.9 &\multicolumn{2}{|c|}{ }&1.26 &0.17 \\
 0.055 & 0.066 & 0.060 &   10.29 &  0.85 &   10.7  &  3.6
                       &   15.2  &  1.4 &    0.78 &  0.26 &  1.48 &  0.18 \\
\cline{10-11}
 0.066 & 0.077 & 0.071 &   9.00  &  0.73 &    9.5  &  3.1
                       &   14.5  &  1.2  &\multicolumn{2}{|c|}{ }&1.61 &0.19 \\
 0.077 & 0.088 & 0.082 &    7.38 &  0.70 &    8.9  &  3.0
                       &   13.4  &  1.2 &    1.13 &  0.28 &  1.82 &  0.23 \\
\cline{10-11}
 0.088 & 0.099 & 0.093 &   6.12  & 0.70  &   10.5  &  3.0
                       &   10.6  &  1.1  &\multicolumn{2}{|c|}{ }&1.73 &0.27 \\
 0.099 & 0.110 & 0.104 &   6.00  & 0.75  &   10.2  &  3.2
                       &    8.4  & 1.2   &    1.72 &  0.40 &  1.40 &  0.26 \\
\cline{10-11}
 0.110 & 0.132 & 0.120 &   4.78  & 0.57  &    8.1  & 2.5
                       &   8.71  & 0.98 &\multicolumn{2}{|c|}{ }&1.82 &0.30 \\
 0.132 & 0.164 & 0.147 &   3.30  & 0.61  &    8.0  & 2.6
                       &   3.65 & 0.94 &    2.06 &  0.54 &  1.11 &  0.35 \\
\cline{10-11}
 0.208 & 0.230 & 0.219 &   2.29  & 0.17  &    2.64 & 0.70
                       &   2.01  & 0.27  &    1.16 &  0.32 &  0.88 &  0.13 \\
\cline{10-11}
 0.230 & 0.274 & 0.251 &   1.498 & 0.089 &    3.29 & 0.37
                       &   1.18  & 0.14  &\multicolumn{2}{|c|}{ }&0.79 &0.10 \\
 0.274 & 0.318 & 0.294 &   1.272 & 0.068 &    1.30 & 0.27
                       &   0.811 & 0.098 &    1.66 &  0.19 &  0.64 &  0.08 \\
\cline{10-11}
 0.318 & 0.384 & 0.348 &   0.925 & 0.046 &    0.66 & 0.17
                       &   0.496 & 0.060 &\multicolumn{2}{|c|}{ }&0.54 &0.07 \\
 0.384 & 0.471 & 0.421 &   0.548 & 0.032 &    0.65 & 0.12
                       &   0.113 & 0.035 &    0.92 &  0.15 &  0.21 &  0.06 \\
\cline{10-11}
 0.471 & 0.603 & 0.529 &   0.266 & 0.020 &   0.229 & 0.073
                       &   0.043 & 0.021 &\multicolumn{2}{|c|}{ }&0.16 &0.08 \\
 0.603 & 0.768 & 0.654 &   0.101 & 0.015 & --0.003 & 0.046
                       &   0.020 & 0.014 &    0.57 &  0.24 &  0.20 &  0.14 \\
  [.1cm] \hline
  \end{tabular}
\caption{
\label{xsfka}
Differential cross sections for the production of $K^\pm$ mesons
per $Z^0$ decay into light, $c$ and $b$ primary flavors.
}
\end{center}
\end{table}

\begin{table}
\begin{center}
 \begin{tabular}{|r@{--}r|c|r@{$\pm$}l|r@{$\pm$}l|
                           r@{$\pm$}l||r@{$\pm$}l|r@{$\pm$}l|} \hline
\multicolumn{2}{|c|}{ } &&
\multicolumn{6}{|c||}{ } &  \multicolumn{4}{|c|}{ } \\[-.3cm]
\multicolumn{2}{|c|}{$x_p$} &&
\multicolumn{6}{|c||}{$K^{*0}/\bar{K}^{*0}$ Production Cross Sections} &
\multicolumn{4}{|c|}{Ratios} \\
  \multicolumn{2}{|c|}{Range} & $<\! x_p\! >$
& \multicolumn{2}{|c|}{$u\bar{u}$, $d\bar{d}$, $s\bar{s}$}
& \multicolumn{2}{|c|}{$c\bar{c}$}
& \multicolumn{2}{|c||}{$b\bar{b}$}
& \multicolumn{2}{|c|}{$c$:$uds$}
& \multicolumn{2}{|c|}{$b$:$uds$} \\[.1cm]\hline
 0.018 & 0.048 & 0.033 & 5.2  & 1.3  & 7.8  & 5.6  &
                         1.3  & 2.1  & 1.51 & 1.15 & 0.25 & 0.41 \\
 0.048 & 0.088 & 0.068 & 4.28 & 0.52 & 1.0  & 2.6  &
                         4.53 & 0.83 & 0.23 & 0.60 & 1.06 & 0.23 \\
 0.088 & 0.149 & 0.118 & 2.14 & 0.29 & 0.5  & 1.6  &
                         3.64 & 0.47 & 0.23 & 0.73 & 1.70 & 0.31 \\
 0.149 & 0.263 & 0.206 & 0.81 & 0.12 & 1.10 & 0.59 &
                         1.43 & 0.24 & 1.35 & 0.76 & 1.75 & 0.40 \\
 0.263 & 0.483 & 0.342 & 0.345& 0.042& 0.29 & 0.20 &
                         0.400& 0.078& 0.85 & 0.58 & 1.16 & 0.27 \\
 0.483 & 1.000 & 0.607 & 0.076& 0.010& 0.026& 0.034&
                         0.012& 0.009& 0.36 & 0.45 & 0.15 & 0.11 \\
  [.1cm] \hline
  \end{tabular}
\caption{
\label{xsfkstar}
Differential cross sections for the production of \kkb\ mesons
per $Z^0$ decay into light, $c$ and $b$ primary flavors.
}
\end{center}
\end{table}

\begin{table}
\begin{center}
 \begin{tabular}{|r@{--}r|c|r@{$\pm$}l|r@{$\pm$}l|
                          r@{$\pm$}l|r@{$\pm$}l|r@{$\pm$}l|} \hline
\multicolumn{2}{|c|}{  } && \multicolumn{6}{|c|}{  }  &
\multicolumn{4}{|c|}{  } \\[-.3cm]
\multicolumn{2}{|c|}{$x_p$} &&
\multicolumn{6}{|c|}{p/$\bar{\rm p}$ Production Cross Sections} &
\multicolumn{4}{|c|}{Ratios} \\
  \multicolumn{2}{|c|}{Range} & $<\! x_p\! >$
& \multicolumn{2}{|c|}{$u\bar{u}$, $d\bar{d}$, $s\bar{s}$}
& \multicolumn{2}{|c|}{$c\bar{c}$}
& \multicolumn{2}{|c|}{$b\bar{b}$}
& \multicolumn{2}{|c|}{$c$:$uds$}
& \multicolumn{2}{|c|}{$b$:$uds$} \\[.1cm]\hline
 0.016 & 0.022 & 0.019 &    8.55 &  1.31 &   17.6  &  5.5
                       &    6.3  &  1.8  &\multicolumn{2}{|c|}{ }&0.74 &0.24 \\
 0.022 & 0.027 & 0.025 &   10.88 &  0.96 &   12.9  & 4.0
                       &    9.0  &  1.3  &    1.57 &  0.38 &  0.83 &  0.14 \\
\cline{10-11}
 0.027 & 0.033 & 0.030 &   12.52 &  0.87 &   15.2  &  3.7
                       &   14.9  &  1.3  &\multicolumn{2}{|c|}{ }&1.19 &0.13 \\
 0.033 & 0.038 & 0.036 &   11.22 &  0.79 &   13.6  &  3.3
                       &   10.6  &  1.1  &    1.21 &  0.23 &  0.94 &  0.12 \\
\cline{10-11}
 0.038 & 0.044 & 0.041 &    8.65 &  0.73 &   10.7  &  3.1
                       &    8.7  &  1.1  &\multicolumn{2}{|c|}{ }&1.00 &0.15 \\
 0.044 & 0.049 & 0.047 &    8.87 &  0.72 &    8.0  &  3.0
                       &    7.9  &  1.02 &    1.07 &  0.26 &  0.89 &  0.13 \\
\cline{10-11}
 0.049 & 0.055 & 0.052 &    6.16 &  0.65 &   10.8  &  2.8
                       &    5.48 &  0.92 &\multicolumn{2}{|c|}{ }&0.89 &0.18 \\
 0.055 & 0.066 & 0.060 &    7.09 &  0.50 &    5.1  &  2.1
                       &    5.97 &  0.75 &    1.04 &  0.27 &  0.84 &  0.12 \\
\cline{10-11}
 0.066 & 0.077 & 0.071 &    4.91 & 0.49  &    7.7  &  2.2
                       &    4.60 & 0.74  &\multicolumn{2}{|c|}{ }&0.94 &0.18 \\
 0.077 & 0.088 & 0.082 &    4.71 & 0.49  &    3.6  &  2.1
                       &    4.37 & 0.76  &    1.18 &  0.34 &  0.93 &  0.19 \\
\cline{10-11}
 0.088 & 0.099 & 0.093 &    3.43 &  0.51 &    4.2  &  2.2
                       &    3.49 &  0.80 &\multicolumn{2}{|c|}{ }&1.02 &0.28 \\
 0.099 & 0.110 & 0.104 &    2.72 &  0.58 &    6.2  &  2.6
                       &    2.99 &  0.88 &    1.72 &  0.61 &  1.10 &  0.40 \\
\cline{10-11}
 0.110 & 0.132 & 0.120 &    2.98 & 0.46  &    0.9  &  1.9
                       &    1.77 & 0.68  &\multicolumn{2}{|c|}{ }&0.59 &0.25 \\
 0.132 & 0.164 & 0.147 &    3.16 & 0.59  &  --0.2  &  2.5
                       &    2.93 & 0.86  &    0.07 &  0.54 &  0.93 &  0.32 \\
\cline{10-11}
 0.230 & 0.274 & 0.251 &   0.738 & 0.085 &    0.84 & 0.34
                       &   0.506 & 0.098 &\multicolumn{2}{|c|}{ }&0.69 &0.15 \\
 0.274 & 0.318 & 0.294 &   0.514 & 0.062 &    0.46 & 0.24
                       &   0.241 & 0.065 &    1.04 &  0.35 &  0.47 &  0.14 \\
\cline{10-11}
 0.318 & 0.384 & 0.348 &   0.338 & 0.037 &    0.16 & 0.14
                       &   0.093 & 0.034 &\multicolumn{2}{|c|}{ }&0.27 &0.10 \\
 0.384 & 0.471 & 0.421 &   0.141 & 0.021 &   0.277 & 0.079
                       &   0.012 & 0.016 &    1.02 &  0.35 &  0.09 &  0.12 \\
\cline{10-11}
 0.471 & 0.603 & 0.529 &   0.088 & 0.010 &   0.040 & 0.034
                       & --0.002 & 0.006 &\multicolumn{2}{|c|}{ }&--.02 &0.07 \\
 0.603 & 0.768 & 0.654 &   0.020 & 0.004 &   0.004 & 0.014
                       &   0.001 & 0.003 &    0.40 &  0.35 &  0.04 &  0.13 \\
  [.1cm] \hline
  \end{tabular}
\caption{
\label{xsfpr}
Differential cross sections for the production of p/$\bar{\rm p}$
per $Z^0$ decay into light, $c$ and $b$ primary flavors.
}
\end{center}
\end{table}

\begin{table}
\begin{center}
 \begin{tabular}{|r@{--}r|c|r@{$\pm$}l|r@{$\pm$}l|
                            r@{$\pm$}l||r@{$\pm$}l|r@{$\pm$}l|} \hline
\multicolumn{2}{|c|}{ } &&
\multicolumn{6}{|c||}{ } &  \multicolumn{4}{|c|}{ } \\[-.3cm]
\multicolumn{2}{|c|}{$x_p$} &&
\multicolumn{6}{|c||}{$\Lambda^0$/$\bar{\Lambda}^0$ Production Cross Sections} &
\multicolumn{4}{|c|}{Ratios} \\
  \multicolumn{2}{|c|}{Range} & $<\! x_p\! >$
& \multicolumn{2}{|c|}{$u\bar{u}$, $d\bar{d}$, $s\bar{s}$}
& \multicolumn{2}{|c|}{$c\bar{c}$}
& \multicolumn{2}{|c||}{$b\bar{b}$}
& \multicolumn{2}{|c|}{$c$:$uds$}
& \multicolumn{2}{|c|}{$b$:$uds$} \\[.1cm]\hline
 0.011 & 0.020 & 0.016 & 4.72 & 0.87 &   1.5  & 3.3  &
                         2.8  & 1.2  &   0.32 & 0.70 & 0.59 & 0.27 \\
 0.020 & 0.030 & 0.025 & 3.87 & 0.49 &   2.5  & 2.0  &
                         4.19 & 0.79 &   0.66 & 0.53 & 1.08 & 0.24  \\
 0.030 & 0.045 & 0.038 & 3.41 & 0.35 &   4.5  & 1.5  &
                         2.39 & 0.50 &   1.32 & 0.46 & 0.70 & 0.16 \\
 0.045 & 0.067 & 0.056 & 2.21 & 0.22 &   3.56 & 0.97 &
                         2.47 & 0.34 &   1.61 & 0.46 & 1.12 & 0.19 \\
 0.067 & 0.100 & 0.082 & 1.14 & 0.16 &   2.89 & 0.72 &
                         1.44 & 0.25 &   2.11 & 0.58 & 1.05 & 0.22 \\
 0.100 & 0.150 & 0.122 & 1.15 & 0.13 &   0.54 & 0.54 &
                         1.10 & 0.17 &   0.47 & 0.48 & 0.96 & 0.18 \\
 0.150 & 0.247 & 0.189 & 0.52 & 0.08 &   0.56 & 0.32 &
                         0.60 & 0.09 &   1.08 & 0.64 & 1.15 & 0.25 \\
 0.247 & 0.497 & 0.319 & 0.24 & 0.05 & --0.13 & 0.19 &
                         0.20 & 0.04 & --0.54 & 0.81 & 0.83 & 0.25
 \\[.1cm] \hline
  \end{tabular}
\caption{
\label{xsflam}
Differential cross sections for the production of \llb\
per $Z^0$ decay into light, $c$ and $b$ primary flavors.
}
\end{center}
\end{table}

\begin{table}
\begin{center}
 \begin{tabular}{|r@{--}r|c|r@{$\pm$}l|r@{$\pm$}l|r@{$\pm$}l||
                                       r@{$\pm$}l|r@{$\pm$}l|} \hline
\multicolumn{2}{|c|}{ } &&
\multicolumn{6}{|c||}{ } &  \multicolumn{4}{|c|}{ } \\[-.3cm]
\multicolumn{2}{|c|}{$x_p$} &&
\multicolumn{6}{|c||}{$K^0/\bar{K}^0$ Production Cross Sections} &
\multicolumn{4}{c|}{Ratios} \\
  \multicolumn{2}{|c|}{Range} & $<\! x_p\! >$
& \multicolumn{2}{|c|}{$u\bar{u}$, $d\bar{d}$, $s\bar{s}$}
& \multicolumn{2}{|c|}{$c\bar{c}$}
& \multicolumn{2}{|c||}{$b\bar{b}$}
& \multicolumn{2}{c|}{$c$:$uds$}
& \multicolumn{2}{|c|}{$b$:$uds$} \\[.1cm]\hline
 0.009 & 0.011 & 0.010 & 19.0  & 4.4  &    6.  &  19.  &
                           6.1 & 3.1  &   0.29 &  0.99 & 0.32 & 0.17 \\
 0.011 & 0.011 & 0.013 & 23.2  & 3.2  &  --3.  &  15.  &
                         23.1  & 5.6  & --0.14 &  0.64 & 0.99 & 0.39 \\
 0.014 & 0.018 & 0.016 & 20.4  & 2.4  &   15.  & 10.   &
                         25.8  & 4.4  &   0.72 &  0.52 & 1.27 & 0.25 \\
 0.018 & 0.022 & 0.020 & 21.2  & 2.3  &  22.7  &  9.7  &
                         21.7  & 3.3  &   1.07 &  0.47 & 1.02 & 0.18 \\
 0.022 & 0.027 & 0.025 & 20.5  & 1.8  &  17.4  &  7.8  &
                         21.4  & 2.6  &   0.85 &  0.39 & 1.04 & 0.15 \\
 0.027 & 0.033 & 0.030 & 17.3  & 1.4  &  12.8  &  6.2  &
                         20.7  & 2.2  &   0.74 &  0.36 & 1.20 & 0.15 \\
 0.033 & 0.041 & 0.037 & 14.1  & 1.2  &  12.8  &  5.1  &
                         19.3  & 1.9  &   0.91 &  0.37 & 1.37 & 0.17 \\
 0.041 & 0.050 & 0.045 & 12.0  & 1.0  &  13.2 &  4.4  &
                         15.6  & 1.5  &   1.10 &  0.38 & 1.30 & 0.16 \\
 0.050 & 0.061 & 0.055 & 10.1  & 0.8  &  10.9  &  3.5  &
                         13.2  & 1.2  &   1.08 &  0.36 & 1.31 & 0.15 \\
 0.061 & 0.074 & 0.067 &  7.73 & 0.69 &  12.8  &  3.2  &
                         13.5  & 1.1  &   1.66 &  0.43 & 1.75 & 0.20 \\
 0.074 & 0.091 & 0.082 &  7.07 & 0.52 &   3.0  &  2.4  &
                         12.3  & 0.9  &   0.42 &  0.33 & 1.74 & 0.17 \\
 0.091 & 0.111 & 0.100 &  5.33 & 0.44 &   7.0  &  2.0  &
                          8.35 & 0.81 &   1.31 &  0.39 & 1.57 & 0.19 \\
 0.111 & 0.142 & 0.126 &  4.17 & 0.34 &   4.6  &  1.5  &
                          5.85 & 0.57 &   1.10 &  0.37 & 1.40 & 0.17 \\
 0.142 & 0.183 & 0.161 &  3.17 & 0.30 &   3.7  &  1.6  &
                          4.26 & 0.55 &   1.18 &  0.53 & 1.35 & 0.21 \\
 0.183 & 0.235 & 0.206 &  2.16 & 0.22 &   2.68 &  0.97 &
                          1.99 & 0.48 &   1.24 &  0.46 & 0.92 & 0.24 \\
 0.235 & 0.301 & 0.262 &  1.12 & 0.16 &   2.62 &  0.72 &
                          0.09 & 0.24 &   2.15 &  0.66 & 0.71 & 0.22 \\
 0.301 & 0.497 & 0.371 &  0.69 & 0.10 &   0.79 &  0.45 &
                          0.10 & 0.10 &   1.44 &  0.70 & 0.14 & 0.14
 \\[.1cm] \hline
  \end{tabular}
\caption{
\label{xsfkzero}
Differential cross sections for the production of $K^0/\bar{K}^0$ mesons
per $Z^0$ decay into light, $c$ and $b$ primary flavors.
}
\end{center}
\end{table}

\begin{table}
\begin{center}
 \begin{tabular}{|r@{--}r|c|r@{$\pm$}l|r@{$\pm$}l|
                           r@{$\pm$}l||r@{$\pm$}l|r@{$\pm$}l|} \hline
\multicolumn{2}{|c|}{ } &&
\multicolumn{6}{|c||}{ } &  \multicolumn{4}{|c|}{ } \\[-.3cm]
\multicolumn{2}{|c|}{$x_p$} &&
\multicolumn{6}{|c||}{$\phi$ Production Cross Sections} &
\multicolumn{4}{|c|}{Ratios} \\
  \multicolumn{2}{|c|}{Range} & $<\! x_p\! >$
& \multicolumn{2}{|c|}{$u\bar{u}$, $d\bar{d}$, $s\bar{s}$}
& \multicolumn{2}{|c|}{$c\bar{c}$}
& \multicolumn{2}{|c||}{$b\bar{b}$}
& \multicolumn{2}{|c|}{$c$:$uds$}
& \multicolumn{2}{|c|}{$b$:$uds$} \\[.1cm]\hline
 0.018 & 0.057 & 0.037 & 0.64  & 0.18  & 1.08  & 0.77  &
                         0.73  & 0.28  & 1.67  & 1.28  & 1.13 & 0.53 \\
 0.057 & 0.079 & 0.068 & 0.48  & 0.18  & 0.31  & 1.02  &
                         0.37  & 0.31  & 0.64  & 2.15  & 0.78 & 0.70 \\
 0.079 & 0.175 & 0.127 & 0.222 & 0.073 & 0.12  & 0.39  &
                         0.42  & 0.11  & 0.56  & 1.75  & 1.88 & 0.81 \\
 0.175 & 0.263 & 0.215 & 0.091 & 0.052 & 0.35  & 0.23  &
                         0.228 & 0.068 & 3.85  & 3.32  & 2.51 & 1.61 \\
 0.263 & 0.483 & 0.357 & 0.052 & 0.021 & 0.185 & 0.085 &
                         0.054 & 0.023 & 3.58  & 2.17  & 1.05 & 0.61 \\
 0.483 & 1.000 & 0.689 & 0.017 & 0.004 &--0.016& 0.013 &
                         0.007 & 0.004 &--0.96 & 0.78  & 0.43 & 0.27 \\
  [.1cm] \hline
  \end{tabular}
\caption{
\label{xsfphi}
Differential cross sections for the production of $\phi$ mesons
per $Z^0$ decay into light, $c$ and $b$ primary flavors.
}
\end{center}
\end{table}

The resulting \pnb s are listed in tables \ref{xsfpi}--\ref{xsfphi}.
The systematic errors listed are only those relevant for the comparison of
different flavors, namely those due to uncertainties in the unfolding procedure;
the systematic errors given in the preceding section are also applicable, but
are common to all three flavor categories.
The flavor unfolding systematic errors were evaluated by varying each element of
the event tagging efficiency matrix $\epsilon_{ii}$ by $\pm$0.01 \cite{mikeh},
varying the heavy quark production fractions $R_b$ and $R_c$ by the errors
on their respective world averages,
and varying each diagonal bias value $B_{ii}^h$ by the larger of
$\pm$0.005 and $\pm$20\% of its difference from unity.
Since the lepton background is strongly flavor-dependent,
the photon conversion rate in the simulation was varied by $\pm$15\%,
and the simulated rates of lepton production from other sources in
light-, $c$, and $b$-flavor events were varied by $\pm$50\%, $\pm$10\%
and $\pm$5\%, respectively.
The unfolding systematic errors are typically small compared with the
statistical errors, and are dominated by the variation in the bias.

\begin{figure}
\vspace{-1.cm}
 \hspace*{0.5cm}   
   \epsfxsize=3.95in
   \begin{center}\mbox{\epsffile{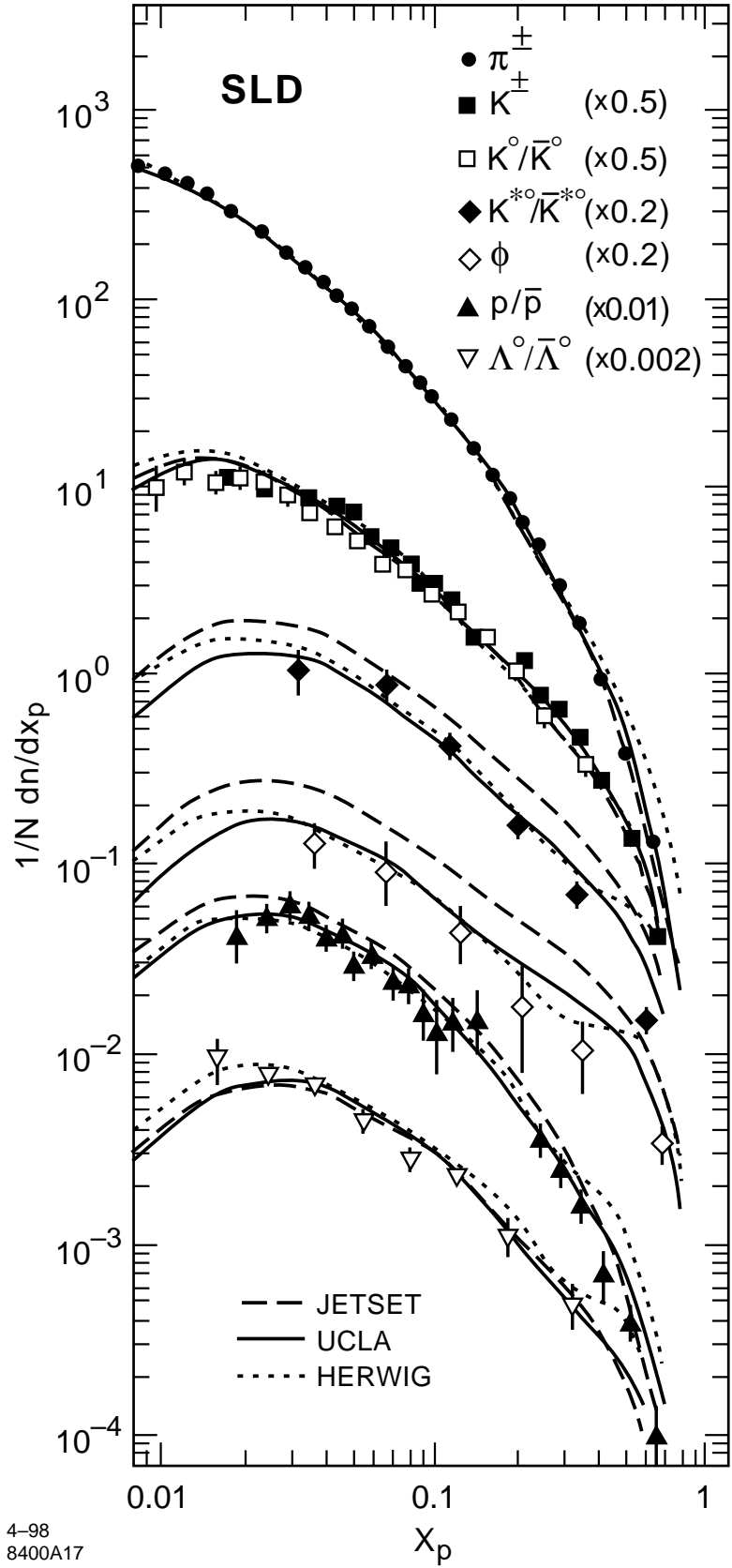}}\end{center}
  \caption{ 
 \label{xsudsmc}
Identified hadron \pnb s in light-flavor events.
Also shown are the predictions of the three fragmentation models;
the prediction of each model for $K^\pm$ is similar to that for $K^0/\bar{K}^0$,
and the two have been averaged.
    }
\end{figure}

In figure~\ref{xsudsmc} we show the \pnb s for the
seven hadron species in light-flavor $Z^0$ decays.
Qualitatively these are similar to those in flavor-inclusive decays
(fig.~\ref{xsall}),
although all \pnb s are larger at high $x_p$ in light flavor events.
The same general features of $\pi$-$K$ and p-$\Lambda^0$ convergence at high
$x_p$ are visible, and the relative suppressions of hadron species with
respect to one another are similar in magnitude and momentum dependence.

Also shown in fig. \ref{xsudsmc} are the predictions of the three simulation
programs.
All models reproduce the shape of each \pnb\ qualitatively.
The JETSET prediction for charged pions is smaller than the data in the range
$x_p<0.015$, and those for the pseudoscalar kaons are larger than the data for
$0.015<x_p<0.03$;
those for the vector mesons and protons reproduce the $x_p$ dependence but show
a larger normalization than the data.
These differences were all seen in the flavor-inclusive results
(figs.~\ref{fallmc}, \ref{xsallmc}), and we can now conclude that they all
indicate problems with the modelling of light-flavor fragmentation,
and cannot be due entirely to mismodelling of heavy hadron production and decay.
The HERWIG prediction for pseudoscalar kaons is also larger than the data at low
$x_p$ and is slightly smaller than the data in the range $0.15<x_p<0.25$.
For all hadron species the HERWIG prediction is larger than the data for
$x_p>0.4$, showing a characteristic shoulder structure.
The UCLA predictions for the baryons and the vector mesons show a similar but
less pronounced structure that is inconsistent with the proton and \kkb\
data.  Otherwise UCLA reproduces the data except for pseudoscalar kaons in the
range $0.015<x_p<0.03$.

\begin{figure}
\vspace{-1.cm}
 \hspace*{0.5cm}   
   \epsfxsize=6.55in
   \begin{center}\mbox{\epsffile{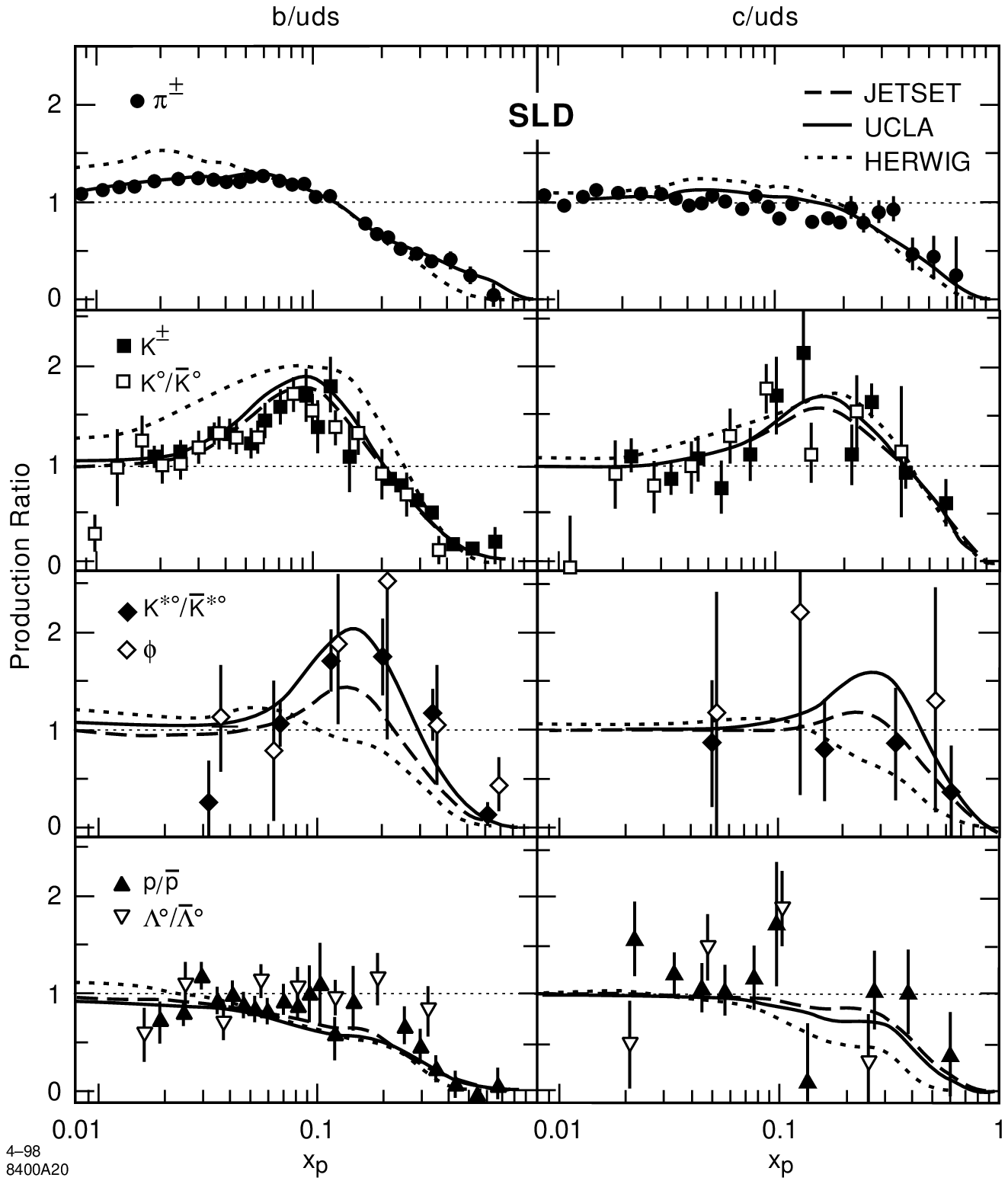}}\end{center}
  \caption{ 
 \label{xsrbumc}
Ratios of production of each hadron species in $b$-flavor events
to that in light-flavor events (left) and in $c$-flavor:light-flavor events
(right).
Also shown are the predictions of the three fragmentation models;
for each model, the predictions for $K^\pm$ and $K^0/\bar{K}^0$ were averaged,
as were those for p/$\bar{\rm p}$ and \llb .
The model predictions for $\phi$ are not shown, but have the same $x_p$
dependence as the corresponding prediction for \kkb , with a peak value
typically higher by 40\%.
    }
\end{figure}

In fig.~\ref{xsrbumc} and tables \ref{xsfpi}--\ref{xsfphi} we give the ratios
of production 
in $b$-flavor to light-flavor events for the seven species.
The systematic errors on the hadron reconstruction and identification largely
cancel in these ratios, and the total errors are predominantly statistical.
There is higher production of charged pions in $b$-flavor events than in
light-flavor events at low $x_p$, with the ratio rising with 
$x_p$ for $0.008<x_p<0.03$ to a plateau value of about 1.25.
The production of both charged and neutral kaons is approximately equal
in the two samples for $x_p<0.03$, but the relative production in $b$-flavor
events then increases with $x_p$, peaking at a value of about 1.7 at
$x_p \approx 0.09$.
The errors on the \kkb\ and $\phi$ ratios are large, but the data are
consistent with behavior similar to that of the pseudoscalar kaon ratios.
There is approximately equal production of baryons in $b$-flavor and
light-flavor events for $x_p<0.15$.
The production of pions and pseudoscalar kaons in $b$-flavor events falls
rapidly with $x_p$ for $x_p>0.1$ relative to that in light-flavor events.
The relative production of the vector mesons and protons also falls at
high $x_p$.

These features are consistent with expectations based on the known properties
of $e^+e^- \rightarrow b\bar{b}$ events, namely that a large fraction of the
event energy (on average about 70\% \cite{bohrer}) is carried by the
leading $B$- and $\bar{B}$-hadrons,
leaving little energy available to produce high momentum fragmentation hadrons.
The $B$ hadrons decay into a large number of lighter particles, including
on average 5.5 stable charged hadrons \cite{pdg}, which are expected to
populate primarily the region $0.02<x_p<0.2$.
Also shown in fig. \ref{xsrbumc} are the
predictions of the three fragmentation models, all of which reproduce these
features qualitatively, although HERWIG overestimates the ratio for pions
in the range $x_p<0.05$ and that for kaons for $x_p<0.3$.
The values of these ratios depend on details of the $B$ and $D$ hadron
energy spectra and decay properties,
and so provide information complementary to that in fig. \ref{xsudsmc}.
However, in drawing conclusions regarding heavy flavor modelling from
these ratios, one must consider how well the model in question reproduces the
light flavor results.
For example, the HERWIG prediction for pion (kaon) production in
light-flavor events (fig. \ref{xsudsmc}) is consistent with (higher than) the
data for $x_p<0.05$, so it is safe to conclude from fig. \ref{xsrbumc}
that HERWIG mismodels pion and kaon production from $B$ decays in this region.
However the fact that the HERWIG ratio for kaons is high in the
region $0.1<x_p<0.3$ is due at least in part to the low HERWIG prediction for
kaon production in light-flavor events in that region.

In fig.~\ref{xsrbumc} we also show the ratios of production 
in $c$-flavor to light-flavor events for the seven species.
The errors are larger than for the $b$:$uds$ comparison and $x_p$ bins have been
combined in some cases for clarity.
Similar qualitative features are observed:
there is higher kaon production in $c$-flavor events than in light-flavor events
at $x_p \sim 0.1$; pion production is slightly higher in $c$-flavor than in
light-flavor events for $x_p< 0.03$, then decreases slowly with $x_p$; both pion
and kaon production appear to fall rapidly with $x_p$ for $x_p>0.3$,
a somewhat higher value than the corresponding $b$:$uds$ ratios.
These features are expected since $c$-jets produce a charmed hadron
with on average about half \cite{bohrer} the beam energy,
a lower fraction than $B$-hadrons,
which leaves more energy available for fragmentation hadrons than in $b$-jets.
The charmed hadron decay products often include a kaon carrying a large
fraction of the charmed hadron momentum, and there are fewer additional charged
pions than in $B$ hadron decays.
Also shown in fig. \ref{xsrbumc} are the $c$:$uds$ ratios predicted by the
three fragmentation models.
All models are consistent with the data, except that HERWIG overestimates
the pion ratio for $0.03<x_p<0.15$.

\section{Comparison with QCD Predictions}

We tested the predictions of Gribov and Lipatov, that, in the limit
$x_p \rightarrow 1$, the momentum distribution for primary leading hadrons
be $(1-x_p)^n$, with $n=2$ for mesons and $n=3$ for baryons.
Since this test benefits from more bins at high $x_p$, we considered
only the charged hadrons.
The cross sections measured in light flavor events provide in principle a
better test than those measured in flavor-inclusive events,
since $c$- and $b$-flavor events cannot contain primary leading pions, kaons or
protons.
However, we have just shown that the contributions from $c$- and $b$-flavor
events are small for $x_p$ greater than about 0.5;  since we have better
statistics for flavor-inclusive events we performed the test on this data set,
as well as on the light-flavor data.
We are limited to $x_p<0.77$ for the charged pions and kaons, but for the
flavor-inclusive analysis of protons we have an additional bin,
obtained from a 2-hypothesis analysis (see sec. 4.1) that also yielded the sum
of meson cross sections ($\pi^{\pm} + K^{\pm}$).
We also considered this meson sum at all momenta, which has smaller statistical
errors than the sum of the individual $\pi^\pm$ and $K^\pm$ cross sections.

\begin{figure}
\vspace{-1.cm}
 \hspace*{0.5cm}   
   \epsfxsize=4.3in
   \begin{center}\mbox{\epsffile{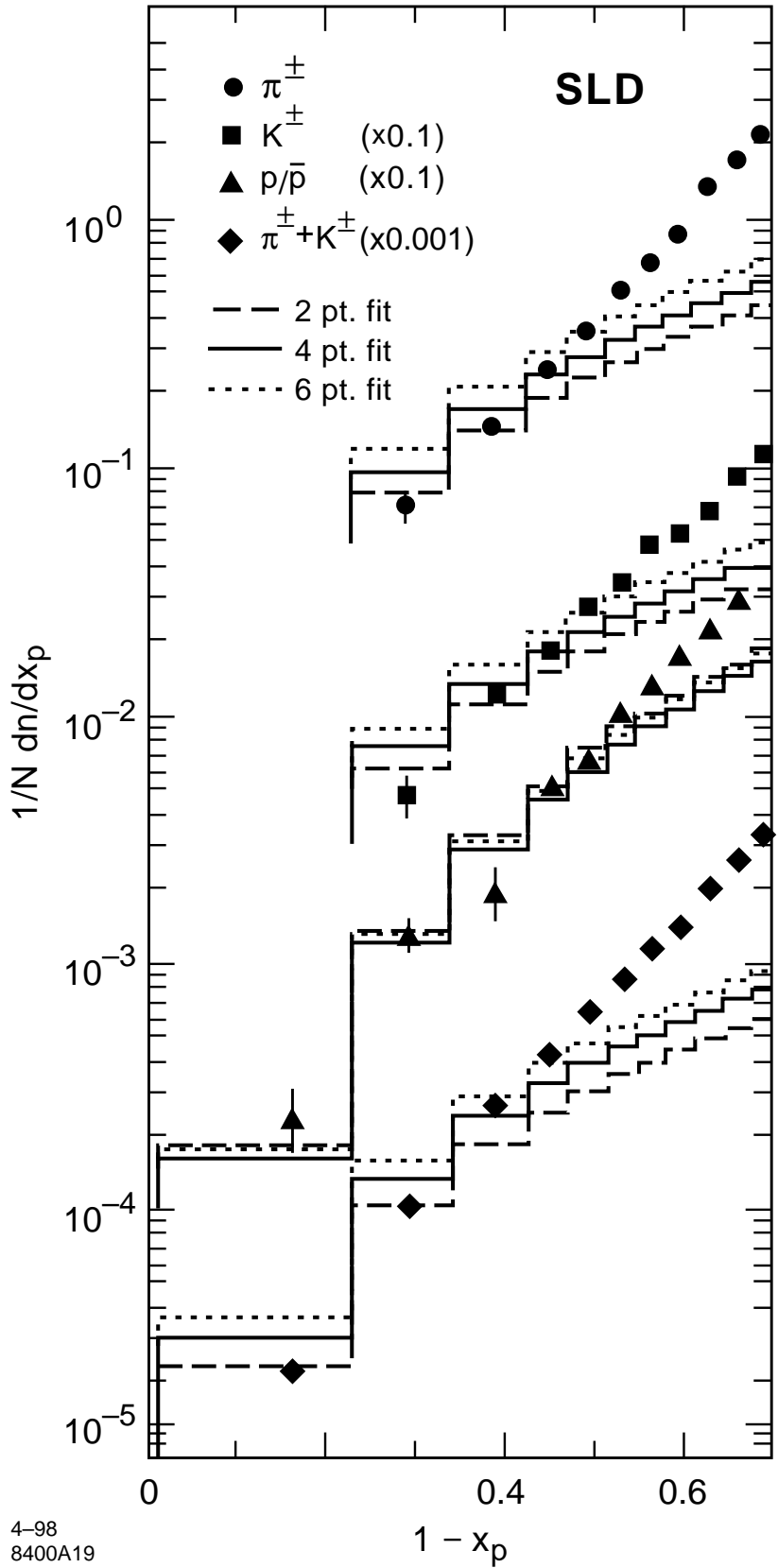}}\end{center}
  \caption{ 
 \label{glall}
Measured \pnb s in flavor-inclusive $Z^0$ decays as a function of ($1-x_p$),
along with the results of polynomial fits, described in the text,
to the data in the 2, 4 and 6 leftmost bins.
Each fitted polynomial has been integrated over each bin and is shown as a
histogram.
    }
\end{figure} 

Figure \ref{glall} shows the $\pi^\pm$, $K^\pm$, p and
($\pi^{\pm} + K^{\pm}$) \pnb s as functions of $(1-x_p)$ in flavor-inclusive
$Z^0$ decays.
Fits of the function $f(x) = A(1-x_p)^n$, with the value of $n$ fixed to 2
(3 for protons), were performed to the first $m$ data points and the
resulting fitted distributions for $m=2,4,6$ are shown in the figure.
In all cases the fit quality is good for $m=2$, but worsens with increasing $m$.
The maximum number of bins for which the confidence level of the $\chi^2$ of the
fit exceeded 0.01 was 3 for $\pi^\pm$ and $K^\pm$, 6 for p/$\bar{\rm p}$, and 2
for the meson sum ($\pi^{\pm} + K^{\pm}$).

Using this criterion, the theoretical prediction is consistent
with our combined meson data for $(1-x_p)<0.34$, with our pion and kaon data
for $(1-x_p)<0.47$, and with our proton data for $(1-x_p)<0.57$.
A similar analysis of the light-flavor sample (not shown) yielded similar
results;
the prediction is consistent with our pion, kaon and combined meson data for
$(1-x_p)<0.53$, and with our proton data for $(1-x_p)<0.62$.

\begin{figure}
\vspace{-1.cm}
 \hspace*{0.5cm}   
   \epsfxsize=6.4in
   \begin{center}\mbox{\epsffile{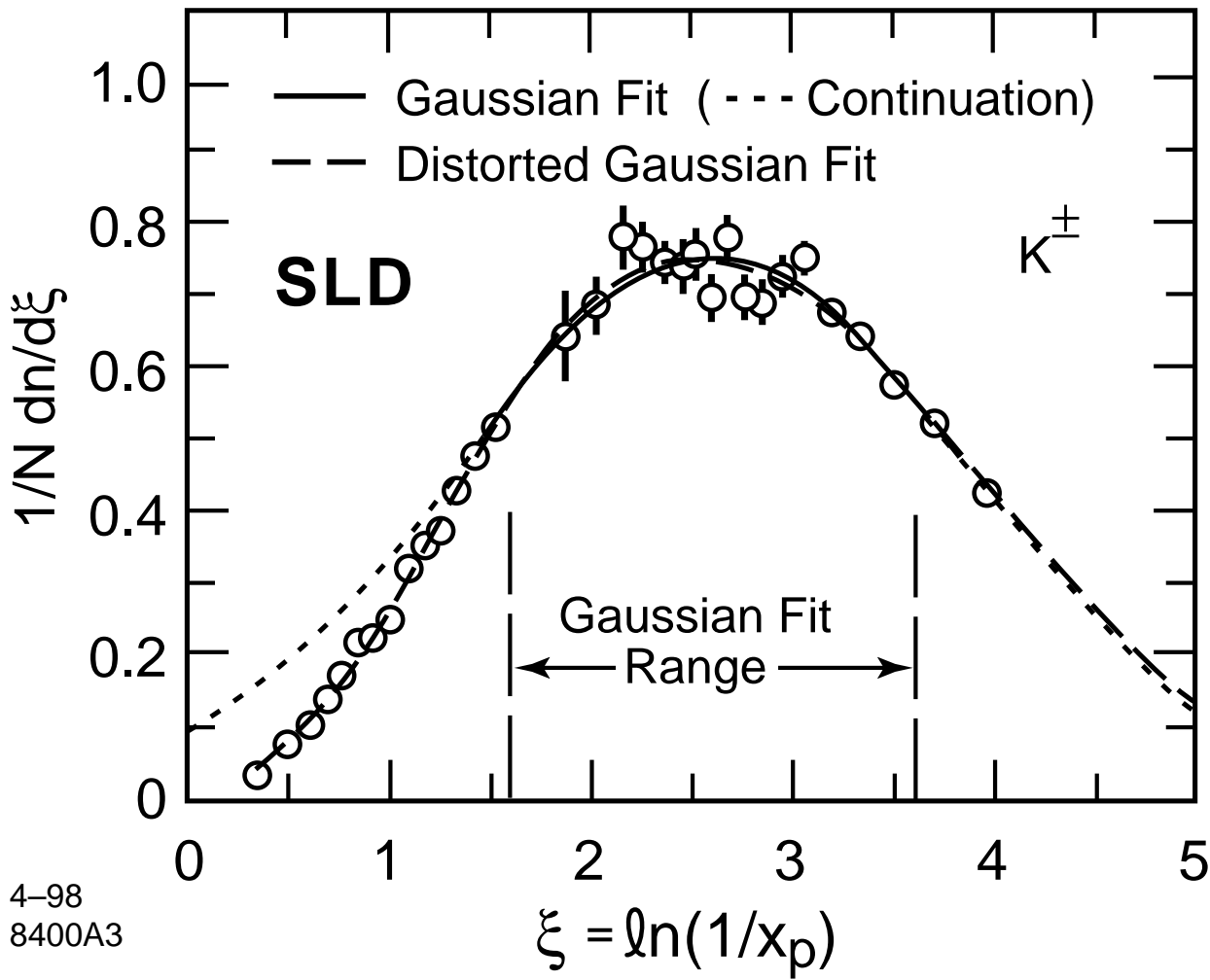}}\end{center}
  \caption{ 
 \label{xione}
Distribution of $\xi = \ln (1/x_p)$ for charged kaons in flavor-inclusive $Z^0$
decays.
The solid and dashed lines indicated the results of fits of the Gaussian and
distorted Gaussian approximations of MLLA QCD described in the text.
The dotted lines indicate the continuations of the fitted Gaussian function.
    }
\end{figure} 

\begin{figure}
\vspace{-1.cm}
 \hspace*{0.5cm}   
   \epsfxsize=6.4in
   \begin{center}\mbox{\epsffile{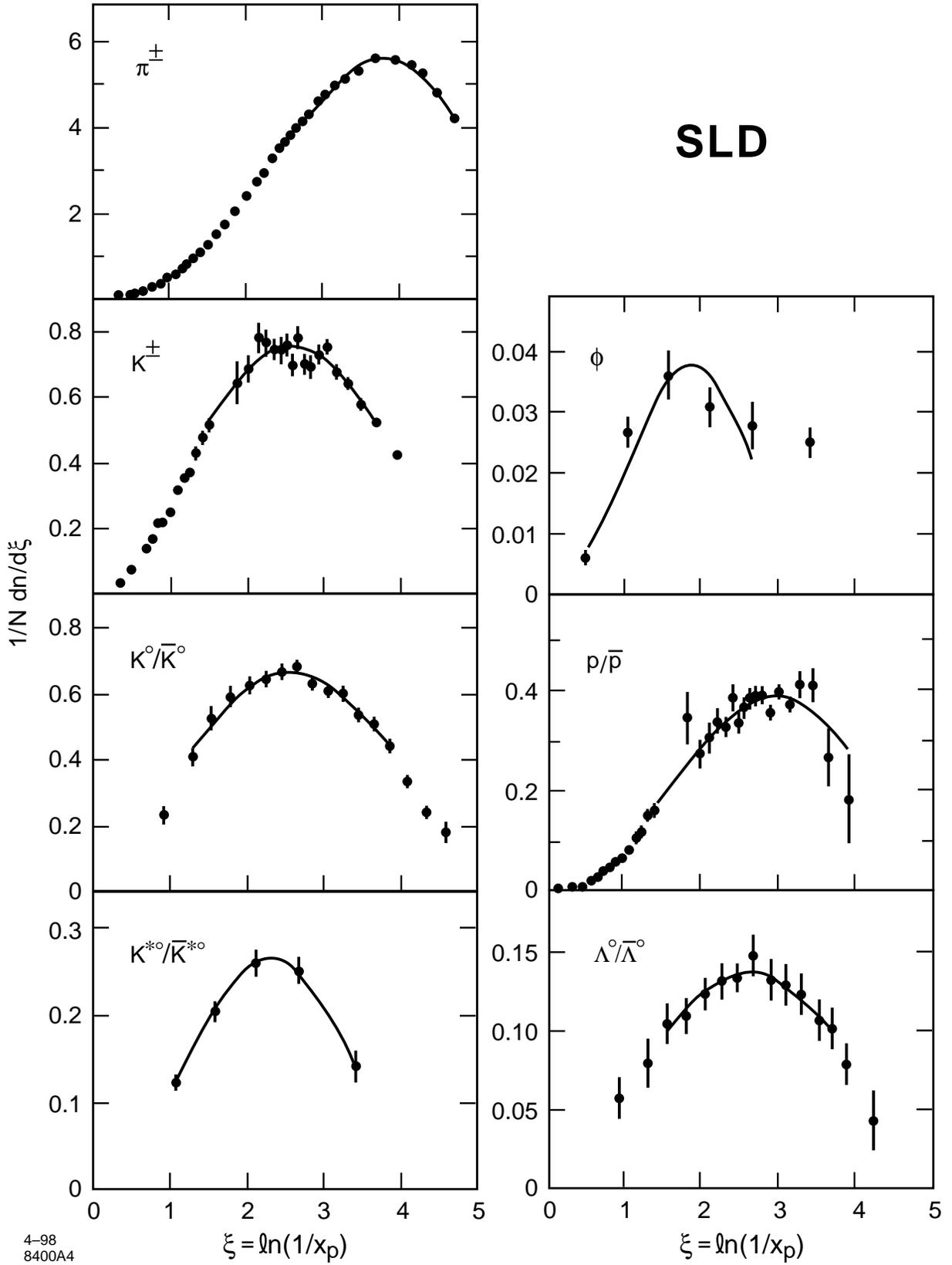}}\end{center}
  \caption{ 
 \label{xiall}
Distributions of $\xi$ for the seven hadron species in flavor-inclusive hadronic
$Z^0$ decays (points),
along with the results of Gaussian fits (solid lines) to the data over a
range of approximately $\pm$1 unit about the peak.
    }
\end{figure} 

In order to test the predictions of QCD in the Modified Leading Logarithm
Approximation (MLLA) combined with the ansatz of 
Local Parton-Hadron Duality (LPHD), 
we converted our measurements into \pnb s in the variable $\xi= \ln(1/x_p)$.
Figure \ref{xione} shows our measured \pnb\ as a function of $\xi$ for the
charged kaons.
Also shown are the results of fits to a simple Gaussian, and a distorted
Gaussian including skewness and kurtosis terms.
The Gaussian fit was performed over a $\xi$ range of width 2 units
positioned near the maximum of the distribution.
The fitted peak position $\xi^*$ was found to be independent of the exact
position of this range within statistical errors, and 
the solid line in fig. \ref{xione} represents the result of a fit over a range
centered on this peak position.
A good fit quality was obtained; the two points above this $\xi$ range could be
added to the fit, as could the first two points below the range, before the
$\chi^2$ began to increase rapidly, indicating that the Gaussian approximation
is consistent with our data over a range of approximately $\pm$1.3 units of
$\xi$ around the peak position.
The distorted Gaussian function is able to describe the data over the full
measured range of $\xi$, as indicated by the dashed line in fig. \ref{xione},
however the distortion terms grow rapidly as points outside the range
described by the simple Gaussian are added.

\begin{figure}
\vspace{-1.cm}
 \hspace*{0.5cm}   
   \epsfxsize=5.2in
   \begin{center}\mbox{\epsffile{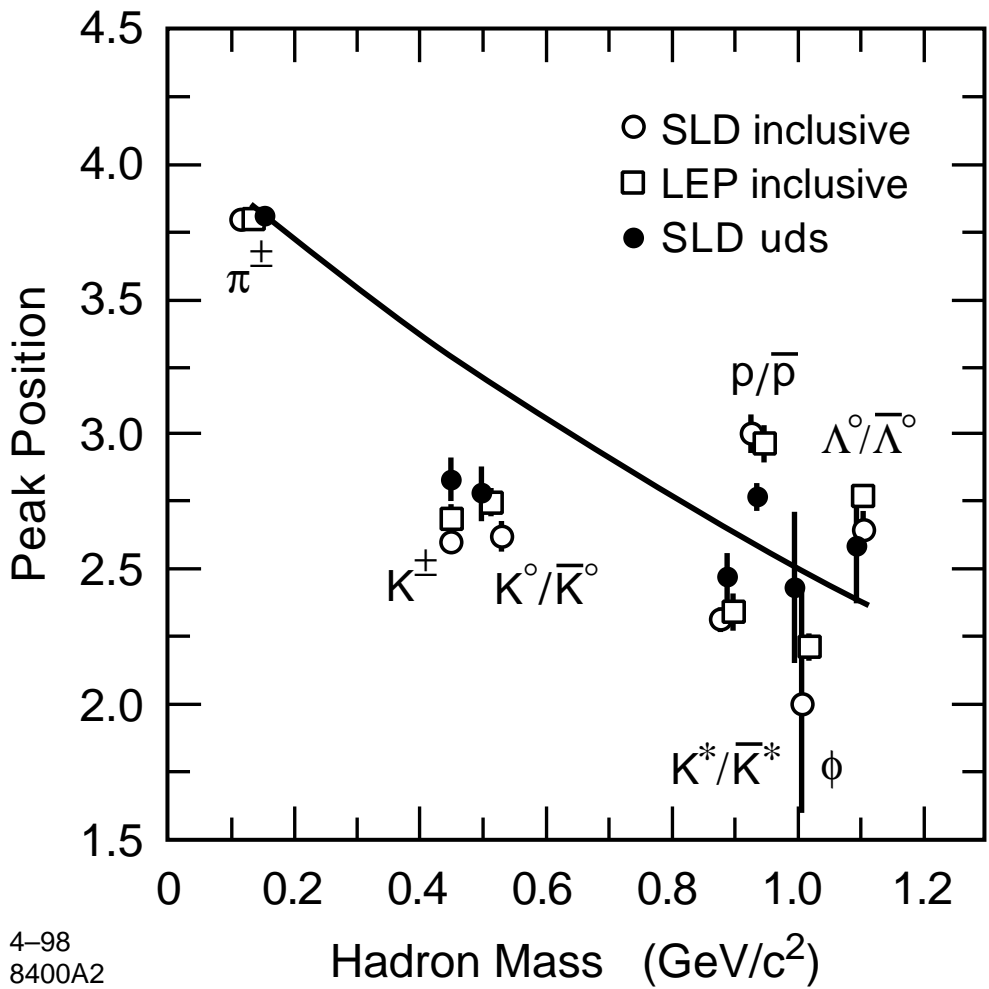}}\end{center}
  \caption{ 
 \label{xiallpk}
Peak positions $\xi^*$ from fits to the $\xi$ distributions in
flavor-inclusive and light-flavor hadronic $Z^0$ decays.
Also shown are averages of similar flavor-inclusive results from experiments
at LEP.  The line is the result of an ad hoc
exponential fit to our light-flavor data.
    }
\end{figure} 

Similar results were obtained for the other hadron species.  Their
$\xi$-distributions are shown in fig. \ref{xiall}.
We fitted a simple Gaussian over a $\xi$ range of approximately $\pm$1 unit
centered on the maximum of each distribution in order to measure
the peak position $\xi^*$ for each hadron species.
Systematic errors on this measurement were evaluated by varying the fit range
and by refitting with each source of correlated experimental systematic error
considered coherently in turn.
Good fit qualities were obtained when the correlated systematic errors were
taken into account.
The peak positions are given in table \ref{xistar} and
shown as a function of hadron mass in fig. \ref{xiallpk}, along with
averages of similar measurements from experiments at LEP \cite{bohrer},
with which they are consistent.
The distribution for pions peaks at a higher $\xi$ value than the those of the
other hadron species,
but otherwise there is no monotonic mass-dependence.

\begin{table}
\begin{center}
\begin{tabular}{|c|l@{$\pm$}l||r@{ $\pm$ }l|r@{ $\pm$ }l|r@{ $\pm$ }l|}\hline
  & \multicolumn{2}{c||}{all flavors} & 
    \multicolumn{2}{c|}{light flavors} & 
    \multicolumn{2}{c|}{c} & 
    \multicolumn{2}{c|}{b} \\[.1cm]  \hline
 $\pi^{\pm}$   & 3.80 & 0.01 & 3.81 & 0.01 & 3.85 & 0.04 & 3.71 & 0.01 \\[.1cm] 
   $K^{\pm}$   & 2.60 & 0.03 & 2.83 & 0.08 & 2.52 & 0.12 & 2.67 & 0.04 \\[.1cm] 
$K^0/\bar{K}^0$& 2.62 & 0.05 & 2.78 & 0.10 & 2.32 & 0.35 & 2.61 & 0.06 \\[.1cm] 
     \kkb      & 2.31 & 0.04 & 2.47 & 0.09 &\multicolumn{2}{c|}{--}&2.11&0.07\\[.1cm] 
   $\phi$      & 2.0  & 0.4  & 2.43 & 0.28 &\multicolumn{2}{c|}{--}&2.18&0.18\\[.1cm] 
p/$\bar{\rm p}$& 3.00 & 0.07 & 2.77 & 0.05 & 3.03 & 0.26 & 2.86 & 0.07 \\[.1cm] 
     \llb      & 2.64 & 0.07 & 2.58 & 0.21 & 2.75 & 0.15 & 2.47 & 0.18
\\[.1cm] \hline 
 \end{tabular}
\caption{
\label{xistar}
Peak positions $\xi^*$ from Gaussian fits to the $\xi$ distributions
for each hadron species measured in flavor-inclusive and flavor-specific
hadronic $Z^0$ decays.  The errors are the sum in quadrature of statistical and
systematic components.}
\end{center}
\end{table}

As discussed in section 1, the MLLA QCD$+$LPHD prediction is valid for
primary fragmentation particles, whereas experiments so far have measured
samples that include decay products of an unknown mix of resonances
as well as of heavy hadrons.
This mix may affect measured $\xi^*$ values differently for different hadron
species.
It is of interest to try to resolve this question experimentally, and we
have therefore applied the same analysis to the three primary event
flavor categories discussed in the previous section.
We expect the light flavor events to be less affected by decay products, as $D$-
and $B$-hadron decays are excluded.

The Gaussian function provides an acceptable description of the $\xi$
distribution for all hadron species in events of each flavor
within about $\pm$1 unit of the peak (not shown), and
the fitted peak positions are listed in table \ref{xistar}.
For the \kkb\ and $\phi$ in $c$-flavor events, the limited sample size
did not allow a reasonable systematic error evaluation, so they are omitted.

The $\xi^*$ values measured in $b$-flavor events are significantly different
from those measured in light-flavor events for $\pi^{\pm}$ and \kkb ;
the difference is 1.5$\sigma$ for $K^\pm$ and $K^0/\bar{K}^0$.
For the other hadron species the $\xi^*$ values measured in events of all
three flavors are consistent.
The $\xi^*$ values measured in light-flavor events 
differ significantly from those measured in flavor-inclusive events for $K^\pm$
and p/$\bar{\rm p}$.
The light-flavor $\xi^*$ values are also shown in fig. \ref{xiallpk}.
The result of an ad hoc exponential fit to the light-flavor
data is shown in fig. \ref{xiallpk} as a reference trajectory, and the
light-flavor data are seen to lie closer to a monotonic trajectory 
than the flavor-inclusive data.

\newpage

\section{Total Production Cross Sections}

We have integrated our differential cross sections over their respective
measurement ranges, taking into account the bin-to-bin correlations in the
systematic errors.  These integrated cross sections per event are listed in
tables \ref{pifraxa}--\ref{xsgneu};
the errors are
dominated by overall normalization uncertainties corresponding to the
uncertainty in our track reconstruction efficiency.
In order to quote total cross sections, we must extrapolate into the unmeasured
regions of $x_p$,
and we have done this using the three MC models discussed above.
From the hadrons of each species generated using each of these models, we
calculated the fraction that were generated with $x_p$ in the range of our
measurement.
For each hadron species the three fractions
were found to be similar, with the UCLA (HERWIG) fraction being
typically 1\% larger (1--2\% smaller) than the JETSET fraction.
The average of the three accepted fractions ranged from 0.812 for $K^\pm$
to 0.945 for $K^0/\bar{K}^0$.
Each integrated measured cross section was divided by the corresponding average 
fraction, and an uncertainty of $\pm$0.01 ($\pm$0.015) was assigned to
the average fraction for $\pi^{\pm}$, $K^{\pm}$, $K^0/\bar{K}^0$,
p/$\bar{\rm p}$
and \llb\ (\kkb\ and $\phi$), corresponding to a typical rms among the three
predictions.
The corrected total cross sections are shown in table \ref{xsintf}, and
were found to be consistent with an average of similar measurements
from experiments at LEP \cite{bohrer}.

\begin{table}
\begin{center}
\begin{tabular}{|c|r@{$\pm$}l||r@{$\pm$}l|r@{$\pm$}l|r@{$\pm$}l||r@{$\pm$}l|r@{$\pm$}l|}\hline
  & \multicolumn{8}{c||}{  } & \multicolumn{4}{c|}{  } \\[-.3cm]
  & \multicolumn{8}{c||}{Total Cross Sections per Event of Flavor} &
    \multicolumn{4}{c|}{Differences} \\[.1cm]
  & \multicolumn{2}{c||}{all}    & \multicolumn{2}{c|}{$uds$}
  & \multicolumn{2}{c|}{$c$}     & \multicolumn{2}{c||}{$b$}
  & \multicolumn{2}{c|}{$c-uds$} & \multicolumn{2}{c|}{$b-uds$} \\[.1cm]  \hline
  & \multicolumn{2}{c||}{  } & \multicolumn{2}{c|}{  }
  & \multicolumn{2}{c|}{  }  & \multicolumn{2}{c||}{  }
  & \multicolumn{2}{c|}{  }  & \multicolumn{2}{c|}{  } \\[-.3cm]
 $\pi^{\pm}$    & 16.84 & 0.37 & 16.46 & 0.47 & 16.30 & 1.01 &
                  18.36 & 0.52 &--0.15 & 0.96 &  1.91 & 0.36 \\[.1cm] 
   $K^{\pm}$    &  2.22 & 0.16 &  2.04 & 0.15 &  2.47 & 0.28 &
                   2.40 & 0.19 &  0.43 & 0.23 &  0.36 & 0.10 \\[.1cm] 
    $K^0$       &  2.01 & 0.08 &  1.86 & 0.09 &  1.86 & 0.21 &
                   2.11 & 0.11 &  0.01 & 0.21 &  0.25 & 0.09 \\[.1cm] 
${K}^{*0}$      & 0.707 & 0.041& 0.727 & 0.081& 0.561 & 0.316&
                  0.768 & 0.124&--0.166& 0.321&  0.041& 0.132\\[.1cm] 
   $\phi$       & 0.105 & 0.008&  0.091& 0.021&  0.131& 0.091&
                  0.121 & 0.026&  0.040& 0.093&  0.030& 0.031\\[.1cm] 
      p          &  1.03 & 0.13 &  1.06 & 0.14 &  1.06 & 0.21 &
                   0.91 & 0.13 &  0.01 & 0.17 &--0.15 & 0.07 \\[.1cm] 
$\Lambda^0$      & 0.395 & 0.022&  0.421& 0.030&  0.341& 0.088&
                  0.383 & 0.032&--0.080& 0.091&--0.038& 0.039 
\\[.1cm] \hline 
 \end{tabular}
\caption{
\label{xsintf}
Corrected total cross sections per hadronic $Z^0$ decay, and per decay into
light, $c$ or $b$ primary flavor.
Differences between the total cross sections for $c$-
and light-flavor and $b$- and light-flavor events.  All errors are the sum in
quadrature of experimental and extrapolation uncertainties}
\end{center}
\end{table}

As a cross check, we fitted the distorted Gaussian function described in section
6 to the $\xi$ distribution for each hadron species,
and calculated the fraction of
the area under the fitted curve that was within the range of our measurement.
An uncertainty was assigned corresponding to the largest variation obtained by
varying the fitted parameter values by all combinations of
$+1\sigma$ and $-1\sigma$.
The resulting fractions are consistent with those obtained using the
fragmentation models, giving
confidence in both the central values and the uncertainties assigned.

We applied the same procedure to our measurements for the three flavor
categories.
The three simulations were found to give similar flavor dependences, with the
accepted fraction in $b$ ($c$) events typically 0.02 (0.01) larger than that in
light-flavor events.  The resulting total cross sections are listed in table
\ref{xsintf} along with differences between flavors, for which some of the
systematic errors cancel.  We observe roughly 15\% more pseudoscalar mesons in
$b$-flavor events than in light-flavor events, and the respective
sums of the charged hadron differences are consistent with our previous
measurement \cite{nchflav} of the differences in total charged multiplicity
between light-, $c$- and $b$-flavor events.
All other differences are consistent with zero.

\section{Leading Particle Effects}

We extended these studies to look for differences between particle and
antiparticle production in light quark (as opposed to antiquark) jets,
in order to address the question of whether e.g. 
a primary $u$-initiated jet contains more hadrons that contain a valence
$u$-quark
(e.g. $\pi^+$, $K^+$, p, $\Lambda^0$) than hadrons that do not
(e.g. $\pi^-$, $K^-$, $\bar{\rm p}$, $\bar{\Lambda}^0$).
To this end we used the light
quark- and antiquark-tagged hemispheres described in section 3.

We measured the \pnb s per light quark jet
\begin{eqnarray}
R^{q}_{h} &=& {1\over{2N_{evts}}}{d\over{dx_{p}}}\left[ N(q\rightarrow
h)+N(\bar{q}\rightarrow\bar{h})\right],\\
R^{q}_{\bar{h}} &=& {1\over{2N_{evts}}}{d\over{dx_{p}}}\left[
N(q\rightarrow\bar{h})+N(\bar{q}\rightarrow h)\right],
\end{eqnarray}
where: $q$ and $\bar{q}$ represent light-flavor quark and antiquark jets
respectively; $N_{evts}$ is the total number of events in the sample; $h$
represents any of the identified hadron species $\pi^{-}$, $K^{-}$,
$\overline{K}^{*0}$, p, or $\Lambda^0$, and $\bar{h}$ indicates the
corresponding antihadron.
Then, for example, $N(q\rightarrow h)$ is
the number of hadrons of species $h$ in light quark jets.
This formulation assumes CP symmetry, i.e.
$N(q\rightarrow h) = N(\bar{q} \rightarrow \bar{h})$, which was found to be
satisfied in the data in all cases.

The charged hadron fractions analysis was repeated on the sample of positively
charged tracks in the quark-tagged jets and negatively charged tracks in the
antiquark-tagged jets, yielding measured values of
$R^{q}_{\pi^{+}}$, $R^{q}_{K^{+}}$, and $R^{q}_{\rm p}$ in the tagged samples.
The same procedure applied to the remaining tracks yielded $R^{q}_{\pi^{-}}$,
$R^{q}_{K^{-}}$, and $R^{q}_{\bar{\rm p}}$.
The \kkb\ and \llb\ analyses were applied similarly to the
quark- and antiquark-tagged jets to yield $R^{q}_{\overline{K}^{*0}}$,
$R^{q}_{K^{*0}}$, $R^{q}_{\Lambda}$ and $R^{q}_{\bar{\Lambda}}$.

The light-tagged event sample contains a residual heavy flavor background of
12\% $c\bar{c}$ and 3\% $b\bar{b}$ events.
The decays of the leading heavy hadrons in simulated heavy flavor background
events give rise to substantial differences between hadron and antihadron
production in the quark-tagged sample over the entire $x_p$ range.
It is essential to understand this contribution, which is typically 15\% of the
observed hadrons for $x_p<0.5$ and decreases at higher $x_p$ (see fig.
\ref{xsrbumc}).
The simulated contribution to each cross section was applied as a correction,
yielding \pnb s per light-quark-tagged jet.

For each hadron species, \pnb s in light quark jets were then extracted by
correcting for the light-tag bias (see sec. 5) and unfolding for the effective
quark (vs. antiquark) purity.
The purity was estimated from the simulation to be 0.76 for the \llb\ and
0.72 for the charged hadrons and \kkb , the latter value reflecting the 
cutoff in acceptance of the CRID at $|\cos\theta|=0.68$.

The measured \pnb s per light quark jet are listed in tables
\ref{xsqpi}--\ref{xsqla}
for the five measured hadron species that are not self-conjugate.
As for the flavor dependent results (sec. 5),
the error given is the sum in quadrature
of the statistical error and those systematic errors arising from the tagging
and correction procedures.  The latter include variation of the event tagging
efficiencies and biases as described in section 5, variation of the electroweak
parameters $R_b$, $R_c$, $A_b$ and $A_c$ by the errors on their respective world
average values \cite{pdg},
and variation of the effective quark purity by $\pm$0.015
to cover the uncertainty in the electron beam polarization and statistical error
on the simulated purity.
The systematic errors are small compared with the statistical errors, and are
typically dominated by the uncertainty on the effective quark purity.
These results supersede those in our previous publication \cite{lpprl}.

\begin{table}
\begin{center}
 \begin{tabular}{|r@{--}r|c|r@{$\pm$}l|r@{$\pm$}l||r@{$\pm$}l|}\hline
\multicolumn{2}{|c|}{ } && \multicolumn{6}{|c|}{ } \\ [-.3cm]
\multicolumn{2}{|c|}{$x_p$} &&
\multicolumn{6}{|c|}{$\pi^+$ and $\pi^-$ Production in u,d,s Jets} \\
\multicolumn{2}{|c|}{Range} & $<x_p>$ & \multicolumn{2}{|c|}{ $\pi^+$ } &
\multicolumn{2}{|c||}{   $\pi^-$   }  &
\multicolumn{2}{|c|}{ $D_{\pi^-}$} \\[.1cm]\hline
 0.016 & 0.022 & 0.019 &  140.9 & 2.5   &  139.0 & 2.6   &--0.007 & 0.016 \\
 0.022 & 0.033 & 0.027 &   98.2 & 1.5   &   96.7 & 1.4   &--0.007 & 0.014 \\
 0.033 & 0.044 & 0.038 &   62.8 & 1.3   &   63.6 & 1.3   &  0.007 & 0.019 \\
 0.044 & 0.055 & 0.049 &   44.2 & 1.4   &   44.9 & 1.4   &  0.007 & 0.029 \\
 0.055 & 0.066 & 0.060 &   33.4 & 1.1   &   33.2 & 1.1   &--0.003 & 0.030 \\
 0.066 & 0.077 & 0.071 &  25.79 & 0.82  &  27.16 & 0.82  &  0.026 & 0.028 \\
 0.077 & 0.088 & 0.082 &  21.66 & 0.71  &  22.34 & 0.71  &  0.016 & 0.029 \\
 0.088 & 0.099 & 0.093 &  17.17 & 0.62  &  18.40 & 0.63  &  0.034 & 0.032 \\
 0.099 & 0.110 & 0.104 &  14.45 & 0.57  &  14.52 & 0.57  &  0.003 & 0.036 \\
 0.110 & 0.121 & 0.115 &  11.44 & 0.50  &  12.84 & 0.52  &  0.057 & 0.038 \\
 0.121 & 0.143 & 0.131 &   9.32 & 0.32  &   9.61 & 0.32  &  0.015 & 0.031 \\
 0.143 & 0.164 & 0.153 &   7.21 & 0.28  &   7.39 & 0.28  &  0.012 & 0.035 \\
 0.164 & 0.186 & 0.175 &   5.40 & 0.24  &   5.49 & 0.25  &  0.008 & 0.041 \\
 0.186 & 0.208 & 0.197 &   4.30 & 0.21  &   4.44 & 0.22  &  0.016 & 0.045 \\
 0.208 & 0.230 & 0.219 &   3.14 & 0.19  &   3.30 & 0.19  &  0.026 & 0.053 \\
 0.230 & 0.274 & 0.251 &   2.37 & 0.12  &   2.59 & 0.12  &  0.043 & 0.043 \\
 0.274 & 0.318 & 0.295 &  1.398 & 0.091 &  1.687 & 0.097 &  0.093 & 0.055 \\
 0.318 & 0.384 & 0.348 &  0.972 & 0.061 &  0.996 & 0.064 &  0.012 & 0.057 \\
 0.384 & 0.471 & 0.423 &  0.456 & 0.040 &  0.504 & 0.042 &  0.050 & 0.077 \\
 0.471 & 0.603 & 0.527 &  0.180 & 0.025 &  0.210 & 0.026 &  0.08  & 0.12 \\
 0.603 & 0.768 & 0.668 &  0.065 & 0.019 &  0.089 & 0.021 &  0.16  & 0.23 \\
  [.1cm] \hline
  \end{tabular}
\caption{
\label{xsqpi}
Differential cross sections for the production of positive and negative pions
in light ($u$, $d$ and $s$) quark jets from hadronic $Z^0$ decays,
along with the normalized difference $D_{\pi^-}$ between the two.
The errors are the sum in quadrature of statistical errors and those systematic
errors arising from the light quark tagging and unfolding procedure.}
\end{center}
\end{table}

\begin{table}
\begin{center}
 \begin{tabular}{|r@{--}r|c|r@{$\pm$}l|r@{$\pm$}l||r@{$\pm$}l|} \hline
\multicolumn{2}{|c|}{ } && \multicolumn{6}{|c|}{ } \\ [-.3cm]
\multicolumn{2}{|c|}{$x_p$} &&
\multicolumn{6}{|c|}{$K^{*0}$ and $\bar{K}^{*0}$ Production in u,d,s Jets} \\
\multicolumn{2}{|c|}{Range} & $<x_p>$ & \multicolumn{2}{|c|}{$K^{*0}$} &
\multicolumn{2}{|c||}{$\bar{K}^{*0}$}  &
\multicolumn{2}{|c|}{ $D_{\bar{K}^{*0}}$} \\[.1cm]\hline
 0.018 & 0.048 & 0.033 & 2.50 & 0.94 & 2.69 & 0.95 &  0.04 & 0.29 \\
 0.048 & 0.088 & 0.068 & 1.64 & 0.36 & 2.40 & 0.38 &  0.18 & 0.14 \\
 0.088 & 0.149 & 0.118 & 1.11 & 0.22 & 0.88 & 0.22 &--0.11 & 0.17 \\
 0.149 & 0.263 & 0.206 & 0.318& 0.087& 0.447& 0.095&  0.17 & 0.19 \\
 0.263 & 0.483 & 0.342 & 0.053& 0.033& 0.264& 0.042&  0.67 & 0.18 \\
 0.483 & 1.000 & 0.607 & 0.022& 0.012& 0.100& 0.015&  0.64 & 0.16
 \\[.1cm] \hline
  \end{tabular}
\caption{
\label{xsqks}
Differential cross sections for the production of $K^{*0}$ and $\bar{K}^{*0}$
mesons in light quark jets, along with their normalized difference.}
\end{center}
\end{table}

\begin{table}
\begin{center}
 \begin{tabular}{|r@{--}r|c|r@{$\pm$}l|r@{$\pm$}l||r@{$\pm$}l|}\hline
\multicolumn{2}{|c|}{ } && \multicolumn{6}{|c|}{ } \\ [-.3cm]
\multicolumn{2}{|c|}{$x_p$} &&
\multicolumn{6}{|c|}{$K^+$ and $K^-$ Production in u,d,s Jets} \\
\multicolumn{2}{|c|}{Range} & $<x_p>$ & \multicolumn{2}{|c|}{  $K^+$  } &
\multicolumn{2}{|c||}{    $K^-$    }  &
\multicolumn{2}{|c|}{  $D_{K^-} $} \\[.1cm]\hline
 0.016 & 0.022 & 0.019 &   8.3  & 1.1  &   14.8 & 1.3  &  0.28 & 0.09 \\
 0.022 & 0.033 & 0.027 &   9.27 & 0.69 &   8.14 & 0.68 &--0.06 & 0.07 \\
 0.033 & 0.044 & 0.038 &   8.05 & 0.68 &   7.70 & 0.68 &--0.02 & 0.08 \\
 0.044 & 0.055 & 0.049 &   8.03 & 0.81 &   7.59 & 0.81 &--0.03 & 0.09 \\
 0.055 & 0.066 & 0.060 &   3.75 & 0.74 &   6.27 & 0.79 &  0.25 & 0.14 \\
 0.066 & 0.088 & 0.077 &   3.44 & 0.45 &   3.90 & 0.47 &  0.06 & 0.11 \\
 0.088 & 0.121 & 0.101 &   3.09 & 0.41 &   2.73 & 0.42 &--0.06 & 0.13 \\
 0.208 & 0.230 & 0.219 &   0.99 & 0.18 &   1.36 & 0.19 &  0.15 & 0.14 \\
 0.230 & 0.274 & 0.251 &   0.595& 0.091&   1.120& 0.099&  0.31 & 0.10 \\
 0.274 & 0.318 & 0.295 &   0.383& 0.072&   0.895& 0.081&  0.40 & 0.11 \\
 0.318 & 0.384 & 0.348 &   0.260& 0.049&   0.665& 0.055&  0.44 & 0.10 \\
 0.384 & 0.471 & 0.423 &   0.163& 0.034&   0.427& 0.039&  0.45 & 0.11 \\
 0.471 & 0.603 & 0.527 &   0.091& 0.023&   0.219& 0.026&  0.42 & 0.14 \\
 0.603 & 0.768 & 0.668 & --0.007& 0.017&   0.120& 0.022&  1.12 & 0.28 \\
  [.1cm] \hline
  \end{tabular}
\caption{
\label{xsqka}
Differential cross sections for the production of positive and negative kaons
in light quark
jets from hadronic $Z^0$ decays, along with their normalized difference.
   }
\end{center}
\end{table}

\begin{table}
\begin{center}
 \begin{tabular}{|r@{--}r|c|r@{$\pm$}l|r@{$\pm$}l||r@{$\pm$}l|}\hline
\multicolumn{2}{|c|}{ } && \multicolumn{6}{|c|}{ } \\ [-.3cm]
\multicolumn{2}{|c|}{$x_p$} &&
\multicolumn{6}{|c|}{p and $\bar{\rm p}$ Production in u,d,s Jets} \\
\multicolumn{2}{|c|}{Range} & $<x_p>$ & \multicolumn{2}{|c|}{    p    } &
\multicolumn{2}{|c||}{$\bar{\rm p}$}  &
\multicolumn{2}{|c|}{ $D_{\rm p}$} \\[.1cm]\hline
 0.022 & 0.033 & 0.027 &   7.1   & 1.1   &   4.7   & 1.4   &  0.20 & 0.21 \\
 0.033 & 0.044 & 0.038 &   5.76  & 0.52  &   4.83  & 0.51  &  0.09 & 0.09 \\
 0.044 & 0.055 & 0.049 &   4.10  & 0.44  &   4.07  & 0.44  &  0.00 & 0.10 \\
 0.055 & 0.066 & 0.060 &   3.65  & 0.44  &   3.20  & 0.44  &  0.07 & 0.12 \\
 0.066 & 0.088 & 0.077 &   2.69  & 0.30  &   2.31  & 0.30  &  0.08 & 0.11 \\
 0.088 & 0.121 & 0.101 &   1.82  & 0.29  &   1.99  & 0.30  &--0.04 & 0.14 \\
 0.230 & 0.274 & 0.251 &   0.618 & 0.078 &   0.292 & 0.072 &  0.36 & 0.15 \\
 0.274 & 0.318 & 0.295 &   0.387 & 0.056 &   0.157 & 0.053 &  0.42 & 0.18 \\
 0.318 & 0.384 & 0.348 &   0.257 & 0.035 &   0.099 & 0.033 &  0.44 & 0.18 \\
 0.384 & 0.471 & 0.423 &   0.117 & 0.020 &   0.076 & 0.019 &  0.21 & 0.19 \\
 0.471 & 0.603 & 0.527 &   0.070 & 0.010 &   0.025 & 0.009 &  0.47 & 0.19 \\
 0.603 & 0.768 & 0.668 &   0.018 & 0.004 &   0.001 & 0.004 &  0.85 & 0.42 \\
  [.1cm] \hline
  \end{tabular}
\caption{
\label{xsqpr}
Differential cross sections for the production of protons and antiprotons in
light quark jets, along with their normalized difference.}
\end{center}
\end{table}

\begin{table}
\begin{center}
 \begin{tabular}{|r@{--}r|c|r@{$\pm$}l|r@{$\pm$}l||r@{$\pm$}l|} \hline
\multicolumn{2}{|c|}{ } && \multicolumn{6}{|c|}{ } \\ [-.3cm]
\multicolumn{2}{|c|}{$x_p$} &&
\multicolumn{6}{|c|}{$\Lambda^0$ and $\bar{\Lambda}^0$ Production in u,d,s Jets} \\
\multicolumn{2}{|c|}{Range} & $<x_p>$ & \multicolumn{2}{|c|}{$\Lambda^0$} &
\multicolumn{2}{|c||}{$\bar{\Lambda}^0$}  &
\multicolumn{2}{|c|}{ $D_{\Lambda^0}$} \\[.1cm]\hline
 0.010 & 0.030 & 0.022 & 0.65 & 0.16 &  1.05 & 0.17 &--0.23 & 0.18 \\
 0.030 & 0.050 & 0.040 & 0.86 & 0.13 &  0.91 & 0.13 &--0.03 & 0.14 \\
 0.050 & 0.070 & 0.060 & 0.529& 0.084&  0.555& 0.084&--0.02 & 0.14 \\
 0.070 & 0.100 & 0.083 & 0.303& 0.057&  0.468& 0.060&--0.21 & 0.14 \\
 0.100 & 0.140 & 0.118 & 0.301& 0.053&  0.319& 0.054&--0.03 & 0.16 \\
 0.140 & 0.180 & 0.158 & 0.190& 0.048&  0.157& 0.047&  0.09 & 0.25 \\
 0.180 & 0.300 & 0.227 & 0.171& 0.034&  0.098& 0.032&  0.27 & 0.23 \\
 0.300 & 0.500 & 0.368 & 0.090& 0.022&  0.013& 0.019&  0.75 & 0.37
 \\[.1cm] \hline
  \end{tabular}
\caption{
\label{xsqla}
Differential cross sections for the production of
$\Lambda^0$ and $\bar{\Lambda}^0$ hyperons in light quark
jets, along with their normalized difference.}
\end{center}
\end{table}

\begin{figure}
\vspace{-1.cm}
 \hspace*{0.5cm}   
   \epsfxsize=3.5in
   \begin{center}\mbox{\epsffile{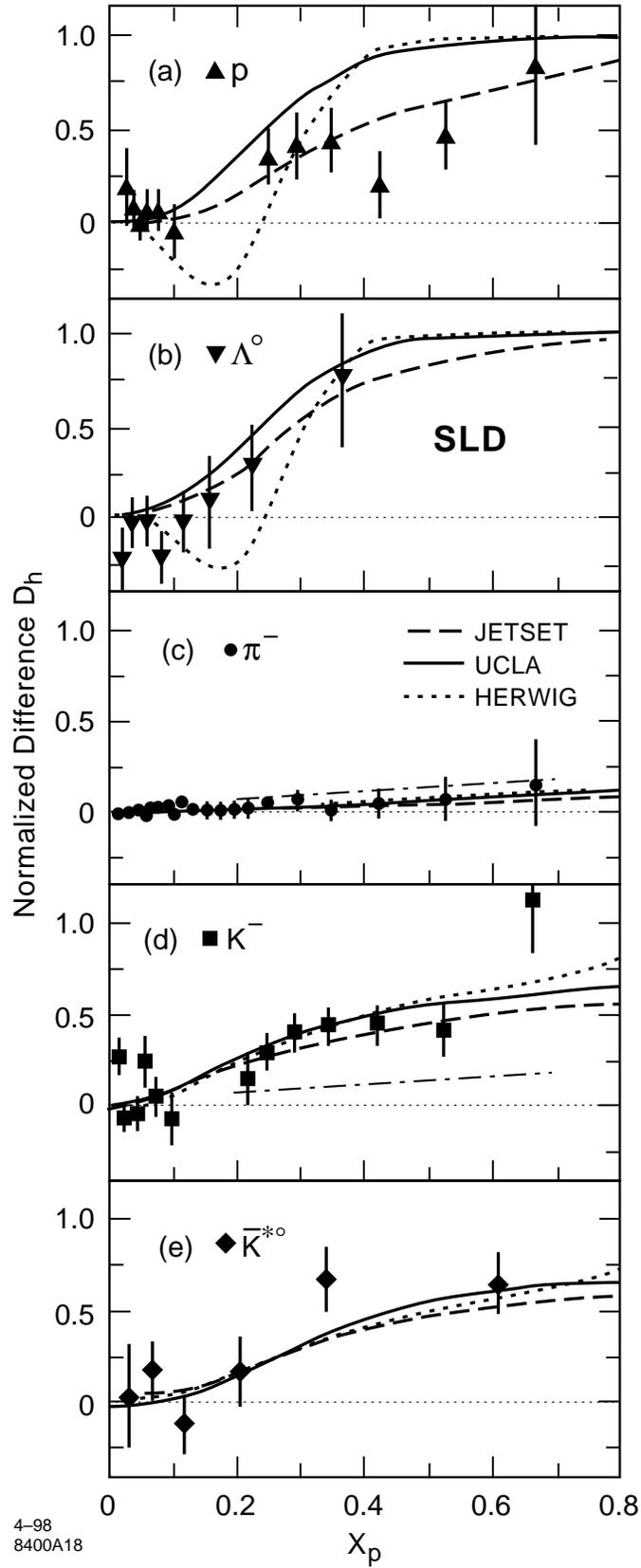}}\end{center}
  \caption{ 
 \label{ndall}
Normalized differences between hadron and antihadron production in light quark
jets.
The thin dot-dashed lines in (c) and (d) represent the fit to the
baryon data scaled by the dilution factor of 0.27 described in the text.
Also shown are the predictions of the three fragmentation models.
    }
\end{figure}

It is convenient to show these results in the form of the difference between
hadron $h$ and antihadron $\bar{h}$ production normalized by the sum:
\begin{equation}
D_{h} =  {R^{q}_{h} - R^{q}_{\overline{h}}\over
          R^{q}_{h} + R^{q}_{\overline{h}}}.
\end{equation}
The common systematic errors cancel explicitly in this variable, which
is shown for each hadron species in fig. \ref{ndall}.
A value of zero corresponds to equal production of hadron and antihadron,
whereas a value of $+$(--)1 corresponds to complete dominance of
(anti)particle production.
In each case the difference is consistent with zero at low $x_p$.
For charged pions it is also consistent with zero at high $x_p$, but for the
other hadrons there are significant positive differences that appear to increase
with increasing $x_p$.

The results for the baryons (fig. \ref{ndall}a,b) afford the most
straightforward interpretation.
Since baryons contain valence quarks and not antiquarks, the observed excess
of both protons and $\Lambda^0$s over their respective antibaryons for $x_p>0.2$
is clear evidence for the production of leading baryons.
The data suggest that the effect increases with $x_p$,
however more data are needed to study the $x_p$ dependence in detail.
For $x_p<0.2$ the data are consistent with equal production of baryons and
antibaryons, however the contribution from fragmentation
is very high in this region and we cannot exclude that leading baryons are also
produced at low $x_p$.

Since a meson
contains one valence quark along with one valence antiquark,
the interpretation of our results for mesons is more complicated.
All down-type quarks are produced equally and with the same
forward-backward asymmetry in $Z^0$ decays in the Standard Model,
so that if a leading neutral
particle such as $\bar{K}^{*0}$ ($s\bar{d}$) were produced equally in $s$ and
$\bar{d}$ jets
(i.e. $D^{d\bar{d}}_{\bar{K}^{*0}} = -D^{s\bar{s}}_{\bar{K}^{*0}}$),
then our measured $D_{\bar{K}^{*0}}$ would be zero.
Our two highest-$x_p$ points are significantly positive,
indicating both that there is leading $\bar{K}^{*0}$ production
{\it and} that more leading $\bar{K}^{*0}$ are produced in $s$ jets than in
$\bar{d}$ jets.
This is an expected consequence of strangeness suppression in the
fragmentation process.  That is, it is expected to be less likely for an
$s\bar{s}$ to be produced from the vacuum and the $s$ to pair up
with an initial $\bar{d}$ than it is for a $d\bar{d}$ to be produced
and the $\bar{d}$ to pair up with the initial $s$.

In the case of charged hadrons such as $\pi^-$ ($d\bar{u}$),
the different $Z^0$ branching ratios and forward-backward asymmetries of up- and
down-type quarks cause a nonzero dilution of leading particle effects.
Assuming Standard Model couplings to the $Z^0$ and equal production of leading
$\pi^+$ in $u$-jets and $\pi^-$ in $d$-jets
(i.e. $D^{d\bar{d}}_{\pi^-} = -D^{u\bar{u}}_{\pi^-}$),
we calculate a dilution factor for our analysis cuts of 0.27.
That is, we would expect to observe $D_{\pi^-} = 0.27 D^{d\bar{d}}_{\pi^-}$.
For purposes of illustration, we have fitted a line to our $D_p$ and
$D_{\Lambda^0}$ points for $x_p > 0.2$, scaled
it by the dilution factor 0.27, and drawn it as the dot-dashed line on figs.
\ref{ndall}c and \ref{ndall}d.
We do not necessarily expect that leading particle effects are identical
for mesons and for baryons, but
this line serves as a basis for a qualitative comparison.

Our measured $D_{\pi^-}$ are consistent with zero everywhere, and consistently
below this line.  This does not rule out leading pion production,
but indicates that nonleading production of pions must be comparable or larger
at all $x_p$.
This could be due to a very soft leading pion momentum
distribution and/or a large ``background" contribution of pions from decays
of excited states such as $\rho^0$, $\omega$, $\eta$, $K^*$.
Our measured $D_{K^-}$ are consistently positive and above the line for
$x_p > 0.2$.
As in the case of $\bar{K}^{*0}/K^{*0}$, this indicates both production of
leading charged kaons and more frequent
production of leading $K^-$ in $s$-jets than in $\bar{u}$-jets.

The quantification of the total number of observed leading particles is
problematic.
For example, in the region $x_p>0.2$ we observe a total of 0.083$\pm$0.005
protons and 0.036$\pm$0.005 antiprotons per light quark jet.
Some of the antiprotons are expected to be ``subleading" antiprotons 
produced in association with a leading baryon, since baryon number is known to
be conserved locally \cite{bcorrl},
whereas others are from a non-leading baryon-antibaryon pair, and provide a
measure of the background of nonleading protons in the high-$x_p$ sample.
We conclude that the number of leading protons we have observed per light
quark jet must lie
between the p-$\bar{\rm p}$ difference and the total number of protons,
i.e. in the range 0.047--0.083 per light quark jet.
Similarly, the number of observed leading $\Lambda^0$ in the range
$0.18<x_p<0.5$ is 0.024--0.039.
For $x_p>0.26$ we measure a total of 0.110$\pm$0.012 $\bar{K}^{*0}$ and
0.023$\pm$0.010 $K^{*0}$ per light quark jet.
In this case, all of these could be leading due to contributions from $s$ and
$d$ jets, and so the sum gives an upper bound on the number of leading
$K^{*0}$/$\bar{K}^{*0}$ produced.
A lower bound is given by the possibility that no leading $K^{*0}$ are produced
in $d$ jets.  In this case all of the observed $K^{*0}$ are nonleading,
we expect an equal number of nonleading $\bar{K}^{*0}$, and the
number of leading $\bar{K}^{*0}$ produced is given by the
$\bar{K}^{*0}$--$K^{*0}$ difference.
Thus we have observed 0.087--0.133 leading \kkb\ per jet with $x_p>0.26$.
Similarly, the number of leading charged kaons produced in the range
$0.21<x_p<0.77$ is 0.141--0.355 per jet.

The measured normalized differences are
compared with the predictions of the three fragmentation models in fig.
\ref{ndall}.  All models reproduce the qualitative features of our data.
For the baryons, the HERWIG prediction drops below zero in the range in which we
have no proton coverage; this behavior might be ruled out with more \llb\ data.
The HERWIG and UCLA predictions rise sharply to unity at $x_p \approx 0.4$ and
are inconsistent with the proton data.
For the mesons all models are consistent with the data.

\section{Production Ratios and Fragmentation Parameters}

\begin{figure}
\vspace{-1.cm}
 \hspace*{0.5cm}   
   \epsfxsize=6.4in
   \begin{center}\mbox{\epsffile{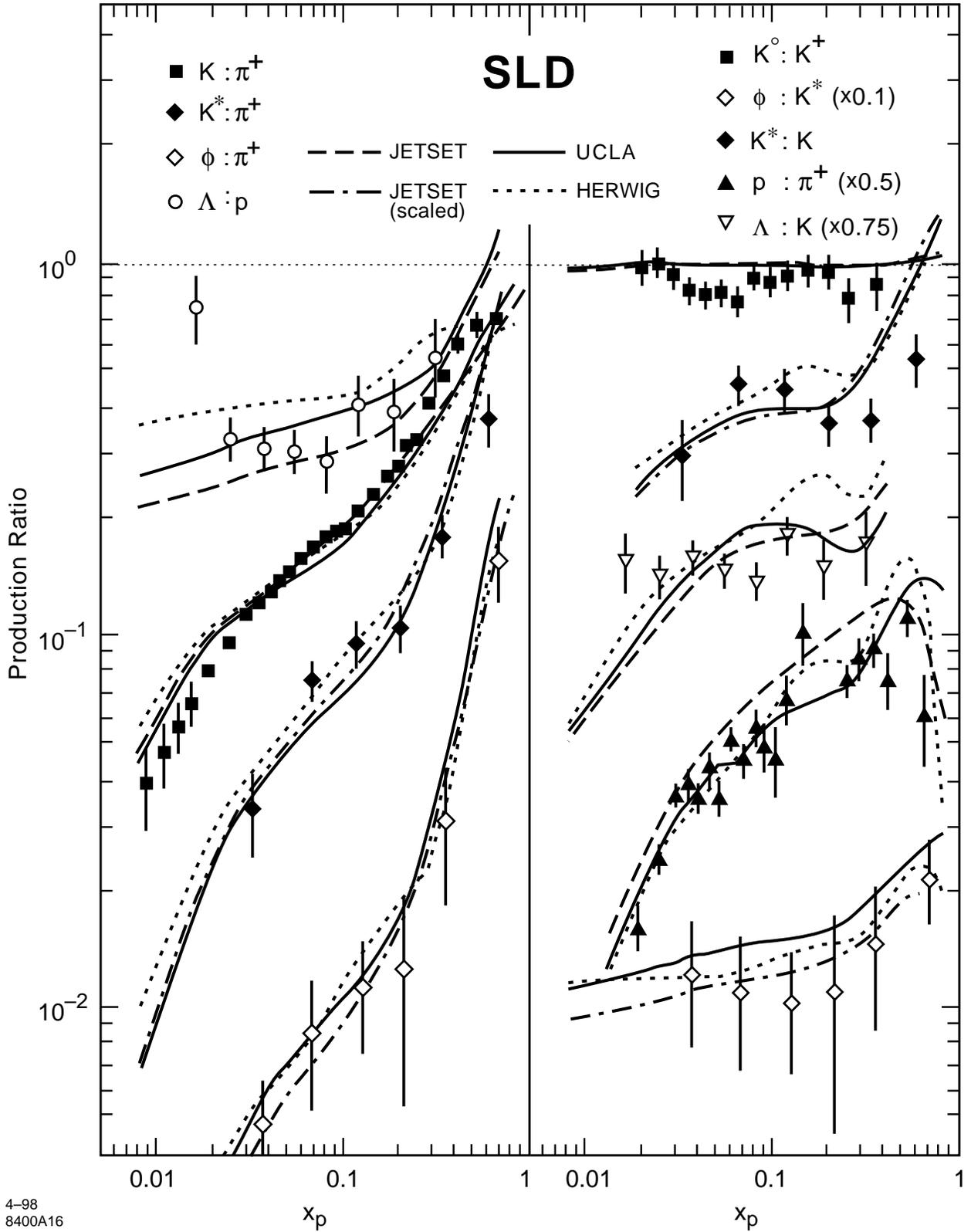}}\end{center}
  \caption{ 
 \label{rtuds}
Ratios of measured \pnb s for various pairs of hadron species in
light-flavor events, along with the predictions of the three fragmentation
models.  In all cases the charge-conjugate states are included in both numerator
and denominator.
Here, ``K" denotes the average of $K^0/\bar{K}^0$ and $K^{\pm}$.
The JETSET predictions for the $K^*$:$\pi^+$, $\phi$:$\pi^+$, $\phi$:$K^*$ and
$K^*$:$K$ ratios have been scaled by factors of 2/3, 1/2, 4/3 and 2/3,
respectively (see text),
in order to clarify the comparison of the momentum dependence.
    }
\end{figure}

Certain aspects of the fragmentation process can be studied more directly by
measuring the relative production of two hadron species that differ
by a single quantum number.
We have calculated the ratios of \pnb s for a number of pairs of hadron species,
for flavor-inclusive and light-flavor events, taking into account any 
systematic errors common to the two species.
The results are shown for light-flavor events in fig. \ref{rtuds}.
In the cases where binning was different for the two hadron species in a pair,
the ratio was obtained by fitting a curve
to the denominator over a region near each $x_p$ value in the numerator.
In some cases charged and neutral pseudoscalar kaons were averaged, and are
denoted simply ``K".
In all cases, charge-conjugate states are included in both numerator and
denominator.

The ratios of the strange mesons to pions vary rapidly with $x_p$.
In flavor-inclusive events (not shown), the values of each of these ratios
vary over a similar range but show less structure,
being consistent with simple powers of $x_p$ for $x_p>0.04$.
The proton:pion ratio also varies rapidly for $x_p<0.1$.
The other ratios shown in fig. \ref{rtuds} are independent of $x_p$ within
our errors.

The $K^0$:$K^{\pm}$ ratio differs significantly from unity over the range
$0.03<x_p<0.09$, averaging 0.86$\pm$0.03;
we observe a similar difference in flavor-inclusive events (not shown), 
as has been observed previously \cite{bohrer}.
Assuming that primary charged and neutral kaons are produced equally in
the fragmentation process, this implies that some hadron species is
produced that decays preferentially into charged kaons.
Our measured cross sections indicate that decays of $\phi$ and $K^*$ mesons
would each account for only $\sim$0.01 of the difference from unity.
Decays of $D$- and $B$-hadrons cannot be the source of this difference since
they have been excluded explicitly.

The predictions of the three fragmentation models are also shown in
fig. \ref{rtuds}, and all describe the qualitative features of the data.
The JETSET prediction for each ratio involving $K^{*}$ or $\phi$ mesons differs
from the data by a large normalization factor, and those predictions have been
scaled by factors derived from fig. \ref{xsudsmc} in order to compare the
momentum dependence with that of the data.
All models underestimate the slope of the $K$:$\pi^+$ ratio, but reproduce
those of the $\phi$:$\pi^+$ and $K^{*}$:$\pi^+$ ratios,
overestimating the latter ratio only at the highest-$x_p$ point.
The $x_p$ dependence of the p:$\pi^+$ ratio is reproduced by all models at
low $x_p$, but only by the JETSET model for $x_p>0.2$.
However the JETSET model shows a normalization difference from the data
of about 20\%.
Similar differences in the model predictions for the $\Lambda$:$K$ ratio cannot
be resolved with the current statistics.
No model reproduces the measured $K^0$:$K^+$ ratio; all predict a
roughly constant value of 0.98 in the range of our measurement.
All models predict a larger value of the $K^*$:$K$ ratio at the
highest-$x_p$ point than is observed in the data.
A similar set of comparisons for flavor-inclusive events (not shown) yielded
the same conclusions.

These ratios can be used to study the suppression of baryons, vector mesons and
strange hadrons in the fragmentation process.
Quantifying such suppression at the primary fragmentation level is problematic
due to possible effects of different masses of the two hadron species in the
ratio and the fact that decay products populate a different $x_p$ region than
their primary parents.
We therefore used the JETSET model, in which there are tunable parameters
controlling the relative production of baryons, strange hadrons and vector
mesons,  to extract suppression parameters in the context of that model.
We first considered the relative production of pseudoscalar ($P$) and vector
($V$) mesons, traditionally expressed in terms of the parameter $P_V = V/(V+P)$.
Since we might expect that measured ratios are not the same at very high $x_p$,
where leading hadron production is important, as they are lower $x_p$, we
defined arbitrarily a ``fragmentation" region, $0.05<x_p<0.25$, and a ``leading"
region, $x_p>0.45$.
In each region we averaged our measured $K^{*}$:$K$ ratio, and compared it with
those obtained in the same region from the JETSET generator run with a series
of input values of the $P_V$ parameter for strange mesons.
We interpolated to find the $P_V$ value at which the model prediction for each
ratio was equal to that measured in the data, and these values are listed in
table \ref{tpv} for the two $x_p$ regions and for both flavor-inclusive and
light-flavor events.
The two measurements in each momentum range are consistent, 
but the $P_V$ value measured in the fragmentation region is significantly
higher than that measured in the leading region for both flavor categories.

\begin{table}
\begin{center}
 \begin{tabular}{|c|c|c|} \hline
 \multicolumn{3}{|c|}{    }  \\[-.3cm]
 \multicolumn{3}{|c|}{Vector:Pseudoscalar Production Parameter $P_V$}  \\[.1cm]
 \hline  \hline 
 &&\\[-.3cm]
 $x_p$ Range  & inclusive  &  light-flavor \\[.1cm]\hline
 &&\\[-.3cm] 
0.055--0.219 & 0.405$\pm$0.020 & 0.433$\pm$0.033 \\
0.439--1.000 & 0.226$\pm$0.029 & 0.279$\pm$0.029
 \\[.1cm] \hline 
 \end{tabular}
\caption{
\label{tpv}
Measurements of the vector-meson fraction $P_V$ extracted from the
measured $K^*$:$K$ production ratio in the context of the JETSET model.}
\end{center}
\end{table}

We next considered the relative production of baryons (B) and
mesons (M), in terms of the parameter $P_B = B / (B+M)$.
A similar set of comparisons of our p:$\pi$ and $\Lambda$:$K$ ratios
with the predictions of the JETSET model as $P_B$ was varied yielded
the measured $P_B$ values listed in table \ref{tpb}.
The four values extracted from the p:$\pi$ ratio are consistent.
The value from the $\Lambda$:$K$ ratio in light-flavor events is consistent
with these four, but that in flavor-inclusive events is slightly larger.

\begin{table}
\begin{center}
 \begin{tabular}{|r@{:}lc|c|c|} \hline
 \multicolumn{5}{|c|}{  }  \\[-.3cm]
 \multicolumn{5}{|c|}{Baryon:Meson Production Parameter $P_B$}
\\[.1cm] \hline\hline
 \multicolumn{2}{|c}{  } &&& \\[-.3cm]
 \multicolumn{2}{|c}{Ratio} & $x_p$ Range & inclusive  &  light-flavor \\[.1cm]\hline
 \multicolumn{2}{|c}{  } &&& \\[-.3cm]
 p & $\pi^\pm$    &  0.055--0.165  &  0.076$\pm$0.003  & 0.074$\pm$0.004 \\
  $\Lambda$ & $K$ &  0.061--0.237  &  0.101$\pm$0.003  & 0.087$\pm$0.005 \\
  p & $\pi^\pm$   &  0.493--0.987  &  0.081$\pm$0.006  & 0.081$\pm$0.009
  \\[.1cm] \hline 
 \end{tabular}
\caption{
\label{tpb}
Measurements of the baryon fraction $P_B$ in the context of the
JETSET model.}
\end{center}
\end{table}

Information on the suppression of strangeness is available from several of our
measurements.
It is conventional to define a suppression factor $\gamma_s$ as the
probability of creating an $s\bar{s}$ from the vacuum, relative to
that of creating a $u\bar{u}$ or $d\bar{d}$, at a given point in the
fragmentation process.
As has been suggested in ref. \cite{lafferty}, the normalized production
difference (see section 8) at high $x_p$ between a strange hadron and its
antihadron in light quark jets provides a robust way of investigating
strangeness suppression for any neutral hadron, such as \kkb , that is
unlikely to be a decay product of a heavier primary particle.
If we assume leading particle dominance, so that $\bar{K}^{*0}$ can be produced
only in $s$ and $\bar{d}$ jets, and that the relative production in $\bar{d}$
jets is suppressed by a factor of $\gamma_s$, then we expect the normalized
difference to be $D_{\bar{K}^{*0}} = (1-\gamma_s)/(1+\gamma_s)$.
From our point in the bin $0.5<x_p<1$ we used this equation to derive a
``direct" measurement of $\gamma_s=0.26\pm0.12$, where we first scaled our
given $D_{\bar{K}^{*0}}$ value by 0.923 to account for the fact that we assumed
contributions from $u$, $d$ and $s$ jets in the original unfolding,
whereas we now assume only $d$ and $s$ contribute.
Similarly, assuming dominant production of leading $K^{\pm}$ and accounting for
the different branching fraction and forward-backward asymmetry of
up- and down-type events,
one expects $1.05 D_{K^-} = (1-0.55\gamma_s)/(1+0.77\gamma_s)$.
From this we derive $\gamma_s=0.41\pm0.17$, using our $D_{K^-}$ data in the
range $0.47<x_p<0.77$.

We also used the JETSET model to predict the normalized differences as a
function of $\gamma_s$, and to extract from our measured $D_{\bar{K}^{*0}}$ and
$D_{K^-}$ the $\gamma_s$ values listed in table \ref{tgs}.
Also listed in table \ref{tgs} are $\gamma_s$ values extracted in the context of
the JETSET model from our measured $K$:$\pi^+$, $\phi$:$K^*$ and
$\Lambda$:p ratios.
For each ratio, the values derived from the flavor-inclusive and light-flavor
events are consistent.
However there is a significant $x_p$ dependence in the values obtained from the
$K$:$\pi^+$ ratio in both flavor categories, and there are several other
significant differences between pairs of values from the same flavor category.
This indicates that the JETSET model cannot
accommodate all of our data with a single $\gamma_s$ value and all other
parameters set to their default values.

\begin{table}
\begin{center}
 \begin{tabular}{|r@{:}l|c|c|c|} \hline
 \multicolumn{5}{|c|}{  }  \\[-.3cm]
 \multicolumn{5}{|c|}{Strangeness Suppression Factor, $\gamma_s$}  \\[.1cm]
 \hline  \hline 
 \multicolumn{2}{|c|}{  } &&& \\[-.3cm]
 \multicolumn{2}{|c|}{Ratio} & $x_p$ Range & inclusive & light-flavor \\[.1cm]\hline
 \multicolumn{2}{|c|}{  } &&& \\[-.2cm]
\multicolumn{2}{|c|}{$D_{\bar{K}^{*0}}$}
                  &  0.482--1.000 &     --           & 0.194$\pm$0.141 \\
\multicolumn{2}{|c|}{$D_{K^{-}}$}
                  &  0.493--0.768 &     --           & 0.249$\pm$0.110
  \\[.1cm] \hline 
 \multicolumn{2}{|c|}{  } &&& \\[-.2cm]
   $K$ & $\pi^+$  &  0.055--0.219 & 0.236$\pm$0.016  & 0.266$\pm$0.014 \\
$\phi$ & $K^*$    &  0.048--0.263 & 0.163$\pm$0.027  & 0.184$\pm$0.052 \\
 $\Lambda$ & p    &  0.050--0.182 & 0.339$\pm$0.014  & 0.311$\pm$0.032
  \\[.1cm] \hline 
 \multicolumn{2}{|c|}{  } &&& \\[-.2cm]
   $K$ & $\pi^+$  &  0.493--0.768 & 0.575$\pm$0.084  & 0.483$\pm$0.091 \\
$\phi$ & $K^*$    &  0.482--1.000 & 0.160$\pm$0.060  & 0.239$\pm$0.075
  \\[.1cm] \hline 
 \end{tabular}
\caption{
\label{tgs}
Measurements of the strangeness suppression factor $\gamma_s$ in the context of
the JETSET model.  The notation $D_h$ refers to the normalized differences
discussed in section 8.}
\end{center}
\end{table}

\section{Summary and Conclusions}

We have measured the production of the seven hadron species $\pi^\pm$,
$K^\pm$, $K^0/\bar{K}^0$, \kkb , $\phi$, p/$\bar{\rm p}$,
and \llb\ as a function of scaled momentum $x_p$ over a wide range in hadronic
$Z^0$ decays.
The SLD Cherenkov Ring Imaging Detector
enabled the clean and efficient identification of stable charged hadrons,
yielding precise measurements of their production cross sections, as well as 
the identification of relatively clean samples of the strange mesons \kkb\
and $\phi$ reconstructed in
decay modes containing charged kaons.
Our measurements of differential production cross sections, total cross sections
and ratios of production of these hadron species in
flavor-inclusive hadronic $Z^0$ decays are consistent 
with averages of those from experiments at LEP.

Using the SLD vertex detector to isolate high-purity light- and $b$-tagged
event samples, we have measured the production of these seven hadron species
in light-, $c$- and $b$-flavor events.
Significant differences between flavors were found,
consistent with expectations based on the known properties of
$B$ and $D$ hadron production and decay.
Our $\pi^\pm$, $K^\pm$ and p/$\bar{\rm p}$ data at high $x_p$ were used to
test the predictions of Gribov and Lipatov for the shape of the $x_p$
distribution of primary leading hadrons as $x_p \rightarrow 1$.
We find the predictions of the theory to be consistent with the flavor-inclusive
(light-flavor) meson data for $x_p>0.66$ ($x_p>0.47$) and with the proton data
for $x_p>0.43$ ($x_p>0.38$).
The shape of the $\xi = -\ln(x_p)$ distribution for each hadron species
in events of each flavor is consistent with the Gaussian form predicted by
MLLA QCD$+$LPHD near its peak.
The peak positions $\xi^*$ for each hadron species in light-flavor events
are more consistent with a monotonic dependence on hadron mass than those in
flavor-inclusive events.

Using the large forward-backward asymmetry induced by the polarized SLC electron
beam to separate light quark from light antiquark hemispheres,
we have updated our measurements of hadron and antihadron production in light
quark jets.
Differences are observed at high $x_p$ between baryon and
antibaryon production, which is evidence for the production of
leading baryons, i.e. baryons that carry the quantum numbers of the initial
quark.
Differences are also observed for both pseudoscalar and vector $K$-mesons, which
indicate not only leading production of these two hadron species but also that
leading strange mesons are produced more often from initial $s$ quarks than from
initial $u$ or $d$ quarks.

Our data were used to test the predictions of three fragmentation
models with default parameters.
In most cases these simulations reproduced the data to within a few percent.
However the JETSET 7.4 model predicts too many p/$\bar{\rm p}$,
\kkb\ and $\phi$ mesons at all $x_p$, and too many $K^\pm$ and $K^0$/$\bar{K}^0$
at low $x_p$.
The UCLA model predicts too many pions in the 2--20 GeV/c range,
a shoulder in the $x_p$ distributions for baryons at high $x_p$, and
larger differences between baryon and antibaryon production at high $x_p$ than
are seen in our light-quark data.
The HERWIG 5.8 model predicts a shoulder in the $x_p$ distribution for most
hadron species at high $x_p$,
a large excess of low-$x_p$ pions and kaons in $b$-flavor events and
of medium-$x_p$ pions in $c$-flavor events, and
a rapid variation in the baryon-antibaryon differences as a function of $x_p$.
All models predict a charged:neutral kaon ratio very close to unity, which is
inconsistent with our light-flavor and flavor-inclusive data.
Also, no model is consistent with the $x_p$ dependence of either our $K$:$\pi$
ratio or our $K^*$:$K$ ratio.

We have studied several parameters of the fragmentation process.
The differences between kaon and antikaon production in light quark jets allow
two new, direct measurements of strangeness suppression at high momentum.
We have also used our ratios of production of pairs of hadron species to extract
fragmentation parameters in the context of the JETSET model.
We find the vector:pseudoscalar meson parameter to be dependent on $x_p$,
and the strangeness suppression parameter to be dependent both on $x_p$ and on
the hadron species used to form the ratio.

\section*{Acknowledgements}

We thank the personnel of the SLAC accelerator department and the
technical
staffs of our collaborating institutions for their outstanding efforts
on our behalf.
We thank S. Brodsky and L. Dixon for useful discussions.

\section*{$^{**}$List of Authors} 
%
%
\begin{center}
\def\iADEL{$^{(1)}$}
\def\iAOMORI{$^{(2)}$}
\def\iBOLO{$^{(3)}$}
\def\iBRUN{$^{(4)}$}
\def\iBU{$^{(5)}$}
\def\iCINC{$^{(6)}$}
\def\iCOLO{$^{(7)}$}
\def\iCOLU{$^{(8)}$}
\def\iCSU{$^{(9)}$}
\def\iFERR{$^{(10)}$}
\def\iFRAS{$^{(11)}$}
\def\iILLI{$^{(12)}$}
\def\iLBL{$^{(13)}$}
\def\iLTU{$^{(14)}$}
\def\iMASS{$^{(15)}$}
\def\iMISSI{$^{(16)}$}
\def\iMIT{$^{(17)}$}
\def\iMOSCOW{$^{(18)}$}
\def\iNAGO{$^{(19)}$}
\def\iOREG{$^{(20)}$}
\def\iOXF{$^{(21)}$}
\def\iPADO{$^{(22)}$}
\def\iPERU{$^{(23)}$}
\def\iPISA{$^{(24)}$}
\def\iRAL{$^{(25)}$}
\def\iRUTG{$^{(26)}$}
\def\iSLAC{$^{(27)}$}
\def\iSOGA{$^{(28)}$}
\def\iSOONG{$^{(29)}$}
\def\iTENN{$^{(30)}$}
\def\iTOHO{$^{(31)}$}
\def\iUCSB{$^{(32)}$}
\def\iUCSC{$^{(33)}$}
\def\iVAND{$^{(34)}$}
\def\iWASH{$^{(35)}$}
\def\iWISC{$^{(36)}$}
\def\iYALE{$^{(37)}$}

  \baselineskip=.75\baselineskip  
\mbox{K. Abe\unskip,\iAOMORI}
\mbox{K.  Abe\unskip,\iNAGO}
\mbox{T. Abe\unskip,\iSLAC}
\mbox{I.Adam\unskip,\iSLAC}
\mbox{T.  Akagi\unskip,\iSLAC}
\mbox{N. J. Allen\unskip,\iBRUN}
\mbox{A. Arodzero\unskip,\iOREG}
\mbox{W.W. Ash\unskip,\iSLAC}
\mbox{D. Aston\unskip,\iSLAC}
\mbox{K.G. Baird\unskip,\iRUTG}
\mbox{C. Baltay\unskip,\iYALE}
\mbox{H.R. Band\unskip,\iWISC}
\mbox{M.B. Barakat\unskip,\iLTU}
\mbox{O. Bardon\unskip,\iMIT}
\mbox{T.L. Barklow\unskip,\iSLAC}
\mbox{J.M. Bauer\unskip,\iMISSI}
\mbox{G. Bellodi\unskip,\iOXF}
\mbox{R. Ben-David\unskip,\iYALE}
\mbox{A.C. Benvenuti\unskip,\iBOLO}
\mbox{G.M. Bilei\unskip,\iPERU}
\mbox{D. Bisello\unskip,\iPADO}
\mbox{G. Blaylock\unskip,\iMASS}
\mbox{J.R. Bogart\unskip,\iSLAC}
\mbox{B. Bolen\unskip,\iMISSI}
\mbox{G.R. Bower\unskip,\iSLAC}
\mbox{J. E. Brau\unskip,\iOREG}
\mbox{M. Breidenbach\unskip,\iSLAC}
\mbox{W.M. Bugg\unskip,\iTENN}
\mbox{D. Burke\unskip,\iSLAC}
\mbox{T.H. Burnett\unskip,\iWASH}
\mbox{P.N. Burrows\unskip,\iOXF}
\mbox{A. Calcaterra\unskip,\iFRAS}
\mbox{D.O. Caldwell\unskip,\iUCSB}
\mbox{D. Calloway\unskip,\iSLAC}
\mbox{B. Camanzi\unskip,\iFERR}
\mbox{M. Carpinelli\unskip,\iPISA}
\mbox{R. Cassell\unskip,\iSLAC}
\mbox{R. Castaldi\unskip,\iPISA}
\mbox{A. Castro\unskip,\iPADO}
\mbox{M. Cavalli-Sforza\unskip,\iUCSC}
\mbox{A. Chou\unskip,\iSLAC}
\mbox{E. Church\unskip,\iWASH}
\mbox{H.O. Cohn\unskip,\iTENN}
\mbox{J.A. Coller\unskip,\iBU}
\mbox{M.R. Convery\unskip,\iSLAC}
\mbox{V. Cook\unskip,\iWASH}
\mbox{R. Cotton\unskip,\iBRUN}
\mbox{R.F. Cowan\unskip,\iMIT}
\mbox{D.G. Coyne\unskip,\iUCSC}
\mbox{G. Crawford\unskip,\iSLAC}
\mbox{C.J.S. Damerell\unskip,\iRAL}
\mbox{M. N. Danielson\unskip,\iCOLO}
\mbox{M. Daoudi\unskip,\iSLAC}
\mbox{N. de Groot\unskip,\iSLAC}
\mbox{R. Dell'Orso\unskip,\iPERU}
\mbox{P.J. Dervan\unskip,\iBRUN}
\mbox{R. de Sangro\unskip,\iFRAS}
\mbox{M. Dima\unskip,\iCSU}
\mbox{A. D'Oliveira\unskip,\iCINC}
\mbox{D.N. Dong\unskip,\iMIT}
\mbox{P.Y.C. Du\unskip,\iTENN}
\mbox{R. Dubois\unskip,\iSLAC}
\mbox{B.I. Eisenstein\unskip,\iILLI}
\mbox{V. Eschenburg\unskip,\iMISSI}
\mbox{E. Etzion\unskip,\iWISC}
\mbox{S. Fahey\unskip,\iCOLO}
\mbox{D. Falciai\unskip,\iFRAS}
\mbox{C. Fan\unskip,\iCOLO}
\mbox{J.P. Fernandez\unskip,\iUCSC}
\mbox{M.J. Fero\unskip,\iMIT}
\mbox{K.Flood\unskip,\iMASS}
\mbox{R. Frey\unskip,\iOREG}
\mbox{T. Gillman\unskip,\iRAL}
\mbox{G. Gladding\unskip,\iILLI}
\mbox{S. Gonzalez\unskip,\iMIT}
\mbox{E.L. Hart\unskip,\iTENN}
\mbox{J.L. Harton\unskip,\iCSU}
\mbox{A. Hasan\unskip,\iBRUN}
\mbox{K. Hasuko\unskip,\iTOHO}
\mbox{S. J. Hedges\unskip,\iBU}
\mbox{S.S. Hertzbach\unskip,\iMASS}
\mbox{M.D. Hildreth\unskip,\iSLAC}
\mbox{J. Huber\unskip,\iOREG}
\mbox{M.E. Huffer\unskip,\iSLAC}
\mbox{E.W. Hughes\unskip,\iSLAC}
\mbox{X.Huynh\unskip,\iSLAC}
\mbox{H. Hwang\unskip,\iOREG}
\mbox{M. Iwasaki\unskip,\iOREG}
\mbox{D. J. Jackson\unskip,\iRAL}
\mbox{P. Jacques\unskip,\iRUTG}
\mbox{J.A. Jaros\unskip,\iSLAC}
\mbox{Z.Y. Jiang\unskip,\iSLAC}
\mbox{A.S. Johnson\unskip,\iSLAC}
\mbox{J.R. Johnson\unskip,\iWISC}
\mbox{R.A. Johnson\unskip,\iCINC}
\mbox{T. Junk\unskip,\iSLAC}
\mbox{R. Kajikawa\unskip,\iNAGO}
\mbox{M. Kalelkar\unskip,\iRUTG}
\mbox{Y. Kamyshkov\unskip,\iTENN}
\mbox{H.J. Kang\unskip,\iRUTG}
\mbox{I. Karliner\unskip,\iILLI}
\mbox{H. Kawahara\unskip,\iSLAC}
\mbox{Y. D. Kim\unskip,\iSOGA}
\mbox{R. King\unskip,\iSLAC}
\mbox{M.E. King\unskip,\iSLAC}
\mbox{R.R. Kofler\unskip,\iMASS}
\mbox{N.M. Krishna\unskip,\iCOLO}
\mbox{R.S. Kroeger\unskip,\iMISSI}
\mbox{M. Langston\unskip,\iOREG}
\mbox{A. Lath\unskip,\iMIT}
\mbox{D.W.G. Leith\unskip,\iSLAC}
\mbox{V. Lia\unskip,\iMIT}
\mbox{C.-J. S. Lin\unskip,\iSLAC}
\mbox{X. Liu\unskip,\iUCSC}
\mbox{M.X. Liu\unskip,\iYALE}
\mbox{M. Loreti\unskip,\iPADO}
\mbox{A. Lu\unskip,\iUCSB}
\mbox{H.L. Lynch\unskip,\iSLAC}
\mbox{J. Ma\unskip,\iWASH}
\mbox{G. Mancinelli\unskip,\iRUTG}
\mbox{S. Manly\unskip,\iYALE}
\mbox{G. Mantovani\unskip,\iPERU}
\mbox{T.W. Markiewicz\unskip,\iSLAC}
\mbox{T. Maruyama\unskip,\iSLAC}
\mbox{H. Masuda\unskip,\iSLAC}
\mbox{E. Mazzucato\unskip,\iFERR}
\mbox{A.K. McKemey\unskip,\iBRUN}
\mbox{B.T. Meadows\unskip,\iCINC}
\mbox{G. Menegatti\unskip,\iFERR}
\mbox{R. Messner\unskip,\iSLAC}
\mbox{P.M. Mockett\unskip,\iWASH}
\mbox{K.C. Moffeit\unskip,\iSLAC}
\mbox{T.B. Moore\unskip,\iYALE}
\mbox{M.Morii\unskip,\iSLAC}
\mbox{D. Muller\unskip,\iSLAC}
\mbox{V.Murzin\unskip,\iMOSCOW}
\mbox{T. Nagamine\unskip,\iTOHO}
\mbox{S. Narita\unskip,\iTOHO}
\mbox{U. Nauenberg\unskip,\iCOLO}
\mbox{H. Neal\unskip,\iSLAC}
\mbox{M. Nussbaum\unskip,\iCINC}
\mbox{N.Oishi\unskip,\iNAGO}
\mbox{D. Onoprienko\unskip,\iTENN}
\mbox{L.S. Osborne\unskip,\iMIT}
\mbox{R.S. Panvini\unskip,\iVAND}
\mbox{H. Park\unskip,\iOREG}
\mbox{C. H. Park\unskip,\iSOONG}
\mbox{T.J. Pavel\unskip,\iSLAC}
\mbox{I. Peruzzi\unskip,\iFRAS}
\mbox{M. Piccolo\unskip,\iFRAS}
\mbox{L. Piemontese\unskip,\iFERR}
\mbox{E. Pieroni\unskip,\iPISA}
\mbox{K.T. Pitts\unskip,\iOREG}
\mbox{R.J. Plano\unskip,\iRUTG}
\mbox{R. Prepost\unskip,\iWISC}
\mbox{C.Y. Prescott\unskip,\iSLAC}
\mbox{G.D. Punkar\unskip,\iSLAC}
\mbox{J. Quigley\unskip,\iMIT}
\mbox{B.N. Ratcliff\unskip,\iSLAC}
\mbox{T.W. Reeves\unskip,\iVAND}
\mbox{J. Reidy\unskip,\iMISSI}
\mbox{P.L. Reinertsen\unskip,\iUCSC}
\mbox{P.E. Rensing\unskip,\iSLAC}
\mbox{L.S. Rochester\unskip,\iSLAC}
\mbox{P.C. Rowson\unskip,\iCOLU}
\mbox{J.J. Russell\unskip,\iSLAC}
\mbox{O.H. Saxton\unskip,\iSLAC}
\mbox{T. Schalk\unskip,\iUCSC}
\mbox{R.H. Schindler\unskip,\iSLAC}
\mbox{B.A. Schumm\unskip,\iUCSC}
\mbox{J. Schwiening\unskip,\iSLAC}
\mbox{S. Sen\unskip,\iYALE}
\mbox{V.V. Serbo\unskip,\iWISC}
\mbox{M.H. Shaevitz\unskip,\iCOLU}
\mbox{J.T. Shank\unskip,\iBU}
\mbox{G. Shapiro\unskip,\iLBL}
\mbox{D.J. Sherden\unskip,\iSLAC}
\mbox{K. D. Shmakov\unskip,\iTENN}
\mbox{C. Simopoulos\unskip,\iSLAC}
\mbox{N.B. Sinev\unskip,\iOREG}
\mbox{S.R. Smith\unskip,\iSLAC}
\mbox{M. B. Smy\unskip,\iCSU}
\mbox{J.A. Snyder\unskip,\iYALE}
\mbox{H. Staengle\unskip,\iCSU}
\mbox{A. Stahl\unskip,\iSLAC}
\mbox{P. Stamer\unskip,\iRUTG}
\mbox{R. Steiner\unskip,\iADEL}
\mbox{H. Steiner\unskip,\iLBL}
\mbox{M.G. Strauss\unskip,\iMASS}
\mbox{D. Su\unskip,\iSLAC}
\mbox{F. Suekane\unskip,\iTOHO}
\mbox{A. Sugiyama\unskip,\iNAGO}
\mbox{S. Suzuki\unskip,\iNAGO}
\mbox{M. Swartz\unskip,\iSLAC}
\mbox{A. Szumilo\unskip,\iWASH}
\mbox{T. Takahashi\unskip,\iSLAC}
\mbox{F.E. Taylor\unskip,\iMIT}
\mbox{J. Thom\unskip,\iSLAC}
\mbox{E. Torrence\unskip,\iMIT}
\mbox{N. K. Toumbas\unskip,\iSLAC}
\mbox{A.I. Trandafir\unskip,\iMASS}
\mbox{J.D. Turk\unskip,\iYALE}
\mbox{T. Usher\unskip,\iSLAC}
\mbox{C. Vannini\unskip,\iPISA}
\mbox{J. Va'vra\unskip,\iSLAC}
\mbox{E. Vella\unskip,\iSLAC}
\mbox{J.P. Venuti\unskip,\iVAND}
\mbox{R. Verdier\unskip,\iMIT}
\mbox{P.G. Verdini\unskip,\iPISA}
\mbox{S.R. Wagner\unskip,\iSLAC}
\mbox{D. L. Wagner\unskip,\iCOLO}
\mbox{A.P. Waite\unskip,\iSLAC}
\mbox{Walston, S.\unskip,\iOREG}
\mbox{J.Wang\unskip,\iSLAC}
\mbox{C. Ward\unskip,\iBRUN}
\mbox{S.J. Watts\unskip,\iBRUN}
\mbox{A.W. Weidemann\unskip,\iTENN}
\mbox{E. R. Weiss\unskip,\iWASH}
\mbox{J.S. Whitaker\unskip,\iBU}
\mbox{S.L. White\unskip,\iTENN}
\mbox{F.J. Wickens\unskip,\iRAL}
\mbox{B. Williams\unskip,\iCOLO}
\mbox{D.C. Williams\unskip,\iMIT}
\mbox{S.H. Williams\unskip,\iSLAC}
\mbox{S. Willocq\unskip,\iSLAC}
\mbox{R.J. Wilson\unskip,\iCSU}
\mbox{W.J. Wisniewski\unskip,\iSLAC}
\mbox{J. L. Wittlin\unskip,\iMASS}
\mbox{M. Woods\unskip,\iSLAC}
\mbox{G.B. Word\unskip,\iVAND}
\mbox{T.R. Wright\unskip,\iWISC}
\mbox{J. Wyss\unskip,\iPADO}
\mbox{R.K. Yamamoto\unskip,\iMIT}
\mbox{J.M. Yamartino\unskip,\iMIT}
\mbox{X. Yang\unskip,\iOREG}
\mbox{J. Yashima\unskip,\iTOHO}
\mbox{S.J. Yellin\unskip,\iUCSB}
\mbox{C.C. Young\unskip,\iSLAC}
\mbox{H. Yuta\unskip,\iAOMORI}
\mbox{G. Zapalac\unskip,\iWISC}
\mbox{R.W. Zdarko\unskip,\iSLAC}
\mbox{J. Zhou\unskip.\iOREG}

\it
  \vskip \baselineskip                   
  \centerline{(The SLD Collaboration)}   
  \vskip \baselineskip        
  \baselineskip=.75\baselineskip   
\iADEL
  Adelphi University,
  South Avenue-   Garden City,NY 11530, \break
\iAOMORI
  Aomori University,
  2-3-1 Kohata, Aomori City, 030 Japan, \break
\iBOLO
  INFN Sezione di Bologna,
  Via Irnerio 46    I-40126 Bologna  (Italy), \break
\iBRUN
  Brunel University,
  Uxbridge, Middlesex - UB8 3PH United Kingdom, \break
\iBU
  Boston University,
  590 Commonwealth Ave. - Boston,MA 02215, \break
\iCINC
  University of Cincinnati,
  Cincinnati,OH 45221, \break
\iCOLO
  University of Colorado,
  Campus Box 390 - Boulder,CO 80309, \break
\iCOLU
  Columbia University,
  Nevis Laboratories  P.O.Box 137 - Irvington,NY 10533, \break
\iCSU
  Colorado State University,
  Ft. Collins,CO 80523, \break
\iFERR
  INFN Sezione di Ferrara,
  Via Paradiso,12 - I-44100 Ferrara (Italy), \break
\iFRAS
  Lab. Nazionali di Frascati,
  Casella Postale 13   I-00044 Frascati (Italy), \break
\iILLI
  University of Illinois,
  1110 West Green St.  Urbana,IL 61801, \break
\iLBL
  Lawrence Berkeley Laboratory,
  Dept.of Physics 50B-5211 University of California-  Berkeley,CA 94720, \break
\iLTU
  Louisiana Technical University,
  , \break
\iMASS
  University of Massachusetts,
  Amherst,MA 01003, \break
\iMISSI
  University of Mississippi,
  University,MS 38677, \break
\iMIT
  Massachusetts Institute of Technology,
  77 Massachussetts Avenue  Cambridge,MA 02139, \break
\iMOSCOW
  Moscow State University,
  Institute of Nuclear Physics  119899 Moscow  Russia, \break
\iNAGO
  Nagoya University,
  Nagoya 464 Japan, \break
\iOREG
  University of Oregon,
  Department of Physics  Eugene,OR 97403, \break
\iOXF
  Oxford University,
  Oxford, OX1 3RH, United Kingdom, \break
\iPADO
  Universita di Padova,
  Via F. Marzolo,8   I-35100 Padova (Italy), \break
\iPERU
  Universita di Perugia, Sezione INFN,
  Via A. Pascoli  I-06100 Perugia (Italy), \break
\iPISA
  INFN, Sezione di Pisa,
  Via Livornese,582/AS  Piero a Grado  I-56010 Pisa (Italy), \break
\iRAL
  Rutherford Appleton Laboratory,
  Chiton,Didcot - Oxon OX11 0QX United Kingdom, \break
\iRUTG
  Rutgers University,
  Serin Physics Labs  Piscataway,NJ 08855-0849, \break
\iSLAC
  Stanford Linear Accelerator Center,
  2575 Sand Hill Road  Menlo Park,CA 94025, \break
\iSOGA
  Sogang University,
  Ricci Hall  Seoul, Korea, \break
\iSOONG
  Soongsil University,
  Dongjakgu Sangdo 5 dong 1-1    Seoul, Korea 156-743, \break
\iTENN
  University of Tennessee,
  401 A.H. Nielsen Physics Blg.  -  Knoxville,Tennessee 37996-1200, \break
\iTOHO
  Tohoku University,
  Bubble Chamber Lab. - Aramaki - Sendai 980 (Japan), \break
\iUCSB
  U.C. Santa Barbara,
  3019 Broida Hall  Santa Barbara,CA 93106, \break
\iUCSC
  U.C. Santa Cruz,
  Santa Cruz,CA 95064, \break
\iVAND
  Vanderbilt University,
  Stevenson Center,Room 5333  P.O.Box 1807,Station B  Nashville,TN 37235,
\break
\iWASH
  University of Washington,
  Seattle,WA 98105, \break
\iWISC
  University of Wisconsin,
  1150 University Avenue  Madison,WS 53706, \break
\iYALE
  Yale University,
  5th Floor Gibbs Lab. - P.O.Box 208121 - New Haven,CT 06520-8121. \break

\rm
\end{center}


\begin{thebibliography}{99}

\bibitem{ert}
See e.g. R.K. Ellis, D.A. Ross, A.E. Terrano, Nucl. Phys. {\bf B178} (1981) 421.

\bibitem{moretti}
S. Moretti, RAL-TR-97-065, hep-ph/9711518.

\bibitem{mlla} T.I.~Azimov, Y.L.~Dokshitzer, V.A.~Khoze and S.I.~Troyan, Z.
Phys. {\bf C27} (1985) 65.

\bibitem{nlla}
G. Marchesini and B.R. Webber, Nucl. Phys. {\bf B238} (1984) 1.

\bibitem{glip}
V.N. Gribov and L.N. Lipatov, Sov. J. Nucl. Phys. {\bf 15}~(1973)~675.

\bibitem{saxon}
D.H. Saxon, {\it High Energy Electron-Positron Physics}, Eds. A. Ali and P.
S\"oding, World Scientific (1988), p. 539.

\bibitem{bohrer}
A. B\"ohrer, Phys. Rep. {\bf 291} (1997) 107.

\bibitem{lpprl} SLD Collab., K. Abe et al.,
Phys. Rev. Lett. {\bf 78} (1997) 3442.

\bibitem{herwig}
G. Marchesini et al., Comp. Phys. Comm. {\bf 67}~(1992)~465.

\bibitem{jetset74}
T. Sj\"ostrand, Comp. Phys. Comm. {\bf 82}~(1994)~74.

\bibitem{ucla}
S. Chun and C. Buchanan, Phys. Rep. {\bf 292} (1998) 239.

\bibitem{sld} SLD Design Report, SLAC-Report 273 (1984).

\bibitem{cdc}
M.D. Hildreth et al., Nucl. Inst. Meth. {\bf A367} (1995) 111.

\bibitem{vxd}
C. J. S. Damerell et al., Nucl. Inst. Meth. {\bf A288}~(1990)~236.

\bibitem{crid}
K. Abe et al., Nucl. Inst. Meth. {\bf A343} (1994) 74.

\bibitem{lac}
D. Axen et al., Nucl. Inst. Meth. {\bf A238} (1993) 472.

\bibitem{thrust}
S. Brandt et al., Phys. Lett. {\bf 12}~(1964)~57;\\
E. Farhi, Phys. Rev. Lett. {\bf 39}~(1977)~1587.

\bibitem{homer}
SLD Collab., K.~Abe et al., Phys. Rev. {\bf D53} (1996) 1023.

\bibitem{davea}
K. Abe et al., Nucl. Inst. and Meth. {\bf A371} (1996) 195.

\bibitem{tomp}
T.J. Pavel, Ph.D. Thesis, Stanford University, January 1997;
SLAC-Report-495.

\bibitem{alr} SLD Collab., K. Abe et al.,
Phys. Rev. Lett. {\bf 73} (1994) 25.

\bibitem{delphi}
DELPHI Collab., P.~Abreu et al., Nucl. Phys. {\bf B444} (1995) 3.

\bibitem{opal}
OPAL Collab., P.D.~Acton et al., Z. Phys. {\bf C63}~(1994)~181.

\bibitem{aleph}
ALEPH Collab., D.~Buskulic et al., Z. Phys. {\bf C66} (1995) 355.

\bibitem {dcone}
SLD Collab., K. Abe et al., Phys. Rev. Lett. {\bf 72}
(1994) 3145.

\bibitem{kenb}
K.G. Baird, Ph.D. Thesis, Rutgers University, December 1995;
SLAC-Report-95-483.

\bibitem{pdg}
Particle Data Group, Phys. Rev. {\bf D54} (1996) 1.

\bibitem{mihaid}
M.O. Dima, Ph.D. Thesis, Colorado State University, March 1997;
SLAC-Report-505.

\bibitem{alephkst}
ALEPH Collab., D.~Buskulic et al., Z. Phys. {\bf C69} (1996) 379.

\bibitem{mikeh}
SLD Collab., K.~Abe et al., Phys. Rev. {\bf D53} (1996) 2271.

\bibitem{nchflav}
SLD Collab., K.~Abe et al., Phys. Lett. {\bf B386} (1996) 475.

\bibitem{bcorrl}
See e.g. DELPHI Collab., P.~Abreu et al., Phys. Lett. {\bf B416} (1998) 247.

\bibitem{lafferty}
G.~D.~Lafferty, Phys. Lett. {\bf B353} (1995) 541.


\end{thebibliography}
\end{document}